\documentclass[10pt,aps,prd,twocolumn,groupedaddress,showpacs]{revtex4-1}
\usepackage{graphicx}
\usepackage{amsmath}
\begin{document}
\title{Gravitational wave background from binary systems}
\author{Pablo A. \surname{Rosado}}
\email[]{pablo.rosado@aei.mpg.de}
\affiliation{Albert Einstein Institute, Max Planck Institute for Gravitational Physics, 30167 Hanover, Germany}
\date{\today}
\begin{abstract}
Basic aspects of the background of gravitational waves and its mathematical characterization are reviewed.
The spectral energy density parameter $\Omega(f)$, commonly used as a quantifier of the background,
is derived for an ensemble of many identical sources emitting at different times and locations.
For such an ensemble, $\Omega(f)$ is generalized to account for the duration of the signals and of the observation,
so that one can distinguish the resolvable and unresolvable parts of the background.
The unresolvable part, often called confusion noise or stochastic background, is made by signals that cannot be either individually identified or subtracted out of the data.
To account for the resolvability of the background, the overlap function is introduced.
This function is a generalization of the duty cycle, which has been commonly used in the literature, in some cases leading to incorrect results.
The spectra produced by binary systems (stellar binaries and massive black hole binaries) are presented over the frequencies of all existing and planned detectors.
A semi-analytical formula for $\Omega(f)$ is derived in the case of stellar binaries (containing white dwarfs, neutron stars or stellar-mass black holes).
Besides a realistic expectation of the level of background, upper and lower limits are given,
to account for the uncertainties in some astrophysical parameters such as binary coalescence rates.
One interesting result concerns all current and planned ground-based detectors (including the Einstein Telescope).
In their frequency range, the background of binaries is resolvable and only sporadically present.
In other words, there is no stochastic background of binaries for ground-based detectors.
\end{abstract}

\maketitle

\newcommand{\om}{\Omega(f)}
\newcommand{\omov}{\Omega(f,\Delta f,\mathcal{N}_0)}
\section{\label{sec:intro} Introduction}
The gravitational wave background \cite{AllenRomano1999,Maggiore2000} is formed by a large number of independent gravitational wave sources.
This paper focuses on the background produced by coalescing binary systems.
These are isolated pairs of massive objects that inspiral towards each other by emitting gravitational radiation until they coalesce.

We review the characterization of the background, for which the spectral energy density parameter, or simply \textit{spectral function}, $\om$, is often used.
This function gives the present fractional energy density (per logarithmic frequency interval) of gravitational radiation at an observed frequency $f$.
A formula for $\om$ is obtained in a clear, self-consistent way, for an ensemble of many identical sources emitting at different times and locations.
This formula is generalized to distinguish whether the signals are resolvable or not, or whether they are observed continuously or sporadically.

The \textit{resolvability} of the signals is an important topic of this work.
Roughly speaking, signals are unresolvable if their waveforms are observed simultaneously at similar frequencies (differing less than the frequency resolution).
Many unresolvable signals form an unresolvable background.
If such a background dominates in a certain frequency interval, one cannot see the waveforms of its components, but a pattern that rather looks like instrumental noise.
For that reason it is often called \textit{confusion noise}.
The other part of the background is \textit{resolvable}.
The waveforms of the resolvable part can be distinguished and in some circumstances subtracted out of the data of a detector \cite{BarackCutler2004,CutlerHarms2006}.

There are many studies in the literature about astrophysical sources that contribute to the background at present.
A few examples of these sources are:
core-collapse supernovae \cite{FerrariEtAl1999},
rotating neutron stars \cite{FerrariEtAl1999b,RegimbauFreitas2001},
formation of neutron stars \cite{CowardEtAl2001,CowardEtAl2002},
inspiralling or coalescing stellar binaries \cite{SchneiderEtAl2001,FarmerPhinney2003,RegimbauFreitas2006,RegimbauMandic2008,RegimbauHughes2009},
inspiralling or coalescing massive black hole binaries \cite{SesanaEtAl2005,SesanaEtAl2008}
and magnetars \cite{RegimbauFreitas2006b}.
But there are inconsistencies in the literature;
for example, according to \cite{AraujoMiranda2005},
the results of some of the previous papers \cite{CowardEtAl2001,CowardEtAl2002} (and also \cite{AraujoEtAl2004,RegimbauMandic2008}, as pointed out in \cite{ZhuEtAl2011b}) are incorrect,
due to a wrong $(1+z)$ factor in the calculations.
Besides, according to \cite{GrishchukEtAl2001}, the definition of the spectral function used in many papers is misleading or misinterpreted.
Finally, as we discuss further on, in some of the mentioned papers, the continuous and unresolvable backgrounds are not properly defined.
To avoid possible misunderstandings or mistakes we tend to present all calculations and definitions as clearly as possible.

We calculate the contributions to the background of the strongest emitting binary systems.
These are the ones composed of white dwarfs, neutron stars, stellar-mass black holes or massive black holes.
The resulting energy spectra are given as maximum, most likely, and minimum expectations, taking into account the present uncertainties in the quantities involved.

We show that ground-based detectors do not encounter any unresolvable background in their \textit{frequency window}
(the frequency range in which they reach their optimal sensitivity).
This applies to present detectors, such as TAMA300 \cite{AndoEtAl2001}, GEO600 \cite{LueckEtAl1997}, Virgo \cite{CaronEtAl1997}, and LIGO \cite{LIGO2009},
but also to planned detectors, such as the advanced versions of LIGO and Virgo \cite{LIGO2009,FlaminioEtAl2005}, LCGT \cite{KurodaEtAl2006}, and ET \cite{PunturoEtAl2010}.
At these frequencies, there is not even a resolvable continuous background, i.e.,
signals are not always present.
Whether or not these signals can be subtracted out of the data is an issue we do not deal with in this paper.

This paper considers a frequency range wider than the frequency windows of ground-based detectors.
It includes the windows of all existing and future detectors,
such as LISA \cite{DanzmannEtAl1996} or BBO \cite{PhinneyEtAl2004},
and also reaches the frequency range of interest for the PTA \cite{HobbsEtAl2010}.

The obtained unresolvable background turns out to be dominated by white dwarf binaries (at frequencies below $\sim0.1$\,Hz)
and by massive black hole binaries (below $\sim 10^{-4}$\,Hz).
This confusion noise could enter the band of LISA and would certainly enter the band of BBO and the complete Parkes PTA \cite{Manchester2008}.


The outline of the paper is as follows:

In Section \ref{sec:bg} we first explain our notation and give some relevant terminology.
We then give a simple heuristic proof of the fact that there is neither an unresolvable nor a continuous background in the frequency window of ground-based detectors.
The formula of the spectral function is derived for an ensemble of many identical sources emitting at different times and locations.
The obtained formula agrees with that of \cite{Phinney2001}.
The concept of resolvability is studied, and the spectral function is generalized to account for it.
To get to that point, we introduce the \textit{overlap function} $\mathcal{N}(f,\Delta f,z)$.
This function gives the average number of signals observed with redshifts smaller than $z$ and frequencies between $f$ and $f+\Delta f$,
where $\Delta f$ is the frequency resolution.
We then use the overlap function to distinguish the continuous and discontinuous parts of the background.

In Section \ref{sec:models} we present the models used to quantify the background of stellar binaries and massive black hole binaries.
The main physical quantities involved in the calculations (such as mass ranges and coalescence rates) are presented in this section.
A semi-analytical formula for the spectral function is derived in the case of stellar binaries.

Section \ref{sec:results} contains the main results of the paper.
The spectral function is shown in the different regimes of resolvability and continuity.
The curves in the plots are given as maximum, most likely, and minimum expected.

In Section \ref{sec:discussion} we justify some of the approximations and assumptions of the previous sections.
We compare our results with others from the literature.
Our notions of continuous background and unresolvable background are compared with the ones of previous work.
We also show that the overlap function, which turns out to be a generalization of the duty cycle,
is a proper quantifier of the resolvability and continuity of the background.

The main conclusions and results are summarized in Section \ref{sec:summary}.
Those readers who are short of time, or primarily interested in the main results, are suggested to go directly to this section.

\section{\label{sec:bg}Characterization of the background}
\subsection{\label{sec:notation} Notation}
All magnitudes are measured in the frame of the cosmological fluid,
since massive objects that are not subject to external forces come quickly to rest with respect to this frame.

We use the index ``$e$'' (for \textit{emission}) for quantities measured close to the system at the time of the emission of the radiation.
For example, $f_e$ (emitted frequency) is the frequency of a wave, measured soon after its emission, before the expansion of the universe stretches its wavelength.
Quantities measured here and now (which are called \textit{observed} quantities) have no index.
The frequency of the wave of the previous example, measured today, is thus denoted by $f$.

We use the indices ``low'', ``upp'', ``min'' and ``max'' to refer to \textit{lower}, \textit{upper}, \textit{minimum} and \textit{maximum}, respectively.

\subsection{\label{sec:definitions} Terminology}
We now introduce some terminology to avoid confusion or ambiguity throughout the paper.

By \textit{system} we mean a certain configuration of physical objects that is a source of gravitational waves.
An example of system is a pair of neutron stars inspiralling towards each other.

We use the term \textit{ensemble} for the collection of systems, all having similar properties and behaviour, formed from the Big Bang until now.
An example is the population of coalescing neutron star binaries in the universe.

By \textit{signal} we refer to the total gravitational radiation emitted by a system.
One system emits only one signal, that can range over a large frequency interval and exist over a long interval of time.
Despite the interval of time it lasts, a signal is assumed to be characterized by a certain redshift, which remains the same from the beginning until the end of the signal
(in Section \ref{sec:minfreq} we discuss the validity of this assumption).
A signal is composed of \textit{signal elements}, each characterized by a certain infinitesimal frequency interval.

The \textit{total (gravitational wave) background} is the collection of all gravitational waves in the present universe.
It can be divided into different parts, according to different criteria.
For example, \textit{primordial} and \textit{contemporary} parts,
\textit{resolvable} and \textit{unresolvable} parts or \textit{continuous} and \textit{discontinuous} parts.
One can also divide the total background into many different parts, each conformed by the contribution of a certain ensemble.
By extension, we use the word \textit{background} when referring to both the total background and to its different parts.
Hence, we can talk about the background of neutron star binaries, which is the collection of signals of the ensemble of neutron star binaries.
The part of this background that fulfills the condition of unresolvability would be the unresolvable background of neutron star binaries.

\subsubsection{Primordial versus contemporary background}
The \textit{primordial background} \cite{Allen1996,Maggiore2000} is composed of gravitational radiation emitted in the early universe, at very large redshifts.
It is analogous to the background formed by the cosmic electromagnetic radiation
\cite{[{First discovery: }]PenziasWilson1965,*[{ COBE mission: }] BoggessEtAl1992, *[{WMAP mission: }]BennettEtAl2003}.
In the case of the latter, the radiation was released (when photons decoupled from matter) roughly a hundred thousand years after the Big Bang.
On the other hand, the primordial gravitational radiation was produced in a tiny fraction of the first second of the universe
\cite{[{}][{, Section 9.4.3 (d) by K. S. Thorne and references therein.}]HawkingIsrael1987}.
In this background might be hidden waves from inflation and cosmic strings \cite{Allen1988,Vilenkin1985}.

The other part of the total background is still being produced at present and thus we refer to it as the \textit{contemporary background}.
It is made by many different systems that formed in the past (at redshifts less than $\sim20$, which is the largest redshift assumed for the systems we study)
and can also form today.
Examples of such systems are coalescing binaries, rapidly-rotating compact objects or core-collapse supernovae
(some references were given in Section \ref{sec:intro}).

In certain frequency ranges one can get a clear view of primordial signals, whereas in others the contemporary signals dominate.
The detection of the primordial background would be the most direct way to observe processes of the very early universe.
But valuable information would also be gained from the detection of the contemporary background, for example about binary formation and coalescence rates.
Furthermore, predictions of the contemporary background can set constraints on the frequency ranges where the primordial one could be detected.
The contemporary background is the main topic of this paper.

In the literature, primordial and contemporary backgrounds are often called \textit{cosmological} and \textit{astrophysical}, respectively.
We do not use this words to avoid ambiguity, since sometimes both terminologies are used together,
for example when talking about \textit{cosmological populations of astrophysical sources} \cite{SchneiderEtAl1999}, which might be confusing for non-specialized readers.

\subsubsection{\label{sec:defunres} Unresolvable versus resolvable background}
It is useful to classify the components of the background depending on their resolvability.
We now briefly comment on this concept;
precise definitions of what we mean by resolvable and unresolvable backgrounds can be found in Section \ref{sec:resolvability}.

Signals spend different intervals of time at different ranges of frequencies.
In the case of binaries, they evolve much more rapidly at higher than at lower frequencies.
At lower frequencies they will thus overlap (i.e., they will be observed at the same time) more frequently than at higher ones.
A frequency bin of width $\Delta f$, which is the frequency resolution allowed by the detector and by the data analysis method,
will often be filled by one or more signals at low frequencies.
On the other hand, a frequency bin at high frequencies will be filled by one or more signals only sporadically, since signals are very short.

An unresolvable part of the background exists as soon as a frequency bin is constantly occupied by an average of one or more signals.
At frequencies where such a background dominates, the waveforms of the signals cannot be distinguished from each other.
When a waveform is not resolvable, one cannot obtain information from it, such as the characteristics of the system that emitted that radiation.
Moreover, such waveforms cannot be subtracted out from the data.

The rest of the background is the resolvable part.
The waveforms of this part can be distinguished from each other.
One can thus obtain information about the system by studying the waveform of the emitted radiation.

For some authors, what we call the unresolvable background is defined as the \textit{stochastic background},
and the remaining gravitational radiation is called the \textit{total gravitational wave signal} \cite{SesanaEtAl2008}.
This is a reasonable definition, but conflicts with what is often called \textit{stochastic background} by many other authors
(for example in \cite{SchneiderEtAl2010} and other papers cited in Section \ref{sec:intro}).
A more precise definition for \textit{stochastic background} can be found in \cite{AllenRomano1999}.

\subsubsection{\label{sec:defcont} Continuous versus discontinuous background}
We now briefly comment on the concept of continuity of the background.
In Section \ref{sec:continuity} we give precise definitions of what we mean by continuous and discontinuous backgrounds.

A continuous background exists in a frequency interval $[f_{\text{low}},f_{\text{upp}}]$ (that can be, for example, the frequency window of a detector)
as soon as this interval is constantly occupied by one or more signals.
If in that interval there are gaps between signals, or the signals occur sporadically, the background is discontinuous.

The condition of continuity tends to that of unresolvability when $f_{\text{upp}}-f_{\text{low}}$ tends to $\Delta f$.
If a background is not continuous in an interval of frequencies, it is not continuous either in a smaller interval.
Therefore, only a continuous background can be unresolvable.

We point out that the continuity of the background is not as relevant as the resolvability.
However, we include it in the paper for two reasons:

First, the continuity has been often used in the literature (for example in the already mentioned papers \cite{RegimbauMandic2008,RegimbauHughes2009})
to define the different regimes of the background.
Once we know how to account for the continuity, we will realize that it is not the right tool to be used.
Instead, the resolvability is the fundamental property of the background.

Second, the continuity can be used to determine how often the background is observed.
Suppose we want to detect a signal of some kind, but there is a background covering the signal.
If the background is discontinuous in a frequency band, sometimes that signal can be clearly seen, without any background.
On the other hand, if the background is continuous, we need to subtract it from the data in order to see that other signal.
As we show in Figure \ref{fig:plotsjointcont}, ET has no continuous background from binaries in its frequency window;
BBO, on the contrary, has a continuous background of binaries crossing its frequency window,
so the subtraction of background signals is necessary in order to detect other kinds of signals (this problem has been treated in \cite{CutlerHarms2006}).

\subsection{\label{sec:heuristic} Heuristic proof of the lack of confusion noise for ground-based detectors}
We now justify, in a simple heuristic way, that there is no continuous background (and therefore no unresolvable background either)
from binary systems at frequencies larger than 10\,Hz.

A neutron star binary takes $\sim 10^3$\,s to evolve from 10\,Hz to the coalescence (which can be proved by using Equation (\ref{eq:timefreq})).
The most realistic coalescence rate for these binaries (see Table \ref{tb:values}) is of $\sim10^5$\,yr$^{-1}$, in the whole observable universe.
This implies $\sim0.003$ coalescences per second.
One could naively say that, on average, one would see $\sim10^3\times 0.003=3$ signals.
But that would only be true if all binaries were close to us, at redshift $\sim 0$.
The farthest binaries (close to redshift $\sim 5$) that we observe today at frequency $\sim 10$\,Hz,
emitted at $\sim 10\times (1+z)=60$\,Hz (using Equation (\ref{eq:fredshift})) and needed just $\sim 8$\,s to coalesce.
An interval of time of $\sim 8$\,s at redshift $\sim 5$ is now observed as an interval of $\sim 8\times (1+z)=48$\,s (using Equation (\ref{eq:timed})).
This implies that an average of $\sim 48\times 0.003\approx0.14$ signals are observed.
The number of signals expected to be observed is thus a number between 3 and 0.14, which, after doing the proper calculation, turns out to be smaller than 1.
Hence, neutron star binaries do not produce a continuous background at frequencies higher than 10\,Hz.

Other binaries whose signals could produce a continuous background in the frequency window of ground-based detectors are those containing a stellar-mass black hole.
But these binaries have a smaller coalescence rate and need less time to coalesce, from an initial frequency of 10\,Hz.
The product \textit{coalescence rate}$\times$\textit{duration of signal} would thus be even smaller.
Therefore they do not produce a continuous background either.

At frequencies larger than 10\,Hz, hence, there is no continuous background from binary systems.
Between 1 and 10\,Hz one could have a continuous background, but it turns out to be well below the realistic sensitivity of a ground-based detector
(see Figure \ref{fig:plotsjointcont}).

\subsection{\label{sec:cosmological} Cosmological model}
\subsubsection{Metric}
We assume a spatially flat, homogeneous and isotropic universe, described by a Friedmann-Robertson-Walker metric,
\begin{equation}
\label{eq:metric}
ds^2=-c^2dt_e^2+a^2(t_e)\left[dr^2+r^2[d\theta^2+\sin^2 (\theta) \, d \phi^2]\right],
\end{equation} 
where $c$ is the speed of light.
The time coordinate $t_e$ is chosen to be, for convenience, the look-back time:
it is 0 at present and $t_0\approx$13.7\,Gyr at the Big Bang.
The usual \textit{look-forward} time would be just $t'=t_0-t_e$, with which the form of the metric would not change.
The dimensionless cosmological scale factor, $a(t_e)$, is chosen to be $a(0)=1$ at present.
The coordinates $r$, $\theta$ and $\phi$ are called comoving coordinates, because they \textit{move} with the cosmological fluid.
For example, two objects at rest with respect to the fluid, at positions $r_1$ and $r_2$ (and equal values of $\theta$ and $\phi$),
have a \textit{comoving distance} $r=r_2-r_1$.
This comoving distance remains the same at every future time.
However, the physical (proper) distance between them is $r_{\text{phys}}(t_e)=a(t_e)r$, and changes with time as the universe expands.
Setting $r=0$ at the Earth, the coordinate $r$ of a distant object is its comoving distance from us.

\subsubsection{\label{sec:redshifting} Redshifting}
The definition of the cosmological redshift $z$ is given by
\begin{equation}
\label{eq:1+z}
1+z=\frac{a(0)}{a(t_e)},
\end{equation}
where, as already said, $a(0)=1$.
This equation gives the value of the scale factor at the time of emission of a graviton (or a photon) that is today observed with a redshift $z$.

We now see how the expansion of the universe affects frequencies and energies of gravitational waves, as well as infinitesimal intervals of time
(a derivation can be found in Section 4.1.4 of \cite{Maggiore2008}).
A frequency $f_e$ of a wave emitted by a system at a redshift $z$ corresponds to
\begin{equation}
\label{eq:fredshift}
f=\frac{f_e}{1+z}
\end{equation} 
at the present time.
Since the energy of a graviton is proportional to its frequency, a graviton emitted with an energy $E_e$ is today observed with
\begin{equation}
\label{eq:eredshift}
E=\frac{E_e}{1+z}.
\end{equation} 
An infinitesimal lapse of time $dt_e$ (emitted interval of time) measured at redshift $z$ is today observed as
\begin{equation}
\label{eq:timed}
dt=dt_e [1+z].
\end{equation}
From (\ref{eq:fredshift}), it follows that an infinitesimal frequency interval $df_e$ emitted at redshift $z$ is today observed as
\begin{equation}
\label{eq:dfdfe}
df=\frac{df_e}{1+z}.
\end{equation} 
Similarly, from (\ref{eq:eredshift}), an infinitesimal energy interval $dE_e$ corresponds to
\begin{equation}
\label{eq:dedee}
dE=\frac{dE_e}{1+z}
\end{equation} 
today.

\subsubsection{Volumes}
Some important quantities in this work are defined as densities, i.e. per unit volume (by which we mean the spatial volume).
Because of the expansion, it is convenient to speak of two different volumes: \textit{physical} and \textit{comoving} volume.

The element of physical volume $d\mathcal{V}$ at fixed time $t_e$ in the metric (\ref{eq:metric}) is given by $a^3(t_e)\,r^2\,\sin (\theta)\,dr\,d\theta\,d\phi$.
We consider only sources uniformly distributed in the sky, so we can integrate for all angles $\theta$ and $\phi$, obtaining
\begin{equation}
\label{eq:dv4pi}
d\mathcal{V}=4\pi a^3(t_e)r^2 dr.
\end{equation} 
The factor $a^3(t_e)$ accounts for the expansion in the three space dimensions.

The element of comoving volume $d\mathcal{V}_c$ is defined by $d\mathcal{V}_c=a^{-3}(t_e)d\mathcal{V}$, which, using (\ref{eq:dv4pi}), gives
\begin{equation}
\label{eq:dv4pir2dr}
d\mathcal{V}_c=4\pi r^2 dr.
\end{equation} 
Suppose eight galaxies (that, at large scales, can be thought as point-like) are placed at the vertices of a cube.
With the expansion, the galaxies separate from each other and the physical volume of the cube increases,
but its comoving volume remains always the same.
Since we are assuming that all massive objects are at rest with respect to the fluid, no system enters or leaves a certain comoving volume.
For this reason it is straightforward to measure densities (for example, the number density of systems) per unit comoving volume.

It is useful to write the element of comoving volume in terms of redshifts, instead of distances.
For that, we have to find a way to transform infinitesimal intervals of comoving distance $dr$ into infinitesimal intervals of redshift $dz$.
Suppose we observe today two gravitons, one emitted at redshift $z$ and the other at $z+dz$.
Since both reach us at the same time, and both travel at the same velocity $c$ with respect to the cosmological fluid,
the one with larger redshift was emitted at a time $dt_e$ before the other,
and thus at a comoving distance $dr$ further away from us than the other.
The path of the gravitons, moving in a radial direction ($d\theta=d\phi=0$), is obtained by setting $ds^2=0$ in (\ref{eq:metric}), which gives
\begin{equation}
\label{eq:drcatdt}
dr=\frac{c}{a(t_e)}dt_e.
\end{equation} 
To write $dt_e$ in terms of redshifts, we use the definition of the redshift.
One can differentiate Equation (\ref{eq:1+z}) with respect to $t_e$, obtaining $dz/dt_e=-\dot{a}(t_e)/a^2(t_e)$.
Using (\ref{eq:1+z}) again and the definition of the Hubble expansion rate, $H(t_e)=-\dot{a}(t_e)/a(t_e)$ (where the minus sign appears because of the use of a look-back time),
one obtains
\begin{equation}
\label{eq:dtedzeq}
dt_e=\frac{1}{[1+z]H(z)}dz.
\end{equation} 
Here, the Hubble expansion rate has been written as a function of the redshift, instead of the time.
The form of $H(z)$ is derived further on in this section.
Introducing (\ref{eq:dtedzeq}) in (\ref{eq:drcatdt}), we obtain a relationship between $dr$ and $dz$,
\begin{equation}
\label{eq:dr-cat}
dr=\frac{c}{a(t_e)}\frac{1}{[1+z]H(z)}dz=\frac{c}{H(z)}dz,
\end{equation} 
where the terms $a(t_e)$ and $[1+z]$ have canceled out, using (\ref{eq:1+z}).
Finally, inserting (\ref{eq:dr-cat}) in (\ref{eq:dv4pir2dr}), the element of comoving volume becomes
\begin{equation}
\label{eq:defcomvol}
d\mathcal{V}_c=4\pi r^2(z) \frac{c}{H(z)}dz.
\end{equation} 
Here, $r(z)$ is obtained by integrating (\ref{eq:dr-cat}),
\begin{equation}
r(z)=\int_0^z \frac{c}{H(z)}dz.
\end{equation} 
Gravitons emitted between redshift $z$ and $z+dz$ define a shell of comoving volume given by $d\mathcal{V}_c$.

The Hubble expansion rate can be written as a function of the redshift.
For that, we use the Friedmann equation (see Chapter 27 of \cite{MisnerEtAl1973}),
\begin{equation}
\label{eq:friedmann}
H^2(t_e)=\frac{8 \pi G}{3} \rho(t_e) -\frac{kc^2}{a^2(t_e)}+\frac{\Lambda}{3},
\end{equation}
where $G$ and $\Lambda$ are the gravitational and cosmological constants, respectively.
This equation is obtained from the Einstein equation,
imposing the metric (\ref{eq:metric}) and the stress-energy tensor of a perfect fluid (see Chapter 5 of \cite{MisnerEtAl1973}).
We assume a spatially flat universe, which means with zero curvature, $k=0$.
The term $\rho(t_e)$ is obtained from the equation of a perfect fluid of density $\rho$ and pressure $p$ (which is also obtained from the Einstein equation),
\begin{equation}
\label{eq:perfectfluid}
\dot{\rho}-3H(t_e)\left[ \rho(t_e) +\frac{p(t_e)}{c^2}\right]=0.
\end{equation} 
We can solve this equation considering a universe dominated by non-relativistic matter (also called dust),
$\rho(t_e)=\rho_m(t_e)$, with the equation of state $p_m=0$.
One obtains $\rho_m(t_e)\propto a^{-3}(t_e)$, which, using (\ref{eq:1+z}), becomes
\begin{equation}
\label{eq:rhomz}
\rho_m(z)=\rho_m^0 (1+z)^3,
\end{equation}
where $\rho_m^0$ is the current value of the density of matter.
One can also solve (\ref{eq:perfectfluid}) using the equation of state of relativistic matter (radiation), $p_r=\rho c^2/3$.
But one can prove that the resulting density, $\rho_r(z)=\rho_r^0(1+z)^4$, dominates in the Friedmann equation only at very large redshifts.
The redshift at which both densities, $\rho_m(z)$ and $\rho_r(z)$ equate is $z_{eq}\approx 3\times 10^3$ (from \cite{JarosikEtAl2011}).
Considering the redshifts involved in this work ($z<20$) we neglect $\rho_r(z)$ compared to $\rho_m(z)$.
Inserting (\ref{eq:rhomz}) in (\ref{eq:friedmann}) and rewriting the latter in terms of the present value of the Hubble expansion rate,
$H_0=[74.2\pm 3.6]$\,km\,s$^{-1}$\,Mpc$^{-1}$ (from \cite{RiessEtAl2009}),
\begin{equation}
\label{eq:defhubblez}
H(z)=H_0\mathcal{E}(z),
\end{equation}
where
\begin{equation}
\label{eq:mathcalez}
\mathcal{E}(z)=\sqrt{\Omega_m[1+z]^3+\Omega_\Lambda}.
\end{equation}
Here, 
\begin{equation}
\Omega_m=\frac{8\pi G \rho_m^0}{3 H_0^2} \quad \text{and} \quad \Omega_\Lambda=\frac{\Lambda}{3H_0^2}
\end{equation} 
are two dimensionless quantities called the density parameters of matter and dark energy, respectively.
The most recent values for the cosmological parameters obtained by the Wilkinson Microwave Anisotropy Probe
after seven years of measurements are given in \cite{JarosikEtAl2011}.
We adopt a density parameter of matter $\Omega_m=0.27$ and of dark energy $\Omega_\Lambda=0.73$.
For simplicity we do not consider any uncertainty in these values.

\begin{figure}
\includegraphics{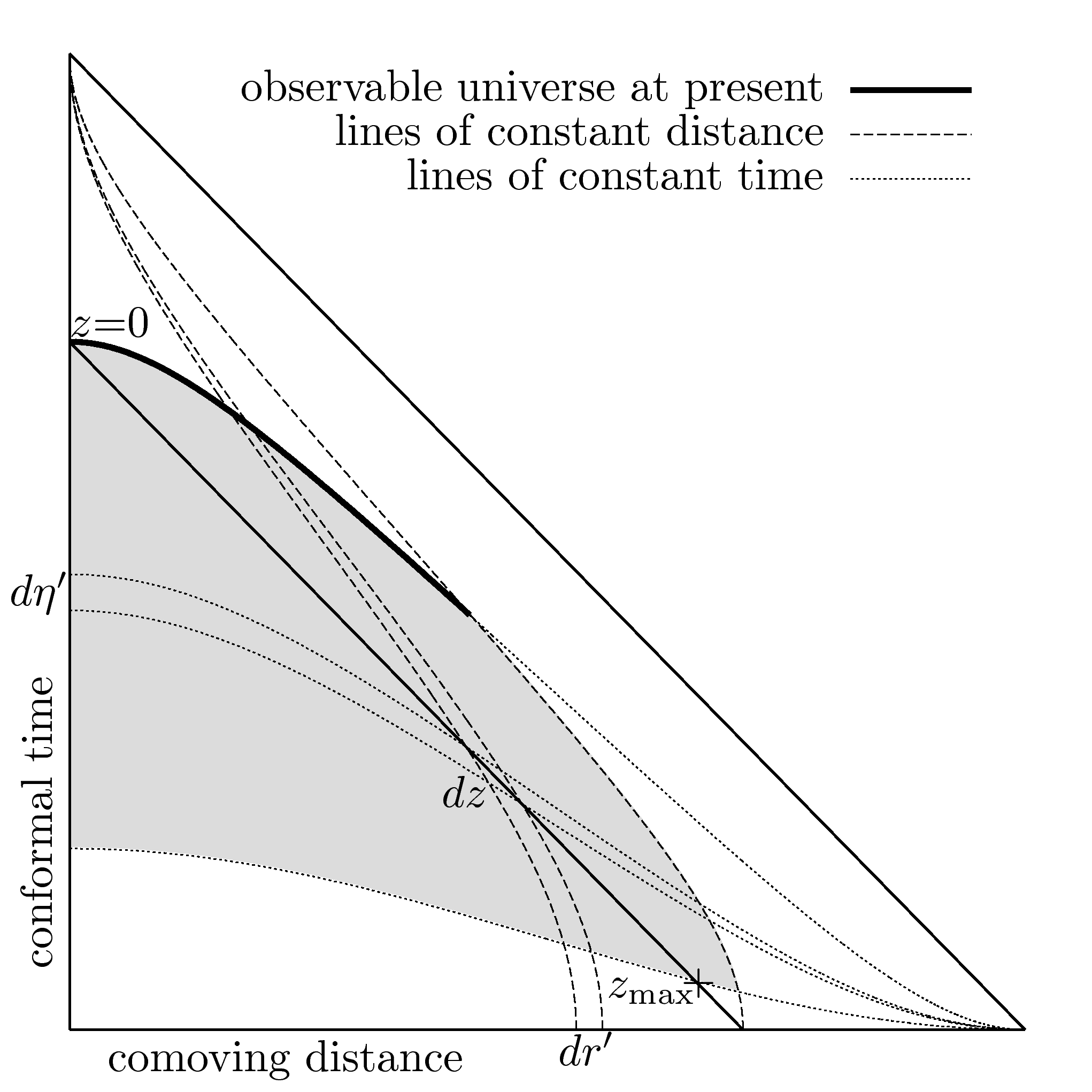}
\caption{Penrose diagram of a universe described by the metric (\ref{eq:metric}).
The straight black line crossing $z=0$ and $z=z_{\text{max}}$ contains all the gravitons that we observe today.}
\label{fig:penrosez}
\end{figure}

For a better understanding of the relationship between volumes and redshifts, we can see Figure \ref{fig:penrosez},
where a Penrose diagram \cite{HawkingEllis1973} for the metric (\ref{eq:metric}) is shown.
Each point of the diagram represents a two-sphere at a certain \textit{conformal time}.
The (look-forward) conformal time is defined by
\begin{equation}
\label{eq:conformaltime}
d\eta = -[1+z] dt_e.
\end{equation} 
The coordinates of Figure \ref{fig:penrosez} are defined by
\begin{equation}
\left\{ \begin{array}{lc}
	   r' =\arctan (\eta +r)-\arctan (\eta -r)\\
	   \eta' = \arctan (\eta +r) + \arctan (\eta -r)
        \end{array}
\right. .
\end{equation} 
Introducing (\ref{eq:conformaltime}) in (\ref{eq:metric}), the path of a graviton fulfills $dr=c\,d\eta$.
In the diagram we use $c=1$, so that $r=\eta$ and thus $r' = \eta'$ for all null paths.
All gravitons that reach us today have traveled along the null path shown (the straight solid line connecting $z=0$ and $z_{\text{max}}$).
This path cuts the horizontal axis at the moment of the Big Bang, fixing the horizon of our observable universe.
For each infinitesimal interval of time $d\eta$ (that describes the difference between the emission of two gravitons that reach us today)
there is a corresponding interval $dz$, along the null path, which represents a shell of infinitesimal comoving volume $d\mathcal{V}_c$.

\subsection{\label{sec:spectralfunction} Spectral function}
Under the assumptions (discussed in \cite{AllenRomano1999}) that the background is Gaussian, stationary, isotropic and unpolarized,
all the information about it is contained in a dimensionless function called the spectral function, defined by
\begin{equation}
\label{eq:defomega}
\om=\frac{\rho_{\ln}(f)}{\rho_c}=\frac{\varepsilon_{\ln}(f)}{c^2\rho_c},
\end{equation}
where $c$ is the speed of light and $f$ is the observed gravitational wave frequency.
The present critical density of the universe is
\begin{equation}
\rho_c=\frac{3H_0^2}{8 \pi G}.
\end{equation} 
This is the density that closes a universe with zero cosmological constant.
This means that $\rho_c$ is the density that, inserted in Equation (\ref{eq:friedmann}) (using $\Omega_\Lambda=0$), gives a zero curvature ($k=0$) at present ($t_e=0$).
The function $\varepsilon_{\ln}(f)$ is defined in such a way that the total energy density of gravitational waves in the present universe is
\begin{equation}
\label{eq:totalenerdens}
\varepsilon_T=\int \varepsilon_{\ln}(f) d \ln f.
\end{equation} 
In the literature, $\varepsilon_{\ln}(f)$ is often written as
\begin{equation}
\label{eq:defepsilonln}
\varepsilon_{\ln}(f)=\frac{d\varepsilon}{d\ln f}.
\end{equation}
Some other authors \cite{GrishchukEtAl2001,[{}][{, section by L. P. Grishchuk in Part III.}] LaemmerzahlEtAl2001} prefer not to use this notation,
arguing that it may lead to misunderstandings (one could mistakenly think that the energy density is differentiated).
Regardless of the notation, Equation (\ref{eq:totalenerdens}) must be fulfilled,
so that $\varepsilon_{\ln}(f)d\ln f$ is the energy per unit physical volume of gravitational waves between $\ln f$ and $\ln f+d\ln f$.
Thus, $\om$ is the fractional energy density of gravitational radiation, per logarithmic frequency interval, in the present universe.

We first describe a system as seen by an observer close to it at the time of emission.
The energy released in gravitational radiation between logarithmic frequencies $\ln f_e$ and $\ln f_e+d\ln f_e$ is
\begin{equation}
\label{eq:deeipei}
dE_e=P_e(f_e)d\ln f_e.
\end{equation} 
This defines $P_e(f_e)$: the energy spectrum of a system at the time of emission.
From (\ref{eq:deeipei}) it follows that $P_e(f_e)=dE_e/d\ln f_e$.

The energy spectrum of a system at the time of emission can be related to the energy spectrum today.
The present energy $dE$ radiated by that system, with a logarithmic observed frequency between $\ln f$ and $\ln f+d\ln f$, is
\begin{equation}
\label{eq:depfdf}
dE=P(f)d\ln f,
\end{equation} 
which defines $P(f)=dE/d\ln f$.
Applying (\ref{eq:dfdfe}) and (\ref{eq:dedee}) to Equation (\ref{eq:depfdf}), one obtains
\begin{equation}
\frac{dE_e}{1+z}=P(f)d\ln f_e.
\end{equation}
Comparing it to Equation (\ref{eq:deeipei}), 
\begin{equation}
\label{eq:pefe}
P(f)=(1+z)^{-1}P_e(f_e)=(1+z)^{-1}P_e(f[1+z]).
\end{equation} 
The function $P_e(f[1+z])$ is explicitly given for the case of a binary system in Equation (\ref{eq:specenergy}).

We now calculate the energy spectrum per unit comoving volume of an ensemble.
The number of systems per unit comoving volume during a time $dt_e$ is
\begin{equation}
\label{eq:defsigcomrate}
dn=\dot{n}(z)dt_e.
\end{equation} 
Here, $\dot{n}(z)=dn/dt_e$ is the signal comoving density rate (number of signals per unit emitted interval of time per unit comoving volume).
The comoving energy density spectrum of an ensemble is
\begin{equation}
\label{eq:pfpf}
p(f)=\int P(f) dn=\int P(f)\dot{n}(z)dt_e.
\end{equation} 
Recall that $P(f)$ in general depends on $z$, according to (\ref{eq:pefe}).
The integrals in (\ref{eq:pfpf}) contain all systems formed from the Big Bang until today.
Thus the limits of the time integral are $0$ (today) and $t_0$ (the beginning of the universe).
We can now change variables to write the previous integral in terms of redshifts, using the paths of gravitons as explained in Section \ref{sec:cosmological},
\begin{equation}
\label{eq:pfeq}
p(f)=\int_0^{\infty} P(f)\dot{n}(z) \frac{dt_e}{dz} dz.
\end{equation} 
Since we have chosen $a(0)=1$, the comoving volume and the physical volume are identical at present.
Therefore,
\begin{equation}
\label{eq:pfeqeps}
p(f)=\varepsilon_{\ln}(f).
\end{equation} 
This means, the comoving energy density spectrum $p(f)$ measured today is what in Equation (\ref{eq:defomega}) was called $\varepsilon_{\ln}$:
the present energy density of gravitational radiation per logarithmic frequency interval (of a certain ensemble).
Using (\ref{eq:defomega}), (\ref{eq:pfeq}), and (\ref{eq:pfeqeps}),
\begin{equation}
\label{eq:firstomega}
\om =\frac{\varepsilon_{\ln}(f)}{\rho_cc^2}=\frac{1}{\rho_cc^2}\int_0^\infty P(f)\dot{n}(z)\frac{dt_e}{dz}dz.
\end{equation} 
In this formula, only the term $dt_e/dz$ depends on the choice of cosmological model.

A similar derivation of (\ref{eq:firstomega}) (using different notation) can be found in \cite{Phinney2001}.
In that paper, the formula for the spectral function, called $\Omega_{\text{gw}}(f)$, is given in Equation (5).
The terms $N(z)$ and $[1+z]^{-1}[f_r\,dE_{\text{gw}}/df_r]|_{f_r=f[1+z]}$ corresponds, with our notation, to $\dot{n}(z)dt_e/dz$ and $P(f)$, respectively.

In Equation (\ref{eq:pfeq}) one can clearly see the assumption of a homogeneous universe, which is implicitly imposed by the metric (\ref{eq:metric}).
At any position within a shell of width $dz$ there is the same number of systems.
In other words, $\dot{n}(z)$ is the same at every point on a line of constant time, in Figure \ref{fig:penrosez}.

The spectral function of an ensemble can be expressed more conveniently.
We write it in terms of the energy spectrum  at the time of emission, $P_e(f[1+z])$, for our particular cosmological model.
Using (\ref{eq:dtedzeq}) and (\ref{eq:pefe}),
\begin{equation}
\label{eq:omegaformula}
\om =\frac{1}{\rho_cc^2H_0}\int_{0}^{\infty} \frac{P_e(f[1+z])\dot{n}(z)}{[1+z]^2\mathcal{E}(z)}dz.
\end{equation} 
The spectral function of the total contemporary background would be the sum of the spectral functions of all different types of ensembles.

But $\om$ does not include all redshifts and frequencies,
since $\dot{n}(z)$ and $P_e(f_e)$ have support only for $z\in [z_{\text{min}},z_{\text{max}}]$ and $f_e\in [f_{\text{min}},f_{\text{max}}]$, respectively.
The maximum frequency $f_{\text{max}}$ is the one above which no more gravitational waves are emitted.
The minimum frequency $f_{\text{min}}$ is the one below which the contribution in gravitational waves is dismissed.
For example, neutron star binaries started to form at a redshift $z_{\text{max}}\sim 5$ ($\sim$12\,Gyr ago), are still forming at present, so $z_{\text{min}}\sim 0$,
and emit in a range of frequencies from $\sim$0.01\,mHz to $\sim$1\,kHz (these ranges are justified in Section \ref{sec:models}).
These limits in redshift and frequency must be taken into account in the integral of (\ref{eq:omegaformula}).

\begin{figure}
\includegraphics{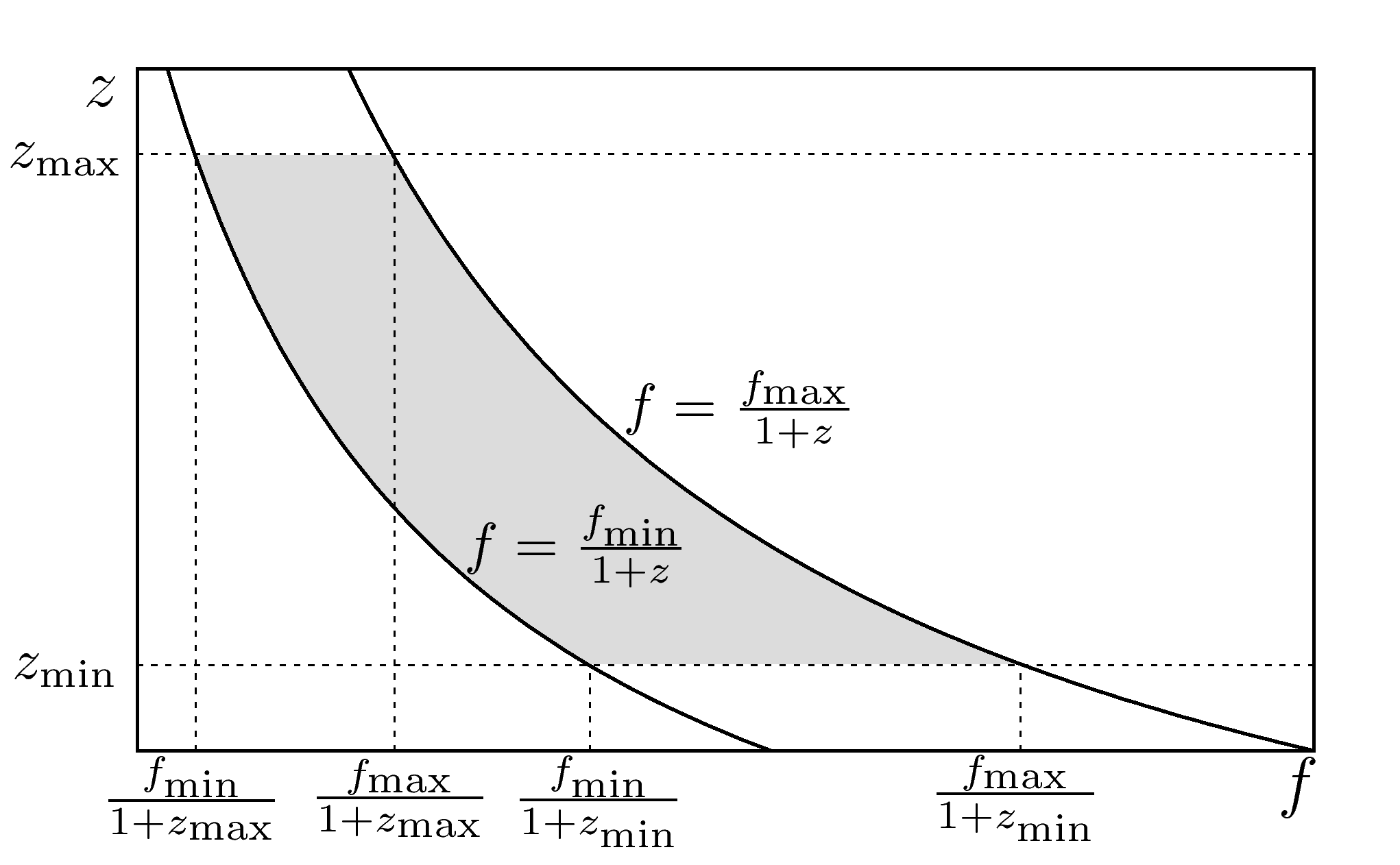}
\caption{Redshift versus observed frequency.
The spectral function of the total background of a certain ensemble has support only within the shaded area.}
\label{fig:pfun}
\end{figure}

To understand how (\ref{eq:omegaformula}) changes with the introduction of these limits, it is helpful to make a plot of redshifts versus frequencies.
Each horizontal line of such a plot gives the range of possible frequencies of a signal at a certain redshift.
If we plotted on the horizontal axis the emitted frequencies $f_e$, the limits $f_{\text{min}}$, $f_{\text{max}}$,
$z_{\text{min}}$, and $z_{\text{max}}$ would define a rectangle,
containing all the points $(f_e,z)$ where both $\dot{n}(z)$ and $P_e(f_e)$ have support.
But representing redshifts versus observed frequencies $f$, one obtains the plot of Figure \ref{fig:pfun} (which is no longer a rectangle).
The shaded area represents the support of $\om$.

We insert two redshift functions $z_{\text{low}}(f)$ and $z_{\text{upp}}(f)$ in the integration limits of (\ref{eq:omegaformula}),
in such a way that the integral is non-zero only in the shaded area of Figure \ref{fig:pfun}.
This is achieved with
\begin{equation}
\label{eq:defzlow}
z_{\text{low}}(f)=\left\{ \begin{array}{lc}
             z_{\text{max}}, & f\le \frac{f_{\text{min}}}{1+z_{\text{max}}} \\
	     \frac{f_{\text{min}}}{f}-1, & \frac{f_{\text{min}}}{1+z_{\text{max}}}< f<\frac{f_{\text{min}}}{1+z_{\text{min}}} \\
	    z_{\text{min}}, & \frac{f_{\text{min}}}{1+z_{\text{min}}}\le f
             \end{array}
\right. ,
\end{equation}
and
\begin{equation}
\label{eq:defzupp}
z_{\text{upp}}(f)=\left\{ \begin{array}{lc}
             z_{\text{max}}, & f\le \frac{f_{\text{max}}}{1+z_{\text{max}}} \\
	     \frac{f_{\text{max}}}{f}-1, & \frac{f_{\text{max}}}{1+z_{\text{max}}}< f<\frac{f_{\text{max}}}{1+z_{\text{min}}} \\
	      z_{\text{min}}, & \frac{f_{\text{max}}}{1+z_{\text{min}}}\le f
             \end{array}
\right. .
\end{equation}
With these limits, only signals emitted with frequencies between $f_{\text{min}}$ and $f_{\text{max}}$,
at redshifts between $z_{\text{min}}$ and $z_{\text{max}}$, contribute to $\om$.

Changing the integration limits in (\ref{eq:omegaformula}), the spectral function becomes
\begin{equation}
\label{eq:omegalim}
\om =\frac{1}{\rho_cc^2H_0} \int_{z_{\text{low}}(f)}^{z_{\text{upp}}(f)} \frac{P_e(f[1+z])\dot{n}(z)}{[1+z]^2\mathcal{E}(z)}dz.
\end{equation} 
This formula gives the spectral function of the total background produced by an ensemble, measured at the present time.
In the next section we generalize this formula to account for the resolvability of the signals.

In several papers, the integral in Equation (\ref{eq:omegalim}) contains an extra $[1+z]^{-1}$ factor (see \cite{AraujoMiranda2005},
where the origin of this misleading factor is explained).
Recall the definition of $P_e(f[1+z])$ in Equation (\ref{eq:deeipei}),
and notice that the integral in (\ref{eq:omegalim}) is not equivalent to, for example, that in Equation (9) of \cite{RegimbauMandic2008},
where the extra $[1+z]^{-1}$ factor is introduced \footnote{Notice also that $P_e(f[1+z])$ has been defined differently in the first version of this paper, in arXiv:1106.5795v1.
The equations for the spectral function are, nevertheless, equivalent in the two versions.}.

Notice that the energy spectrum $P_e(f_e)$ does not depend on time.
We are thus adding the contribution of each system as if it were instantaneous (this means, as if it were a point in the Penrose diagram of Figure \ref{fig:penrosez}).
This is justified if the inspiral times are much smaller than the cosmic timescales,
so the time a system needs to evolve from emitting at $f_{\text{min}}$ to $f_{\text{max}}$ is much less than the Hubble time, $H_0^{-1}\approx$13\,Gyr.
In Section \ref{sec:minfreq} we point out that this assumption is not always fulfilled, but it turns out to be irrelevant in practice.

\subsection{\label{sec:resolvability} Resolvability of the background}
In this section we introduce the \textit{overlap function}, $\mathcal{N}(f,\Delta f,z)$,
that allows us to define and quantify the resolvability of the background.

We first define some quantities that are necessary for the definitions of the different parts of the background.
Let $\mathcal{B}(f,z_1,z_2)$ be the collection of signal elements with observed frequencies between $f$ and $f+df$
and with redshifts between $z_1$ and $z_2$.
Let $\tau_e(f_e,\Delta f_e)$ be the interval of time (measured close to the system at the moment of emission)
that a system at $z$ spends emitting between $f_e$ and $f_e+\Delta f_e$.
Written in terms of observed frequencies, this interval of time is $\tau_e(f,\Delta f,z)$.
We define $\dot{N}(z)$ in such a way that $\dot{N}(z)dz$ is the number of signals produced per unit emitted interval of time between $z$ and $z+dz$.
Since $\dot{n}(z)$ is the number of signals per unit emitted interval of time per unit comoving volume at redshift $z$, $\dot{N}(z)$ is given by
\begin{equation}
\dot{N}(z)=\dot{n}(z) \frac{d\mathcal{V}_c}{dz}.
\end{equation} 
The value of $\dot{N}(z)$ at a certain redshift $z$ can be considered an average over an interval of time that is much longer than a typical observation time,
but much shorter than the Hubble time.
For the sake of simplicity let us assume that we know precisely this function,
and that it gives the exact number of signals produced per unit emitted interval of time.
For instance, if we have $\int_0^z\dot{N}(z)dz=1$\,hour$^{-1}$, one signal is assumed to be produced between redshift $0$ and $z$ exactly every hour.

Let us illustrate the resolvability with the following example:
One signal is produced every hour between $z$ and $z+dz$, i.e. $\dot{N}(z)dz=1$\,hour$^{-1}$.
Each signal spends one hour between $f$ and $f+\Delta f$, i.e. $\tau_e(f,\Delta f,z)=1$\,hour.
Thus, whenever we see that frequency bin, it will be occupied by $\tau_e(f,\Delta f,z) \times \dot{N}(z)dz=1$ signal produced between $z$ and $z+dz$.
If, for the same $\dot{N}(z)dz$, we consider a different range of frequencies, where $\tau_e(f,\Delta f,z)=2$\,hours,
we will always see in that frequency bin 2 overlapping signals, which will not be distinguishable.
We can perform a similar calculation, considering all redshifts between $z_1$ and $z_2$:
$\int_{z_1}^{z_2}\tau_e(f,\Delta f,z) \times \dot{N}(z)dz$ gives the number of signals between redshift $z_1$ and $z_2$ that overlap in a frequency bin.
If that number is larger than 1, those signals cannot be resolved.
This leads us to the definition of the overlap function.

The overlap function is defined by
\begin{equation}
\label{eq:defoverlapfun}
\mathcal{N}(f,\Delta f,z)= \int_{z_{\text{low}}(f)}^z \tau_e (f,\Delta f,z') \dot{N}(z') dz'.
\end{equation}
It thus gives the expected number of signals with redshifts smaller than $z$ and frequencies between $f$ and $f+\Delta f$.
For example, $\mathcal{N}(f,\Delta f,z)=1$ implies that, as soon as one signal leaves a frequency bin, another signal enters it,
so the bin is constantly occupied by one signal.
Hence, $\mathcal{N}(f,\Delta f,z)>1$ implies that signals overlap in a frequency bin.
We can impose $\mathcal{N}(f,\Delta f,z)=\mathcal{N}_0$ and invert this equation with respect to the redshift $z$.
The obtained function,
\begin{equation}
\mathcal{N}^{-1}=\mathcal{N}^{-1}(f,\Delta f,\mathcal{N}_0)
\end{equation}
is the redshift such that all signals between $f$ and $f+\Delta f$ with redshifts smaller than $\mathcal{N}^{-1}(f,\Delta f,\mathcal{N}_0)$ sum $\mathcal{N}_0$.
To obtain an overlap larger than $\mathcal{N}_0$ at a certain frequency $f$,
one has to consider only signals from redshifts larger than $\mathcal{N}^{-1}(f,\Delta f,\mathcal{N}_0)$.
In Section \ref{sec:ovfun} we give a formula for $\mathcal{N}^{-1}(f,\Delta f,\mathcal{N}_0)$ for an ensemble of binary systems.

We now give some relevant definitions:

The \textit{total background} of an ensemble between frequency $f$ and $f+df$ is $\mathcal{B}(f,z_{\text{low}}(f),z_{\text{upp}}(f))$.
One can assign a spectral function to it, $\Omega_{\text{total}}(f)$.

The total background can be divided into two parts: \textit{resolvable} and \textit{unresolvable}.
If there exists a certain $z_*$ such that $z_{\text{low}}(f)<z_*<z_{\text{upp}}(f)$ and $\mathcal{N}(f,\Delta f,z_*)=1$,
the \textit{unresolvable part} is $\mathcal{B}(f,z_*,z_{\text{upp}}(f))$,
and the \textit{resolvable part} is $\mathcal{B}(f,z_{\text{low}}(f),z_*)$.
If there is no $z_*$ such that $z_{\text{low}}(f)<z_*<z_{\text{upp}}(f)$ and $\mathcal{N}(f,\Delta f,z_*)=1$,
the resolvable part coincides with the total background and the unresolvable part is the empty set.
One can assign a spectral function to the resolvable part of the background, $\Omega_{\text{resolvable}}(f)$,
and to the unresolvable part, $\Omega_{\text{unresolvable}}(f)$.

The resolvable part \textit{dominates} at a frequency $f$ when $\Omega_{\text{resolvable}}(f)>\Omega_{\text{unresolvable}}(f)$.
When this happens, even if there is an unresolvable background present, it is weak compared to the background of the closer (stronger) signals,
and thus the latter can still be resolved.
On the other hand, the unresolvable part \textit{dominates} at a frequency $f$ when $\Omega_{\text{unresolvable}}(f)>\Omega_{\text{resolvable}}(f)$.
In this case, even if there are some close resolvable signals, they cannot be resolved in practice,
since they are obscured by the superposition of many weak distant signals.

In Section \ref{sec:unresolvable} other possible criteria for the resolvability of the background are commented on.

\begin{figure}
\includegraphics{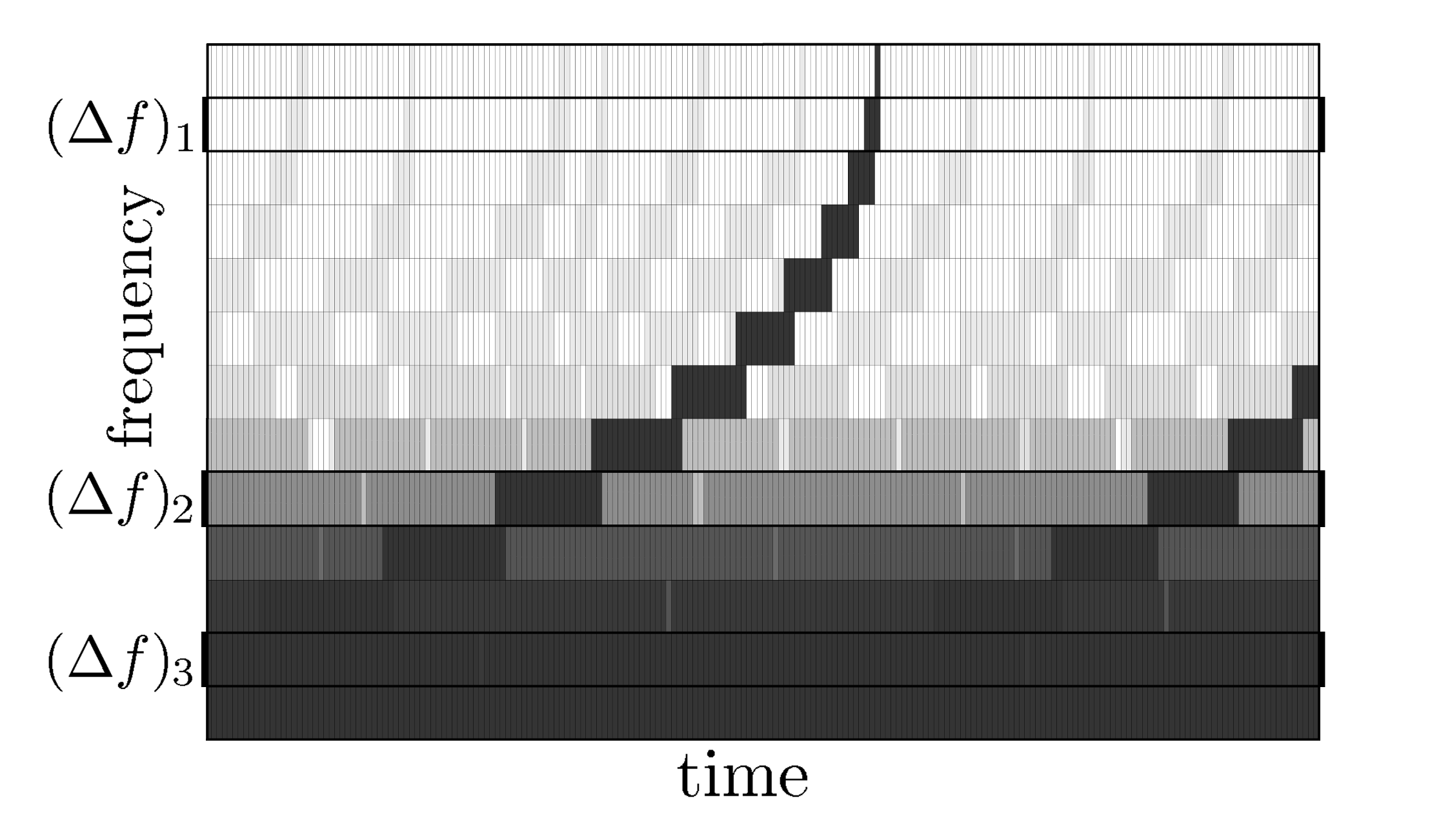}
\caption{Observed frequency versus time.
Each line represents the evolution of one signal (like the one produced by a binary).
Closer signals, as well as the superposition of many signals, are plotted darker than distant individual signals.
Three frequency bins are distinguished:
in $(\Delta f)_1$ the total background is completely resolvable,
in $(\Delta f)_2$ there is an unresolvable part and a dominating resolvable part,
and in $(\Delta f)_3$ there is a dominating unresolvable part and a resolvable part.}
\label{fig:ftplot5}
\end{figure}

In Figure \ref{fig:ftplot5} we give an illustrative example to understand the definitions of the different parts of the background.
There we plot the evolution in time of the observed frequency of many similar signals, like the ones produced by an ensemble of binaries.
The horizontal axis range is an interval of time of the order of a typical observation time.
This axis is divided in small intervals $\Delta t$, which is the time resolution.
The vertical axis can be considered the frequency window of a hypothetical detector,
with such a low instrumental noise that allows us to observe signals emitted at very large redshifts.
This axis is divided into small intervals $\Delta f$, the frequency resolution.
Darker \textit{pixels} in the plot represent stronger backgrounds, i.e. with larger spectral function.
The bin $(\Delta f)_1$ is in a range of frequencies where the total background is completely resolvable: all signals can be clearly distinguished from each other.
In $(\Delta f)_2$, an unresolvable part starts to contribute, but close binaries can still be clearly distinguished from each other, since the resolvable part dominates.
Finally, in $(\Delta f)_3$ the unresolvable part of the background dominates over the resolvable one.
One should keep in mind that this example does not accurately follow the definition of unresolvability,
since the spectral function does not account for individual signals.

We now generalize the formula of the spectral function to account for the resolvability of the background.
We solve the integral in (\ref{eq:omegalim}) for the signals that fulfill the condition $\mathcal{N}(f,\Delta f,z)\ge \mathcal{N}_0$.
For that, we can retain the same upper limit of the integral, $z_{\text{upp}}(f)$, and change the lower one, replacing $z_{\text{low}}(f)$ by
\begin{equation}
\label{eq:zoverlap}
\overline{z}(f,\Delta f,\mathcal{N}_0)=\left\{ \begin{array}{lc}
	    z_{\text{upp}}(f), & f< f_{\text{p,min}} \\
            \mathcal{N}^{-1}(f,\Delta f,\mathcal{N}_0), & f_{\text{p,min}}\le f\le f_{\text{p,max}} \\
	    z_{\text{upp}}(f), & f_{\text{p,max}}<f
             \end{array}
\right. .
\end{equation}
We have introduced four \textit{limiting frequencies}:
$f_{\text{p,max}}$ ($f_{\text{p,min}}$) represents the maximum (minimum) frequency at which the unresolvable part of the background is present,
and $f_{\text{d,max}}$ ($f_{\text{d,min}}$) represents the maximum (minimum) frequency at which the unresolvable part dominates over the resolvable.
Using \eqref{eq:zoverlap} we obtain the spectral function of an ensemble with more than $\mathcal{N}_0$ signals per frequency bin $\Delta f$,
\begin{align}
\label{eq:omgensol}
\omov=&\frac{1}{\rho_cc^2H_0} \int_{\overline{z}(f,\Delta f,\mathcal{N}_0)}^{z_{\text{upp}}(f)} P_e(f[1+z]) \nonumber\\
&\times \frac{\dot{n}(z)}{[1+z]^2\mathcal{E}(z)}dz.
\end{align}
This is the main equation of the paper and a generalization of Equation (\ref{eq:omegalim}) with which we can distinguish the different regimes of the background.

The unresolvable background is fully characterized by the spectral function $\omov$.
It is not easy to determine whether the assumptions mentioned at the beginning of Section \ref{sec:spectralfunction} are always fulfilled for such a background.
But it is clear that valuable information is lost when using the spectral function to characterize a resolvable background, where signals can be individually distinguished.

The spectral function of the unresolvable part of the background is, according to the definitions given at the beginning of this section,
\begin{equation}
\label{eq:omunres}
\Omega_{\text{unresolvable}}(f)=\Omega (f,\Delta f,1),
\end{equation} 
where $\Delta f$ can be chosen as the inverse of the observation time.
On the other hand, the spectral function of the resolvable part is
\begin{equation}
\label{eq:omres}
\Omega_{\text{resolvable}}(f)=\Omega_{\text{total}} (f)-\Omega_{\text{unresolvable}}(f),
\end{equation} 
where
\begin{equation}
\label{eq:omtotal}
\Omega_{\text{total}}(f)=\Omega (f,\Delta f,0).
\end{equation} 
Here, $\Omega (f,\Delta f,0)$ coincides with the $\Omega (f)$ given in Equation (\ref{eq:omegalim}), and the value of $\Delta f$ becomes irrelevant.

\begin{figure*}
\includegraphics{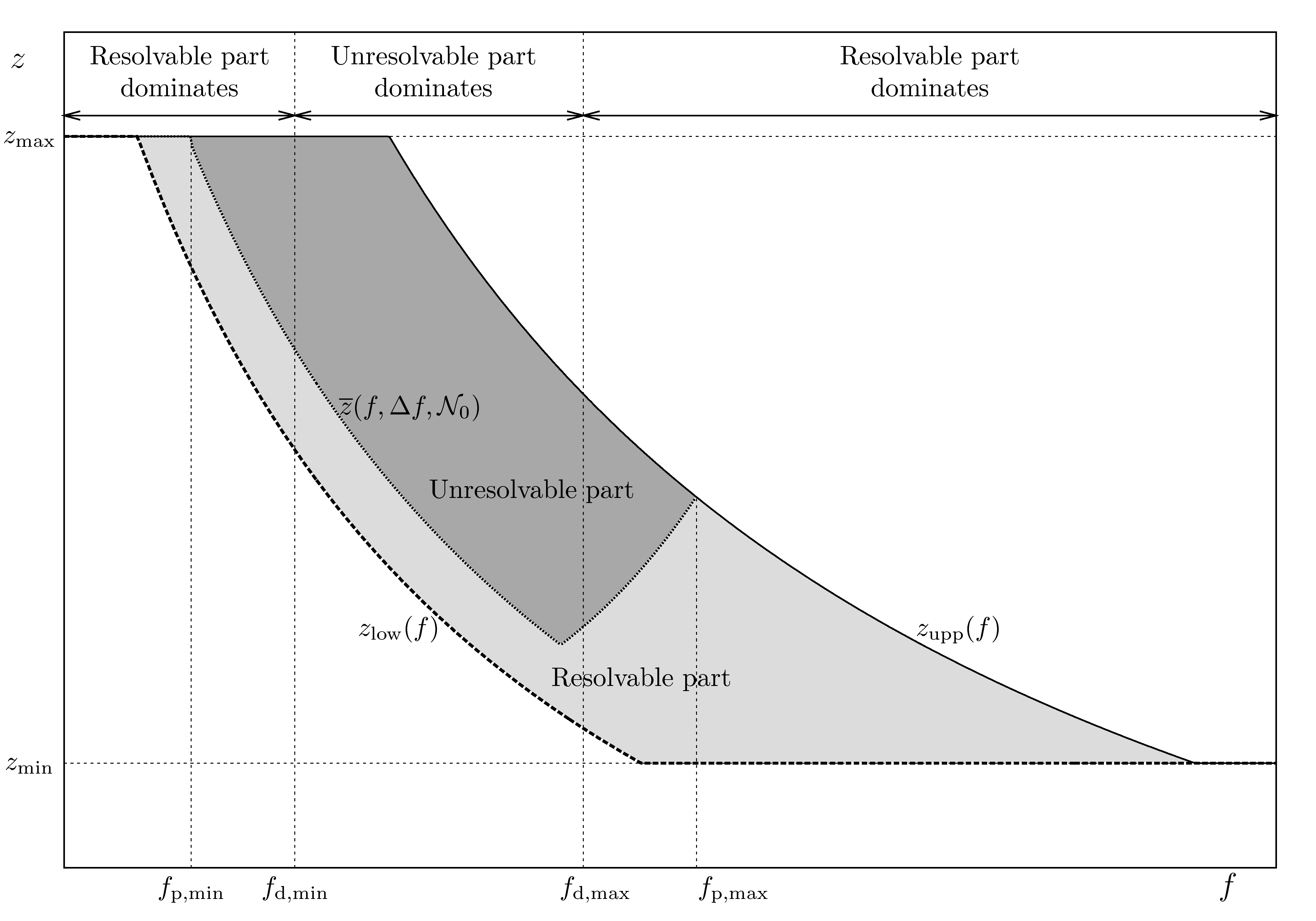}
\caption{Redshift versus observed frequency.
Each horizontal line contains the possible observed frequencies of a signal.
The light-shaded (dark-shaded) area represents the resolvable (unresolvable) part of the background.
The redshift functions $z_{\text{low}}(f)$, $\overline{z}(f,\Delta f,\mathcal{N}_0)$, and $z_{\text{upp}}(f)$ are shown with dashed, dotted, and solid lines, respectively.
The frequencies $f_{\text{p,min}}$ and $f_{\text{p,max}}$ delimit the interval where the unresolvable part is present.
The frequencies $f_{\text{d,min}}$ and $f_{\text{d,max}}$ delimit the interval where the unresolvable part dominates.}
\label{fig:pfun2}
\end{figure*}

Another picture that illustrates the distinct parts of the background is in Figure \ref{fig:pfun2}.
This graph is the same as that in Figure \ref{fig:pfun}, but also represents the redshift function $\overline{z}(f,\Delta f,\mathcal{N}_0)$
that defines the frontier between the resolvable (light-shaded area) and unresolvable (dark-shaded) parts of the background.

The mathematical definitions of the limiting frequencies can be understood by looking at the graph in Figure \ref{fig:pfun2}.
The frequencies $f_{\text{d,min}}$ and $f_{\text{d,max}}$ are the ones at which the resolvable and the unresolvable parts have equal spectral function, so
\begin{equation}
\label{eq:fdmaxdef}
\Omega_{\text{unresolvable}}(f_{\text{d,min/max}})=\Omega_{\text{resolvable}}(f_{\text{d,min/max}}),
\end{equation} 
The frequencies $f_{\text{p,min}}$ and $f_{\text{p,max}}$ are the ones at which the function $\mathcal{N}^{-1}(f,\Delta f,\mathcal{N}_0)$ intersects $z_{\text{upp}}(f)$,
so
\begin{equation}
\label{eq:deffpmax}
\mathcal{N}^{-1}(f_{\text{p,min/max}},\Delta f,\mathcal{N}_0)=z_{\text{upp}}(f_{\text{p,min/max}}).
\end{equation}
In Section \ref{sec:limfreqs} we calculate these limiting frequencies for an ensemble of binary systems.

So far we have distinguished the regimes of resolvability by using the frequency resolution $\Delta f$, but not the time resolution $\Delta t$.
In Section \ref{sec:discovfun} we show how to redefine the overlap function to account for the time resolution.
In practice, the effect of introducing a realistic $\Delta t$ in the calculations turns out to be irrelevant for our work.

In Section \ref{sec:ovvsdc} we show that the overlap function is a generalization of what in the literature is often called the \textit{duty cycle}, $D(z)$.
The latter is proven to be a good quantifier of the unresolvability of the background only for very short signals (\textit{bursts}).
Furthermore, we use the name \textit{overlap function} and not \textit{duty cycle},
because the latter may be confusing: $D(z)$ can be greater than unity,
unlike the typical duty cycles used in electronics or in gravitational wave detectors.

\subsection{\label{sec:continuity} Continuity of the background}
The overlap function can be used to characterize not only the resolvability but also the continuity of the background.
We now give some definitions, similar to the ones given in the previous section:

Given a frequency interval $[f_{\text{low}},f_{\text{upp}}]$ (that can be the frequency window of a detector),
the total background $\mathcal{B}(f,z_{\text{low}}(f),z_{\text{upp}}(f))$ of an ensemble between frequency $f$ and $f+df$ (where $f_{\text{low}}\le f\le f_{\text{upp}}$)
can be divided into two parts: \textit{discontinuous} and \textit{continuous}.
If there exists a certain $z_*$ such that
$z_{\text{low}}(f)<z_*<z_{\text{upp}}(f)$ and $\mathcal{N}(f,\min(f_{\text{max}},f_{\text{upp}})-f,z_*)=1$,
the \textit{continuous part} is $\mathcal{B}(f,z_*,z_{\text{upp}}(f))$,
and the \textit{discontinuous part} is $\mathcal{B}(f,z_{\text{low}}(f),z_*)$.
If there is no $z_*$ such that $z_{\text{low}}(f)<z_*<z_{\text{upp}}(f)$ and $\mathcal{N}(f,\Delta f,z_*)=1$,
the discontinuous part coincides with the total background and the continuous part is the empty set.
One can assign a spectral function to the discontinuous part of the background, $\Omega_{\text{discontinuous}}(f)$,
and to the continuous part, $\Omega_{\text{continuous}}(f)$.

The definitions of resolvable and unresolvable backgrounds are valid both for signals which frequency increases in time, such as binaries,
and for signals which frequency decreases (for which one could change $\Delta f$ by $-\Delta f$ in the definitions).
On the other hand, the given definitions of discontinuous/continuous backgrounds assume that the frequency increases in time.
For signals with decreasing frequency, the condition of continuity would be $\mathcal{N}(f,f-\max(f_{\text{low}},f_{\text{min}}),z_*)\ge1$.

In the following, when talking about the continuous background, we will assume $f_{\text{low}}=0$ and $f_{\text{upp}}=\infty$.
This implies that any part of the background that is not continuous in this circumstance is definitely discontinuous,
for any other choice of $f_{\text{low}}$ and $f_{\text{upp}}$.
Besides, the unresolvable background is necessarily continuous.

So, the spectral function of the continuous background is
\begin{equation}
\label{eq:omcont}
\Omega_{\text{continuous}}(f)=\Omega (f,f_{\text{max}}-f,1).
\end{equation} 
The spectral function of the discontinuous background is
\begin{equation}
\label{eq:omdisc}
\Omega_{\text{discontinuous}}(f)=\Omega_{\text{total}}(f)-\Omega_{\text{continuous}}(f),
\end{equation} 
where $\Omega_{\text{total}}(f)$ is the same of Equations (\ref{eq:omtotal}) and (\ref{eq:omegalim}).

\section{\label{sec:models} Models for the ensembles}
Our work is focused on the contemporary background produced by coalescing binary systems.
These are systems composed of two objects that inspiral towards each other, producing gravitational waves with an increasing frequency until they coalesce.
In order to emit gravitational waves significantly, they must be sufficiently massive and/or compact.
Each binary is assumed to be isolated and describing an orbit of zero eccentricity.
Its components are assumed to be non-spinning.

We sort the binary systems into two classes: stellar binaries and massive black hole binaries.

By \textit{stellar binary} we mean a system whose components have masses of the order of a solar mass (or tens of it).
We consider those stellar binaries formed by two stellar-mass black holes (from now on we call this type of binary BH-BH),
a stellar-mass black hole and a neutron star (BH-NS),
two neutron stars (NS-NS), a neutron star and a white dwarf (NS-WD) or two white dwarfs (WD-WD).
The majority of the star formation rates in the literature vanish at redshifts larger than 5 or 6
(see for example \cite{ConnollyEtAl1997,PorcianiMadau2001,2dFGRS2001,HopkinsBeacom2006,NagamineEtAl2006,FardalEtAl2007,WilkinsEtAl2008}).
If we expect no star formation at higher redshifts, no coalescence from stellar binaries is expected either.
We choose then a maximum redshift for binary coalescences of $z_{\text{max}}=5$.
The minimum redshift is chosen $z_{\text{min}}=0$, since these binaries can also form at present.

\textit{Massive black hole binaries} (from now on, MBH-MBH)
are systems believed to exist in the center of many galaxies \cite{[{ }][{ and references therein, for a review on this topic.}]KormendyRichstone1995,MagorrianEtAl1998}.
Their components have masses that range from $\sim10^2\,M_{\odot}$ to $\sim10^{10}\,M_{\odot}$.
We consider four different models for massive black hole formation, presented in \cite{ArunEtAl2009}:
two of them (called SE/SC, for \textit{small seeds} and \textit{efficient/chaotic accretion}) with light-seed black holes produced as remnants of Population III stars,
and two with heavy-seed black holes formed from dynamical instabilities in the nuclei of protogalaxies
(called LE/LC, for \textit{large seeds} and \textit{efficient/chaotic accretion}).
These formation models allow coalescences at redshifts reaching $z\approx20$.

\subsection{\label{sec:enerspec} Energy spectrum}
We assume that the energy lost by a system when emitting gravitational radiation between $f_e$ and $f_e+df_e$ is of the form
\begin{equation}
\label{eq:deesys}
dE_{e,\text{sys}}=-\kappa [f_e]^b df_e,
\end{equation} 
for real constants $\kappa$ and $b$.
This formula is valid for all systems considered in this work (binaries) and also other systems
(see for example the emission model of magnetars in \cite{RegimbauFreitas2006b}).
In particular, for a binary system,
\begin{equation}
\label{eq:paramener}
\kappa =\frac{1}{3}[G^2 \pi^2 \mathcal{M}^5]^{1/3} \quad \text{and} \quad b=-\frac{1}{3}.
\end{equation}
Here we have introduced the chirp mass $\mathcal{M}$ of the binary, defined by
\begin{equation}
\label{eq:chirpmass}
\mathcal{M}=\frac{[m_1 m_2]^{3/5}}{[m_1+m_2]^{1/5}},
\end{equation} 
where $m_1$ and $m_2$ are the masses of the two components of the binary.

This energy spectrum is obtained by assuming that the energy of the system, as well as the separation of the bodies, varies slowly with time.
This is valid as long as the orbit is far wider than the last stable one (see Equation (\ref{eq:defslso})).
In these circumstances, the system is well described by the Newtonian equations of motion of two point masses in a circular orbit.

We thus derive Equation (\ref{eq:deesys}) for binary systems using Newtonian mechanics.
The energy of the system, in terms of the separation $s$ between the stars of masses $m_1$ and $m_2$ is
\begin{equation}
\label{eq:enersys}
E_{e,\text{sys}}=-\frac{1}{2}\frac{Gm_1m_2}{s}.
\end{equation} 
We reduce the two-body problem to one fictitious body, of mass equal to the reduced mass of the system, $m_1m_2/[m_1+m_2]$,
suffering the same force as each of the real bodies.
Applying Newton's second law,
\begin{equation}
\label{eq:newton}
\frac{Gm_1m_2}{s^2}=\frac{m_1 m_2}{m_1+m_2} [2\pi f_{e,\text{orbit}}]^2 s,
\end{equation}
where $f_{e,\text{orbit}}$ is the orbital frequency, that is related to the frequency of the gravitational waves in the quadrupolar approximation by
\begin{equation}
\label{eq:fgwforbit}
f_e=2f_{e,\text{orbit}}.
\end{equation} 
Introducing it in Equation (\ref{eq:newton}) one obtains a formula that relates the separation of the masses with the frequency of the gravitational waves,
\begin{equation}
\label{eq:radfreq}
s=\left[ \frac{G[m_1+m_2]}{\pi^2f_e^2}\right]^{1/3}.
\end{equation}
Replacing (\ref{eq:radfreq}) in (\ref{eq:enersys}) and differentiating, one finally obtains the energy spectrum (\ref{eq:deesys})
with the values of $\kappa$ and $b$ given in (\ref{eq:paramener}).
A more detailed derivation of Equation (\ref{eq:deesys}) can be found in Chapter 4 of \cite{Maggiore2008}.

What we need is the energy spectrum of the gravitational radiation at the time of emission, in terms of observed frequencies,
$P_e(f[1+z])$ (recall the formula of $\omov$ in Equation (\ref{eq:omgensol})).
Using (\ref{eq:fredshift}), (\ref{eq:deeipei}), and (\ref{eq:deesys}),
\begin{align}
\label{eq:specenergy}
P_e(f[1+z])&=\frac{dE_e}{d\ln f_e}=\bigg| \frac{dE_{e,\text{sys}}}{d\ln f_e} \bigg|=f_e \kappa [f_e]^{-1/3} \nonumber \\
&=\kappa f^{2/3} [1+z]^{2/3}.
\end{align} 
The absolute value has been introduced because $P_e(f[1+z])$ is a positive quantity;
$dE_e$ is the amount of gravitational wave energy within a frequency interval, regardless of whether the energy of the system increases or decreases with the frequency.
Equation (\ref{eq:specenergy}) must be inserted in (\ref{eq:omgensol}),
\begin{equation}
\label{eq:omgenbin}
\omov=\frac{\kappa f^{2/3}}{\rho_cc^2H_0} \int_{\overline{z}(f,\Delta f,\mathcal{N}_0)}^{z_{\text{upp}}(f)} \frac{\dot{n}(z)}{[1+z]^{-4/3}\mathcal{E}(z)}dz,
\end{equation}
to obtain the spectral function of an ensemble of binary systems.

\subsection{Interval of time per frequency bin}
To calculate the overlap function we need the interval of time that a system spends emitting in a frequency bin.
The frequency of the radiation emitted by a binary evolves in (look-forward) time following the relation
\begin{equation}
\label{eq:fdot}
\frac{df_e}{dt_e}=\frac{96}{5}\pi^{8/3}\left[\frac{G\mathcal{M}}{c^3}\right]^{5/3}f_e^{11/3},
\end{equation} 
the derivation of which can be found in Chapter 4 of \cite{Maggiore2008}.
Integrating (\ref{eq:fdot}) between $f_e$ and $f_e+\Delta f_e$ one obtains the interval of time that the signal spends in that frequency interval,
\begin{equation}
\label{eq:timefreq}
\tau_e(f_e,\Delta f_e)=\delta_2 \left[f_e^{-8/3}-[f_e+\Delta f_e]^{-8/3}\right],
\end{equation} 
where
\begin{equation}
\delta_2= \frac{5 c^5}{256 \pi^{8/3} [G\mathcal{M}]^{5/3}}.
\end{equation}
The interval of time $\tau_e(f_e,\Delta f_e)$ can be written in terms of observed frequencies,
\begin{equation}
\label{eq:timefreqz}
\tau_e(f,\Delta f,z)=\delta_2 Q[f,\Delta f][1+z]^{-8/3},
\end{equation} 
where we have defined
\begin{equation}
Q(f,\Delta f)=f^{-8/3}-[f+\Delta f]^{-8/3}.
\end{equation} 
The function $\tau_e(f,\Delta f,z)$ gives the interval of time that a signal, produced at a redshift $z$, needs to change from an observed frequency $f$ to $f+\Delta f$.
We remark that this is an absolute (positive) interval of time, and not a look-back time.

In certain conditions $Q(f,\Delta f)$ can be simplified, by performing a Taylor expansion around $\Delta f=0$,
\begin{equation}
\label{eq:aproxq}
Q(f,\Delta f)\approx \frac{8}{3} \Delta f\,f^{-11/3},
\end{equation} 
for $\Delta f\ll f$.

\subsection{Maximum frequencies}
The energy spectrum of binary systems (Equation (\ref{eq:deesys})) is assumed to be zero outside a certain frequency range $[f_{\text{min}},f_{\text{max}}]$.
We now present our choices of $f_{\text{max}}$ for each type of system.
We omit the index $e$ to simplify the notation, but one should keep in mind that $f_{\text{max}}$ is an emitted frequency.

For all binaries that do not contain a white dwarf, $f_{\text{max}}$ is reached when both stars are as close to each other as $s_{\text{lso}}$.
This is the separation at the last stable orbit (see for example Box 25.6 in \cite{MisnerEtAl1973}),
\begin{equation}
\label{eq:defslso}
s_{\text{lso}}=\frac{6G[m_1+m_2]}{c^2},
\end{equation} 
which is 3 times the Schwarzschild radius of each star.
Using Equation (\ref{eq:radfreq}), the frequency of the last stable orbit is
\begin{equation}
\label{eq:flsoformula}
f_{\text{max}}^{\text{lso}}=\frac{1}{6\sqrt{6}\pi}\frac{c^3}{G [m_1+m_2]},
\end{equation}
where we have used the index ``lso'' to distinguish this maximum frequency from the following ones.

For WD-WD, since the radius of a white dwarf is much bigger than its Schwarzschild radius,
we assume that the maximum frequency is reached when both stars \textit{touch} each other.
This happens when the separation between them is $r_1+r_2$, the sum of their radii.
This separation corresponds to a frequency (using again Equation (\ref{eq:radfreq}))
\begin{equation}
f_{\text{max}}^{\text{WD-WD}}=\frac{1}{\pi}\sqrt{\frac{G[m_1+m_2]}{[r_1+r_2]^3}}.
\end{equation} 
For $r_1$ and $r_2$ one can use
\begin{equation}
\label{eq:radiomasa}
r_i=0.0112 R_\odot \sqrt{\left[\frac{m_i}{m_{\text{Cha}}}\right]^{-2/3}-\left[\frac{m_i}{m_{\text{Cha}}}\right]^{2/3}},
\end{equation}
where $m_{\text{Cha}}\approx$1.44\,$M_\odot$ is Chandrasekhar's mass.
This formula gives the approximate radius $r_i$ of a white dwarf as a function of its mass $m_i$.
It is obtained from Equations (27) and (28) (where there is an extra factor $M_\odot$ on the right side) in \cite{Nauenberg1972}.

For NS-WD, we use the same criterion as for WD-WD, but considering that the radius of the neutron star is negligible with respect to the radius of the white dwarf.
Hence, the maximum frequency is
\begin{equation}
f_{\text{max}}^{\text{NS-WD}}=\frac{1}{\pi}\sqrt{\frac{G[m_1+m_2]}{r_1^3}},
\end{equation} 
where $r_1$ is the radius of the white dwarf, that can be calculated with (\ref{eq:radiomasa}).

\subsection{\label{sec:fmin} Minimum frequencies}
There is a certain minimum frequency, $f_{\text{min}}$, such that the gravitational radiation emitted by a system below this frequency is disregarded,
because other mechanisms of energy loss are more effective.
It is difficult to find a precise description of these mechanisms for each type of system.
We adopt a simple criterion, for all stellar binaries, that fixes the value of $f_{\text{min}}$:
the interval of time between the beginning of the inspiral phase (when the binary emits at frequency $f_{\text{min}}$)
and the coalescence (when it emits at $f_{\text{max}}$) cannot be larger than a certain interval of time $T_{\text{max}}$.
This condition is equivalent to $\tau_e(f_{\text{min}},f_{\text{max}}-f_{\text{min}})=T_{\text{max}}$.
Using Equation (\ref{eq:timefreq}), one obtains
\begin{equation}
f_{\text{min}}=\left[\frac{T_{\text{max}}}{\delta_2}+f_{\text{max}}^{-8/3}\right]^{-3/8}.
\end{equation}  
For all systems considered in this work we can reasonably perform the approximation
\begin{equation}
\label{eq:deffminapprox}
f_{\text{min}}\approx \left[\frac{T_{\text{max}}}{\delta_2}\right]^{-3/8}.
\end{equation}
As we did with the maximum frequencies, we omit the index $e$ to simplify the notation, but $f_{\text{min}}$ is always an emitted frequency.
The assumption of a maximum inspiral time is justified in Section \ref{sec:minfreq}.
The maximum inspiral times chosen are $T_{\text{max}}=12$\,Gyr for stellar binaries and 75\,Myr for massive black hole binaries.
These choices are now explained.

\subsubsection{\label{sec:maxinsptime} Maximum inspiral time for stellar binaries}
For stellar binaries, $T_{\text{max}}=t_5-0$, where $t_5$ is the look-back time at which the first stellar binaries coalesced (at $z\approx5$).
Integrating (\ref{eq:dtedzeq}),
\begin{equation}
t_5=\int_0^{t_5}dt=\int_0^5 \frac{1}{[1+z]H(z)}dz\approx 12\,\text{Gyr}.
\end{equation} 
This choice of $T_{\text{max}}$ is somewhat arbitrary and even leads to an inconsistency:
only binaries that coalesced at small redshifts could have that much time to evolve from an initial frequency $f_{\text{min}}$ until the coalescence.
Moreover, at frequencies close to the minimum one,
the approximation of short inspiral times compared to the Hubble time, commented at the end of Section \ref{sec:spectralfunction}, is not valid anymore.
In Section \ref{sec:minfreq} we justify our choice of $T_{\text{max}}$, the exact value of which turns out to be unimportant in practice.

\subsubsection{Maximum inspiral time for massive black hole binaries}
The process that leads to two massive black holes coalescing can be briefly summarized in three main phases \cite{SesanaEtAl2004}:
\textit{dynamical friction}, \textit{gravitational slingshot} and \textit{gravitational radiation}.
When two dark matter halos containing black holes merge, the black holes suffer dynamical friction \cite{Quinlan1996} with the environment and sink to the center,
forming a wide binary (with large orbital period).
At a certain distance the dynamical friction phase ceases to be effective.
Then the binary can continue to shrink because of three-body interactions with surrounding stars \cite{MilosavljeviMerritt2003}.
These stars are ejected from the center and subtract some energy from the binary in the process.
This phase is called \textit{gravitational slingshot} because of the ejection of stars.
Eventually the dynamical friction plus the slingshot phases shrink the orbit enough,
so that the binary can continue inspiralling until a coalescence in a finite interval of time by only emitting \textit{gravitational radiation},
which constitutes the third phase.
Other possible evolutions involving interaction with surrounding gas have been investigated in the literature \cite{DottiEtAl2006,EscalaEtAl2005,KocsisEtAl2010}.

We impose that the minimum frequency is the one at which the gravitational slingshot phase ends and the gravitational radiation starts to dominate
(see the discussion in Section \ref{sec:minfreqmbhb}).
As we now show, this condition is reasonably well fulfilled by imposing the same maximum inspiral time $T_{\text{max}}=75$\,Myr for all masses.

The frequency at which the slingshot and radiation phases overlap is the one at which the energy spectra of the two phases are equal,
\begin{equation}
\label{eq:energyphases}
\frac{dE_e}{df_e}\bigg|_{S}=\frac{dE_e}{df_e}\bigg|_{R}.
\end{equation} 
The variation of the energy of the gravitational waves with their frequency, in any of the two phases, can be written as
\begin{equation}
\label{eq:dedfdedt}
\frac{dE_e}{df_e}=\frac{dE_e}{dt_e}\frac{dt_e}{ds}\frac{ds}{df_e}=\frac{dE_e}{dt_e}\left[\frac{ds}{dt_e}\right]^{-1}\frac{ds}{df_e}.
\end{equation} 
Here, $dt_e$ is an interval of time and $s$ is the separation of the two black holes,
which is the semi-major axis of the ellipse described.
Since the orbit is assumed circular, $s$ corresponds to the radius of the orbit.
The term $dE_e/dt_e$ is the same in both phases.
Also $ds/df_e$ has the same form in the two phases.
Only the evolution of the semi-major axis in time, $ds/dt_e$, is different.
Thus, instead of finding the frequency that fulfills Equation (\ref{eq:energyphases}), we can obtain the separation $s$ at which
\begin{equation}
\label{eq:dadtsr}
\frac{ds}{dt_e}\bigg|_{S}=\frac{ds}{dt_e}\bigg|_{R},
\end{equation} 
and then calculate the corresponding frequency using (\ref{eq:radfreq}).

Following \cite{SesanaEtAl2004} (or similarly \cite{Quinlan1996}),
we write the evolution in time of the semi-major axis of a binary in the two phases.
In the gravitational radiation phase, this evolution fulfills
\begin{equation}
\label{eq:dadtr}
\frac{ds}{dt_e}\bigg|_R=-\frac{64G^3m_1 m_2 [m_1+m_2]}{5 c^5 s^3},
\end{equation} 
whereas in the gravitational slingshot phase,
\begin{equation}
\label{eq:slingshotphase}
\frac{ds}{dt_e} \bigg|_{S}=-\frac{\mathcal{H}\sigma_* s^2}{2 \pi r_c^2}.
\end{equation} 
In the latter, $\mathcal{H}$ is the hardening rate, $\sigma_*$ is the velocity dispersion of the stars in the bulge of the galaxy, and
$r_c$ is the core radius (see \cite{Quinlan1996} for more details).
We use the value of $\mathcal{H}$ reached in the limit of a very hard binary, $\mathcal{H}\approx 15$.
It is known that there is a correlation between $\sigma_*$ and the mass of the massive black hole $m_{\text{BH}}$ hidden in the bulge
(see \cite{FerrareseMerritt2000} and \cite{GebhardtEtAl2000}).
This relation (from the most recent fits, by \cite{GueltekinEtAl2009}) is
\begin{equation}
\log_{10} \left(\frac{m_{\text{BH}}}{M_\odot}\right) =c_1+c_2 \log_{10} \left(\frac{\sigma_*}{200 \,\mbox{km s}^{-1}}\right),
\end{equation} 
with $(c_1,c_2)=(8.12\pm 0.08,\,4.24\pm 0.41)$.
From this equation we obtain $\sigma_*(m_{\text{BH}})$ and use $m_{\text{BH}}=m_1+m_2$ to account for the two components of the binary.
The core radius $r_c$, in the limit of a very hard binary, grows during the gravitational slingshot phase until it reaches
\begin{equation}
r_c\approx \frac{3G[m_1+m_2]}{4\sigma_*^2}\ln \left( \frac{G m_2}{4 \sigma_*^2 s} \right),
\end{equation} 
where $m_2$ is the mass of the lighter black hole.

We now calculate the separation $s$ at which both phases overlap.
Replacing (\ref{eq:dadtr}) and (\ref{eq:slingshotphase}) in (\ref{eq:dadtsr}),
\begin{equation}
\label{eq:a5ln2}
s^5\ln^{-2} \left(\frac{Gm_2}{4\sigma_*^2 s}\right)=\frac{72 \pi G^5 m_1m_2[m_1+m_2]^3}{5\mathcal{H}c^5\sigma_*^5}.
\end{equation} 
This equation can be numerically solved for each pair of equal masses $m_1=m_2=m$,
obtaining the separation (let us call it $s_R$) at which the gravitational radiation phase starts to dominate.
Using (\ref{eq:radfreq}) one can calculate the frequency $f_R$ that corresponds to $s_R$.
It turns out that the obtained dependence of $f_R$ with $m$ is very accurately fitted by $f_{\text{min}}(m)$, defined in Equation (\ref{eq:deffminapprox}),
using $T_{\text{max}}\approx 75$\,Myr.
This is a numerical coincidence that eases further calculations.
The origin of this coincidence is the following:
omitting the logarithm on the left side of Equation (\ref{eq:a5ln2}), $s_R\propto m/\sigma_*$ while $\sigma_*\propto m^{1/4.24}$.
This leads to $s_R\propto m^{0.764}$.
According to Equation (\ref{eq:radfreq}), $f\propto m^{1/2}s^{-3/2}$ and therefore $f_R\propto m^{-0.646}$.
On the other hand, according to Equation (\ref{eq:deffminapprox}), $f_{\text{min}}\propto m^{-5/8}=m^{-0.64}$.
Therefore, the dependences of $f_R$ and $f_{\text{min}}$ with $m$ are almost the same.
As a conclusion, setting a maximum inspiral time of 75\,Myr is (almost) equivalent to considering only waves emitted during the gravitational radiation phase.

\subsection{\label{sec:calcstelbin} Calculations for stellar binaries}

\subsubsection{\label{sec:modelscoalrate} Coalescence rate}
The signal comoving density rate $\dot{n}(z)$, that was defined in Equation (\ref{eq:defsigcomrate}),
represents, in the case we study now, the number of binaries per unit emitted interval of time per unit comoving volume that coalesce at a redshift $z$.
We can thus call it the \textit{coalescence rate} or simply \textit{rate}.

To obtain $\dot{n}(z)$, one could choose a star formation rate from the literature (which is usually a function of the redshift)
and transform it into a coalescence rate, for which a coalescence probability distribution is necessary.
This procedure is followed for example in \cite{Freitas1997}.
In Section \ref{sec:discrate} we show that the use of a constant coalescence rate is well justified, given the large uncertainties in the local coalescence rate.
Therefore, to simplify calculations, we assume a rate of the form
\begin{equation}
\label{eq:ratecons}
\dot{n}(z)=\left\{ \begin{array}{lc}
             0, & 0<z<z_{\text{min}} \\
	     R, & z_{\text{min}}\le z \le z_{\text{max}} \\
	    0, & z_{\text{max}}<z
             \end{array}
\right. ,
\end{equation}
for a real constant $R$.
The values of $R$ for each ensemble are given in Table \ref{tb:values}.

Some of the coalescence rates in the literature are estimated only within our galaxy.
We need to extrapolate those coalescence rates, given per Milky Way equivalent galaxy, $\mbox{MWEG}^{-1}$, to the rest of the universe.
One simple way to translate galactic rates into rates per cubic megaparsec, Mpc$^{-3}$, is explained in Section 3 of \cite{Phinney1991}.
We use the same conversion factor of \cite{LIGO2010}, which is referred to \cite{KopparapuEtAl2008},
\begin{equation}
\label{eq:convfact}
1\text{ MWEG}^{-1}=0.0116 \text{ Mpc}^{-3}.
\end{equation} 
A similar factor is given in Equation (4) of \cite{PostnovYungelson2006}.
The conversion (\ref{eq:convfact}) assumes that the blue-light luminosity of the Milky Way is $1.7\times 10^{10}$\,$L_{B,\odot}$,
where $L_{B,\odot}$ is the blue luminosity of the Sun,
while that of the close universe is $0.0198\times 10^{10}$\,$L_{B,\odot}$ per cubic megaparsec.
All these factors are very uncertain, as discussed, for example, in \cite{KalogeraEtAl2001}.
We assume no uncertainty in (\ref{eq:convfact}) but then round the coalescence rates to one significant figure.

\renewcommand{\tabcolsep}{0.4cm}
\renewcommand{\arraystretch}{2.1}

\begin{table*}
\begin{tabular}{ c | c | c | c | c | c |}
\cline{2-6}
& BH-BH & BH-NS & NS-NS & NS-WD & WD-WD \\
\hline
\multicolumn{1}{| c |}{Minimum $R/[$Myr$^{-1}$\,Mpc$^{-3}]$} & $1 \times 10^{-4}$ & $6\times 10^{-4}$ & $1\times 10^{-2}$ & $2\times 10^{-2}$ & $2 \times 10^1$\\
\hline
\multicolumn{1}{| c |}{Most likely $R/[$Myr$^{-1}$\,Mpc$^{-3}]$} & $5\times 10^{-3}$ & $3\times 10^{-2}$ & 1 & $4 \times 10^{-1}$ & $1 \times 10^2$\\
\hline
\multicolumn{1}{| c |}{Maximum $R/[$Myr$^{-1}$\,Mpc$^{-3}]$} & $3 \times 10^{-1}$ & $1$ & 9 & 9 & $5\times 10^2$\\
\hline
\end{tabular}
\caption{\label{tb:values} Minimum, most likely and maximum coalescence rates assumed for each type of ensemble.
The coalescence rates of BH-BH and BH-NS are taken from \cite{MandelShaughnessy2010},
where they refer to \cite{KalogeraEtAl2007} and \cite{ShaughnessyEtAl2008}, respectively.
For NS-NS, the values are taken from \cite{KalogeraEtAl2004Erratum} (our minimum and maximum values are the minimum and maximum ones allowed by the uncertainties).
The rates of NS-WD and WD-WD are taken from Table 1 in \cite{Nelemans2003}.
In Section \ref{sec:results} we consider also the recent coalescence rates of BH-BH predicted in \cite{BulikEtAl2011}, of $R=0.36^{+0.50}_{-0.26}$\,Mpc$^{-3}$\,Myr$^{-1}$.
The values given in the literature per Milky Way equivalent galaxy are converted using (\ref{eq:convfact}).
All coalescence rates are rounded to one significant figure.}
\end{table*}

\renewcommand{\tabcolsep}{6pt}
\renewcommand{\arraystretch}{1}

\subsubsection{Spectral function}
We now rewrite $\omov$ in a simple way.
Introducing the constant rate $R$ in (\ref{eq:omgenbin}),
\begin{equation}
\label{eq:omsol}
\omov =\delta_1 f^{2/3} \left[g(z_{\text{upp}}(f))-g(\overline{z}(f,\Delta f,\mathcal{N}_0)) \right].
\end{equation}
Here,
\begin{equation}
\label{eq:defgamma1}
\delta_1=\frac{R \kappa}{\rho_cc^2H_0}
\end{equation}  
and $g(z)$ is the solution of the integral
\begin{equation}
\label{eq:gdef}
g(z)=\int [1+z]^{-4/3}\mathcal{E}^{-1}(z)dz.
\end{equation} 
We solve this integral semi-analytically in Section \ref{sec:semian}.

\subsubsection{\label{sec:ovfun} Overlap function}
We obtain an explicit formula for the overlap function of binary systems with a constant coalescence rate.
Introducing (\ref{eq:defcomvol}) and (\ref{eq:timefreqz}) in (\ref{eq:defoverlapfun}),
\begin{align}
\label{eq:ovfunexp}
\mathcal{N}&(f,\Delta f,z)= \int_{z_{\text{low}}(f)}^z \left[\delta_2 Q(f,\Delta f) [1+z']^{-8/3}\right] R \nonumber \\
&\times \left[4 \pi  \left[\frac{c}{H_0}\int_0^{z'} \mathcal{E}^{-1}(z'')dz''\right]^2 \frac{c}{H_0} \mathcal{E}^{-1}(z')\right] dz'.
\end{align}
This intricate equation can be rewritten to obtain a simple expression for the overlap function,
\begin{equation}
\label{eq:ovfunlong}
\mathcal{N}(f,\Delta f,z)= \delta_2 \delta_3 Q(f,\Delta f)[\overline{g}(z)-\overline{g}(z_{\text{low}}(f))].
\end{equation}
Here we have defined
\begin{equation}
\delta_3=4 \pi R \frac{c^3}{H_0^3}
\end{equation}
and
\begin{equation}
\label{eq:gbardef}
\overline{g}(z)= \int [1+z]^{-8/3}\left[ \int_0^z \mathcal{E}^{-1}(z') dz' \right]^2 \mathcal{E}^{-1}(z) dz.
\end{equation} 
This integral cannot be analytically solved.
One can invert (\ref{eq:ovfunlong}) with respect to the redshift, obtaining
\begin{equation}
\label{eq:redshiftfunc}
\mathcal{N}^{-1}(f,\Delta f,\mathcal{N}_0)=\overline{g}^{-1}\left(\frac{\mathcal{N}_0}{\delta_2 \delta_3 Q(f,\Delta f)}+\overline{g}(z_{\text{low}}(f))\right).
\end{equation} 
In Section \ref{sec:semian} we give a semi-analytical solution for $\mathcal{N}(f,\Delta f,z)$ and $\mathcal{N}^{-1}(f,\Delta f,\mathcal{N}_0)$.

\subsubsection{\label{sec:limfreqs} Limiting frequencies}
The limiting frequencies $f_{\text{p,min}}$, $f_{\text{d,min}}$, $f_{\text{d,max}}$ and $f_{\text{p,max}}$
are defined in Section \ref{sec:resolvability}.
For the systems we study, $f_{\text{p,min}}$ and $f_{\text{d,min}}$ turn out to be close to $f_{\text{min}}/[1+z_{\text{max}}]$,
which is the minimum frequency at which the spectral function has support.
For simplicity, we assume
\begin{equation}
f_{\text{p,min}}=f_{\text{d,min}}=\frac{f_{\text{min}}}{[1+z_{\text{max}}]}.
\end{equation}
On the other hand, the frequencies $f_{\text{d,max}}$ and $f_{\text{p,max}}$ must be calculated using Equations (\ref{eq:fdmaxdef}) and (\ref{eq:deffpmax}), respectively.

We now show how to calculate $f_{\text{p,max}}$.
Inserting Equation (\ref{eq:redshiftfunc}) in (\ref{eq:deffpmax}),
\begin{equation}
\label{eq:fpmaxdefalt}
Q(f_{\text{p,max}},\Delta f)=\frac{\mathcal{N}_0}{\delta_2 \delta_3 [\overline{g}(z_{\text{upp}}(f_{\text{p,max}}))-\overline{g}(z_{\text{low}}(f_{\text{p,max}}))]}.
\end{equation}
One can obtain $f_{\text{p,max}}$ by solving this equation.
However, one can use a more convenient formula for $f_{\text{p,max}}$, that we present now.
All stellar binaries satisfy $f_{\text{min}}/[1+z_{\text{min}}]<f_{\text{p,max}}$, so $z_{\text{low}}(f_{\text{p,max}})=z_{\text{min}}$.
Adopting a frequency resolution $\Delta f=1$\,yr$^{-1}$, the condition $\Delta f\ll f_{\text{p,max}}$ is fulfilled by all stellar binaries.
We can thus use the approximation of Equation (\ref{eq:aproxq}) in (\ref{eq:fpmaxdefalt}), obtaining
\begin{equation}
\label{eq:deffpmaxalt}
f_{\text{p,max}}\approx \left\{ \begin{array}{lc}
\left[\frac{8 \Delta f \delta_2 \delta_3 [\overline{g}(z_{\text{max}})-\overline{g}(z_{\text{min}})]}{3\mathcal{N}_0} \right]^{3/11}, & \Xi\le 1 \\
\left[\frac{8 \Delta f \delta_2 \delta_3 \left[\overline{g}\left(\frac{f_{\text{max}}}{f_{\text{p,max}}}-1\right)
-\overline{g}(z_{\text{min}})\right]}{3\mathcal{N}_0} \right]^{3/11}, & \Xi>1\\
\end{array} \right. ,
\end{equation} 
where the dimensionless parameter $\Xi$ is defined by
\begin{equation}
\Xi=\frac{Q(\frac{f_{\text{max}}}{1+z_{\text{max}}},\Delta f)\delta_2 \delta_3 [\overline{g}(z_{\text{max}})-\overline{g}(z_{\text{min}})]}{\mathcal{N}_0}.
\end{equation} 
If $\Xi \le 1$, we have a simple formula for $f_{\text{p,max}}$.
The condition $\Xi \le 1$ is fulfilled by all stellar binaries that do not contain a white dwarf.
For NS-WD and WD-WD, $\Xi > 1$, and one has to solve Equation (\ref{eq:deffpmaxalt}) numerically.

Similarly, one can obtain a formula for the limiting frequency $f_{\text{d,max}}$,
using Equations (\ref{eq:fdmaxdef}), (\ref{eq:omsol}) and (\ref{eq:aproxq}).
We point out that $f_{\text{d,max}}$ is by definition smaller than $f_{\text{p,max}}$.
In addition, one can show that $f_{\text{d,max}}/f_{\text{p,max}}$ cannot be smaller than a certain factor $F$, so
\begin{equation}
F \le \frac{f_{\text{d,max}}}{f_{\text{p,max}}}<1.
\end{equation} 
The value of this factor is
\begin{equation}
F=\left(\frac{\overline{g}\left(g^{-1}\left(\frac{1}{2}[g(z_{\text{max}})+g(z_{\text{min}})]\right)\right)-
\overline{g}(z_{\text{min}})}{\overline{g}(z_{\text{max}})-\overline{g}(z_{\text{min}})}\right)^{3/11}.
\end{equation} 
For $z_{\text{max}}=5$ and $z_{\text{min}}=0$, one obtains $F\approx 0.6$.
All stellar binaries except WD-WD fulfill that $f_{\text{d,max}}\approx 0.6\times f_{\text{p,max}}$.
For WD-WD, $f_{\text{d,max}}$ and $f_{\text{p,max}}$ are almost equal; moreover, they are almost as large as $f_{\text{max}}$.
This means that the total background of WD-WD is almost entirely dominated by its unresolvable part.

One should notice that $\Omega(f,\Delta f,0)$ is equivalent to the old definition of the spectral function, $\om$, in Equation (\ref{eq:omegalim}).
Setting $\mathcal{N}_0=0$, the function $\mathcal{N}^{-1}(f,\Delta f,0)$ becomes $z_{\text{low}}(f)$ (using Equation (\ref{eq:redshiftfunc})).
Then, the limiting frequencies $f_{\text{p,min}}$ and $f_{\text{p,max}}$ become $f_{\text{min}}/[1+z_{\text{max}}]$ and $f_{\text{max}}/[1+z_{\text{min}}]$,
respectively (see Figures \ref{fig:pfun} and \ref{fig:pfun2}).
Using Equation (\ref{eq:zoverlap}), $\overline{z}(f,\Delta f,0)$ becomes identically $z_{\text{low}}(f)$,
and thus Equations (\ref{eq:omegalim}) and (\ref{eq:omgensol}) become equivalent.

\subsubsection{\label{sec:semian} Semi-analytical solutions}
In order to obtain a semi-analytical solution for $\omov$, we need two functions,
$g(z)$ and $\overline{g}(z)$, that fit accurately the numerical solutions of the integrals in Equations (\ref{eq:gdef}) and (\ref{eq:gbardef}).

A possible choice of the functions $g(z)$ and $\overline{g}(z)$ is
\begin{equation}
\label{eq:defg}
g(z)=a_1 \, \arctan^{a_4}\left(a_2\,z^{a_3} \right)
\end{equation} 
and
\begin{equation}
\label{eq:defgbarra}
\overline{g}(z)=\overline{a}_1\, \arctan^{\overline{a}_4} \left(\overline{a}_2\, z^{\overline{a}_3}\right),
\end{equation} 
for certain parameters $(a_1,a_2,a_3,a_4)$ and $(\overline{a}_1,\overline{a}_2,\overline{a}_3,\overline{a}_4)$ that must be numerically calculated.
The optimal parameters between $z_{\text{min}}=0$ and $z_{\text{max}}=5$ are
\begin{equation}
(a_1,a_2,a_3,a_4)=(0.5604,1.235,1.0047,0.8364)
\end{equation}
and
\begin{equation}
(\overline{a}_1,\overline{a}_2,\overline{a}_3,\overline{a}_4)=(0.07024,0.8658,1.3236,1.511).
\end{equation}
These values can be used for all ensembles of stellar binaries,
since the integrals in Equations (\ref{eq:gdef}) and (\ref{eq:gbardef}) depend only on cosmological parameters.

The semi-analytical formula for the overlap function, using (\ref{eq:defgbarra}), becomes
\begin{equation}
\mathcal{N}(f,\Delta f,z)=\delta_2 \,\delta_3 \,Q(f,\Delta f) \,\overline{a}_1\,\arctan^{\overline{a}_4} \left(\overline{a}_2\,z^{\overline{a}_3} \right).
\end{equation} 
We invert it with respect to the redshift, obtaining
\begin{equation}
\label{eq:zrayadef}
\mathcal{N}^{-1}(f,\Delta f,\mathcal{N}_0)=\left[ \frac{1}{\overline{a}_2}
\tan \left( \left[\frac{\mathcal{N}_0}{\overline{a}_1\delta_2 \delta_3 Q(f,\Delta f)}\right]^{1/\overline{a}_4}\right) \right]^{1/\overline{a}_3}.
\end{equation}
Introducing it in (\ref{eq:zoverlap}) we obtain a formula for $\overline{z}(f,\Delta f,\mathcal{N}_0)$.

Finally, using (\ref{eq:defg}), the semi-analytical formula for the spectral function of binary systems is
\begin{align}
\label{eq:omsemian}
\omov =& \delta_1 a_1 f^{2/3}\left[\arctan^{a_4} (a_2 z_{\text{upp}}^{a_3}(f)) \right.\nonumber \\
&\left.-\arctan^{a_4} (a_2 \overline{z}^{a_3}(f,\Delta f,\mathcal{N}_0))\right].
\end{align} 
The redshift function $z_{\text{upp}}(f)$ is given in (\ref{eq:defzupp}).
The limiting frequency $f_{\text{p,max}}$ can be calculated as explained in Section \ref{sec:limfreqs},
using the semi-analytical formula (\ref{eq:defgbarra}) for $\overline{g}(z)$.

\subsubsection{\label{sec:valuesstellar} Mass ranges}
We calculate the spectral function of an ensemble assuming that all similar objects have equal mass.
For example, in the ensemble of NS-WD, all neutron stars have equal mass $m_{\text{NS}}$ and all white dwarfs have equal mass $m_{\text{WD}}$.
For this reason, given a range of possible masses for an object, we should not consider values of masses too different from the mean one.

For a neutron star, we assume a mass in the range $1.3\le m_{\text{NS}}/M_\odot\le 1.7$.
This interval is taken from \cite{StrobelWeigel2001}, where the lower limit predicted is $(0.878-1.284)\,M_\odot$, and the upper limit, $(1.699-2.663)\,M_\odot$.
We use the largest mass of the lower limit and the smallest mass of the upper limit and round all values to two significant figures.
The most likely value is the average of both limits of the interval.
So, our choice for the masses of neutron stars is
\begin{equation}
\label{eq:nsmasses}
(m_{\text{NS}}^{\text{min}}, m_{\text{NS}}^{\text{med}},m_{\text{NS}}^{\text{max}})=(1.3,1.5,1.7)\,M_\odot.
\end{equation}

The mass distribution of white dwarfs of spectral type DA, according to \cite{LimogesBergeron2010},
is described by a Gaussian distribution with mean $\mu=$0.606\,$M_\odot$ and standard deviation $\sigma=$0.135\,$M_\odot$.
The distribution of white dwarfs of spectral type DB has $\mu=$0.758\,$M_\odot$ and $\sigma=$0.192\,$M_\odot$.
Since we do not make a distinction between DA and DB white dwarfs, we calculate the Gaussian distribution that best fits the average of both distributions,
obtaining $\mu=$0.663\,$M_\odot$ and $\sigma=$0.177\,$M_\odot$.
Similar results can be obtained using, for example, the distributions given in \cite{KeplerEtAl2007}.
We assume a minimum mass of $\mu-\sigma=$0.49\,$M_\odot$ and a maximum one of $\mu+\sigma=$0.84\,$M_\odot$.
Thus,
\begin{equation}
\label{eq:wdmasses}
(m_{\text{WD}}^{\text{min}}, m_{\text{WD}}^{\text{med}}, m_{\text{WD}}^{\text{max}})=(0.49,0.66,0.84)\,M_\odot.
\end{equation}

For stellar-mass black holes, we calculate the mean $\mu$ and standard deviation $\sigma$ of the list of masses given in Table 1 of \cite{Ziolkowski2008},
obtaining $\mu=7.8$\,$M_\odot$ and $\sigma=3.7$\,$M_\odot$.
We assume for the minimum mass $\mu -\sigma=$4.1\,$M_\odot$ and for the maximum one $\mu+\sigma=$12\,$M_\odot$.
Again, the most likely value is the average of both.
Similar results can be achieved with the masses of Table 1 of \cite{Casares2007}.
The masses we use are, therefore,
\begin{equation}
\label{eq:bhmasses}
(m_{\text{BH}}^{\text{min}}, m_{\text{BH}}^{\text{med}}, m_{\text{BH}}^{\text{max}})=(4.1,7.8,12)\,M_\odot.
\end{equation}


\subsection{\label{sec:mbhb} Calculations for massive black hole binaries}
The masses of MBH-MBH range several orders of magnitude.
It is reasonable to expect a very different number of signals produced by binaries of chirp mass 10$^2$\,M$_{\odot}$ than by binaries of 10$^{10}$\,M$_{\odot}$.
To be consistent with the given definition of ensemble (a population of many systems with similar properties and behaviour),
MBH-MBH form a \textit{superensemble} composed of many ensembles, each one characterized by an infinitesimal range of chirp masses.

The coalescence rate now depends on the chirp mass and the redshift.
Instead of $\dot{n}(z)$ one now has a signal comoving density rate of the form $\dot{\overline{n}}(\mathcal{M},z)$.
This gives the number of signals per unit emitted interval of time per unit comoving volume per unit chirp mass.
We do not have an analytical formula for $\dot{\overline{n}}(\mathcal{M},z)$
\footnote{The numerical values of the functions $\dot{\overline{n}}(\mathcal{M},z)$ (for each of the four models mentioned at the beginning of Section \ref{sec:models})
were kindly provided by A. Sesana and M. Volonteri in a private communication}.

The spectral function of the total background of the superensemble is
\begin{equation}
\label{eq:ommbhbnocor}
\Omega_{\text{total}}(f)=f^{2/3}\int_{z_{\text{min}}}^{z_{\text{max}}}
\int_{\mathcal{M}_{\text{low}}(z',f)}^{\mathcal{M}_{\text{upp}}(z',f)} I_1(\mathcal{M}',z') d\mathcal{M}'dz',
\end{equation} 
where
\begin{equation}
I_1(\mathcal{M},z)= \frac{8[G \pi \mathcal{M}]^{5/3}}{9c^2H_0^3}\dot{\overline{n}}(\mathcal{M},z)[1+z]^{-4/3}\mathcal{E}^{-1}(z).
\end{equation} 
One can notice that (\ref{eq:ommbhbnocor}) is the same as (\ref{eq:omgenbin}),
just changing $\dot{n}(z)$ by $\dot{\overline{n}}(\mathcal{M},z)d\mathcal{M}$ and integrating over chirp mass.
The functions $\mathcal{M}_{\text{low}}(z,f)$ and $\mathcal{M}_{\text{upp}}(z,f)$ give, at every frequency and redshift,
the minimum and maximum chirp masses that can contribute, respectively.
In other words, the interval $[\mathcal{M}_{\text{low}}(z,f),\mathcal{M}_{\text{upp}}(z,f)]$ contains the chirp masses of those binaries which, at redshift $z$,
have minimum frequency $f_{\text{min}}\le f$ and maximum frequency $f_{\text{max}}\ge f$.
They are obtained by inverting $f_{\text{min}}$ (Equation (\ref{eq:deffminapprox})) and $f_{\text{max}}$ (Equation (\ref{eq:flsoformula})), respectively,
with respect to $\mathcal{M}$.
Hence,
\begin{equation}
\mathcal{M}_{\text{low}}(z,f)=\left[ \frac{5 c^5}{256 \pi^{8/3}G^{5/3} T_{\text{max}}}\right]^{3/5}\left[f[1+z]\right]^{-8/5},
\end{equation} 
and
\begin{equation}
\mathcal{M}_{\text{upp}}(z,f)=\frac{c^3}{6\,\sqrt{6}\,2^{6/5}\,\pi \, G}[f\,[1+z]]^{-1}.
\end{equation} 
In the last equation we have used that, if the two masses of the binary are equal, then $m_1=m_2=m=2^{1/5}\mathcal{M}$.

The overlap function of the total background of the superensemble is
\begin{align}
\label{eq:ovmbhbnocor}
\mathcal{N}(f,\Delta f,z)=&\int_{z_{\text{min}}}^{z} \int_{\mathcal{M}_{\text{low}}(z',f)}^{\mathcal{M}_{\text{upp}}(z',f)} Q(f,\Delta f) \nonumber \\
&\times I_2(\mathcal{M}',z')d\mathcal{M}'dz',
\end{align} 
where
\begin{align}
I_2&(\mathcal{M},z)= \left[\frac{5 c^5}{256\pi^{8/3}[G\mathcal{M}]^{5/3}} [1+z]^{-8/3}\right] \dot{\overline{n}}(\mathcal{M},z) \nonumber \\
&\times\left[4 \pi  \left[\frac{c}{H_0}\int_0^{z} \mathcal{E}^{-1}(z')dz'\right]^2 \frac{c}{H_0} \mathcal{E}^{-1}(z)\right].
\end{align}
Equation (\ref{eq:ovmbhbnocor}) is the same as (\ref{eq:ovfunexp}), just changing $\dot{n}(z)$ by $\dot{\overline{n}}(\mathcal{M},z)d\mathcal{M}$
and integrating over chirp mass.


Section 4 of \cite{SesanaEtAl2008} describes a discrepancy between a semi-analytical calculation of the unresolvable background of MBH-MBH and a Monte Carlo simulation.
The discrepancy occurs because the semi-analytical approach does not take into account the discrete nature of the systems.
To account for it, they change the range of masses considered in the semi-analytical calculation.
We now proceed in a similar way, to calculate the unresolvable part of the background.

The average number of signals with frequency equal or larger than $f$ and chirp mass equal or larger than $\mathcal{M}$ is
\begin{align}
\label{eq:nfeq}
\overline{\mathcal{N}}(f,\mathcal{M})=&\int_{\mathcal{M}}^{\mathcal{M}_{\text{max}}}
\int_{z_{\text{low}}(\mathcal{M}',f)}^{z_{\text{upp}}(\mathcal{M}',f)} Q(f,f_{\text{max}}(\mathcal{M}')-f) \nonumber \\
&\times I_2(\mathcal{M}',z')dz'd\mathcal{M}'.
\end{align}
We impose that a signal, emitted at frequency $f$ by a binary with chirp mass $\mathcal{M}$,
can contribute to the continuous or the unresolvable background only if $\overline{\mathcal{N}}(f,\mathcal{M})\ge 1$.
The largest chirp mass $\overline{\mathcal{M}}(f)$ that contributes at frequency $f$ is obtained by solving
$\overline{\mathcal{N}}(f,\overline{\mathcal{M}}(f))=1$.
We calculate $\overline{\mathcal{M}}(f)$ numerically, use it as the upper limit of the integral over chirp mass in Equation (\ref{eq:ovmbhbnocor}),
and equate $\mathcal{N}(f,\Delta f, z)$ to 1:
\begin{align}
\mathcal{N}(f,\Delta f,z)=&\int_{z_{\text{min}}}^{z} \int_{\mathcal{M}_{\text{low}}(z',f)}^{\overline{\mathcal{M}}(f)} Q(f,\Delta f) \nonumber \\
&\times I_2(\mathcal{M}',z')d\mathcal{M}'dz'=1.
\end{align} 
Inverting the result of this equation with respect to the redshift, one obtains $\mathcal{N}^{-1}(f,\Delta f,1)$.
Signals with frequency $f$ emitted by binaries with chirp masses in the range $[\mathcal{M}_{\text{low}}(z,f),\overline{\mathcal{M}}]$
form an unresolvable background if their redshifts are larger than $\mathcal{N}^{-1}(f,\Delta f,1)$.

The spectral function of the unresolvable background is, therefore, 
\begin{align}
\label{eq:omsmbhbcorov1}
\Omega_{\text{unresolvable}}=&f^{2/3}\int_{\mathcal{N}^{-1}(f,\Delta f,1)}^{z_{\text{max}}}
\int_{\mathcal{M}_{\text{low}}(z',f)}^{\overline{\mathcal{M}}(f)} I_1(\mathcal{M}',z') \nonumber\\
&\times d\mathcal{M}'dz'.
\end{align} 
Similarly, one can solve
\begin{align}
\mathcal{N}(f,f_{\text{max}}-f,z)=&\int_{z_{\text{min}}}^{z} \int_{\mathcal{M}_{\text{low}}(z',f)}^{\overline{\mathcal{M}}(f)} Q(f,f_{\text{max}}-f) \nonumber \\
&\times I_2(\mathcal{M}',z')d\mathcal{M}'dz'=1
\end{align}
and invert it with respect to the redshift, obtaining a function $\mathcal{N}^{-1}(f,f_{\text{max}}-f,1)$.
Replacing $\mathcal{N}^{-1}(f,\Delta f,1)$ by $\mathcal{N}^{-1}(f,f_{\text{max}}-f,1)$ in Equation (\ref{eq:omsmbhbcorov1}), one gets
\begin{align}
\Omega_{\text{continuous}}=&f^{2/3}\int_{\mathcal{N}^{-1}(f,f_{\text{max}}-f,1)}^{z_{\text{max}}}
\int_{\mathcal{M}_{\text{low}}(z',f)}^{\overline{\mathcal{M}}(f)} I_1(\mathcal{M}',z') \nonumber\\
&\times d\mathcal{M}'dz',
\end{align} 
which is the spectral function of the continuous background.

\section{\label{sec:results}Results}
\begin{figure*}
\includegraphics{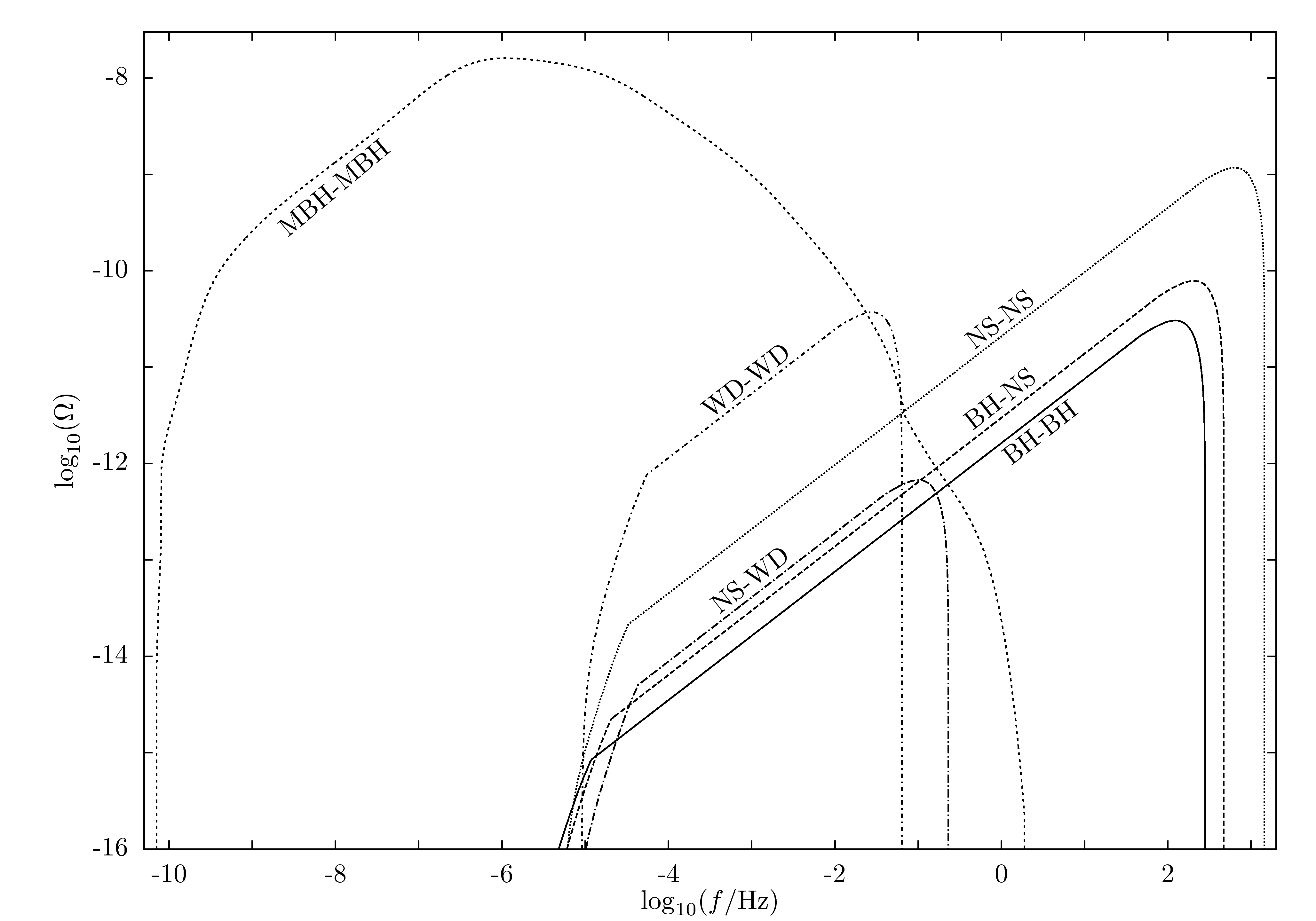}
\caption{Spectral function of the total background versus observed frequency.
The contributions of the different ensembles are calculated with the most likely values of masses and coalescence rates.
No restrictions in the duration of the signals are assumed in this plot, which means that also very short and sporadic signals are taken into account.
As discussed in the text, the spectral function in such circumstances should not be compared to the sensitivity curves of a detector.}
\label{fig:plotsjoint}
\end{figure*}

\begin{figure*}
\includegraphics{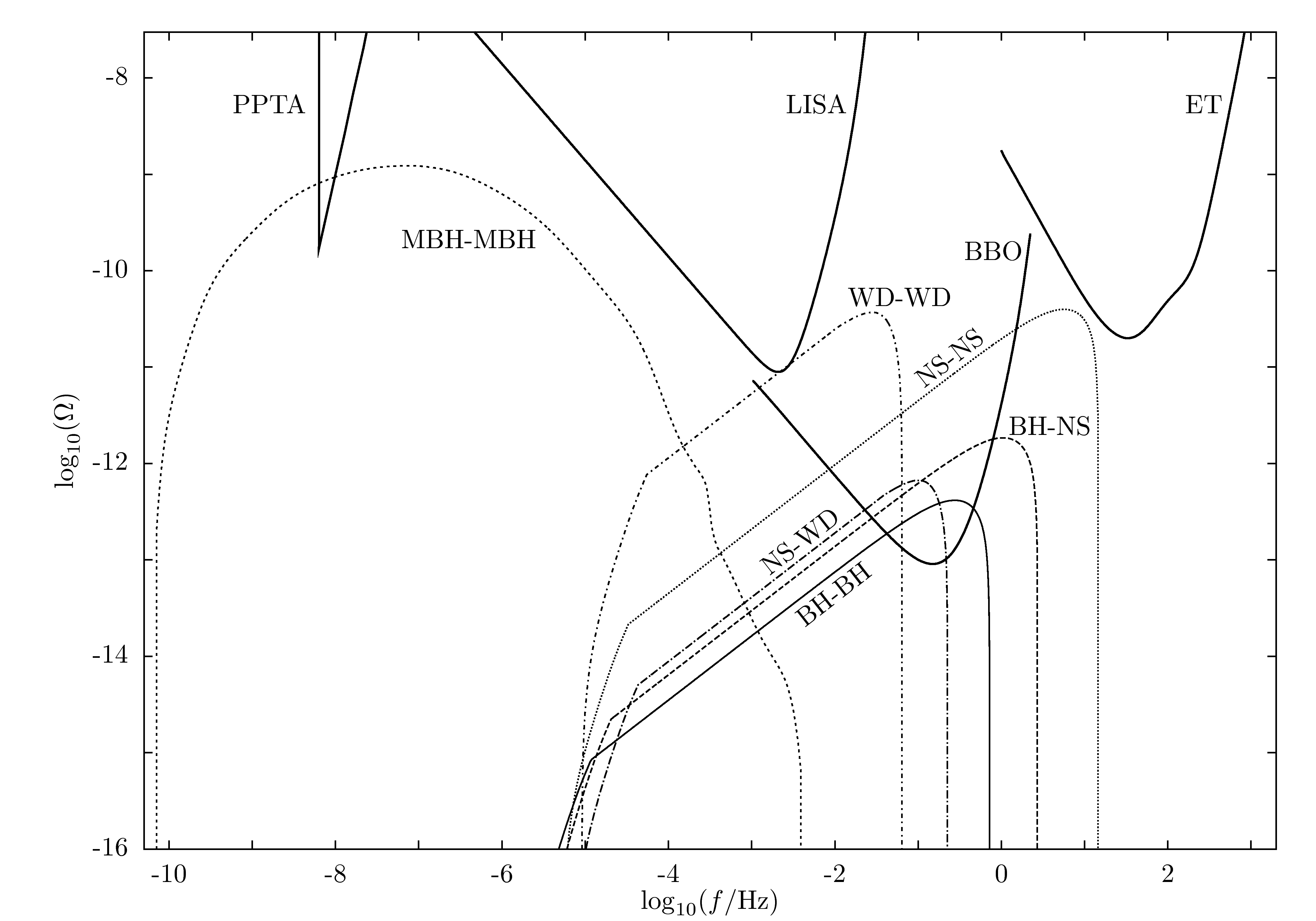}
\caption{Spectral function of the continuous background versus observed frequency.
The contributions of the different ensembles are calculated with the most likely values of masses and coalescence rates.
The sensitivity curves of LISA\footnote{Obtained using \url{http://www.srl.caltech.edu/~shane/sensitivity/MakeCurve.html} with the standard parameters.},
ET (from \cite{Regimbau2011}), BBO (from \cite{CutlerHarms2006}) and the complete Parkes PTA (from \cite{SesanaEtAl2008}) are plotted for comparison.}
\label{fig:plotsjointcont}
\end{figure*}

\begin{figure*}
\includegraphics{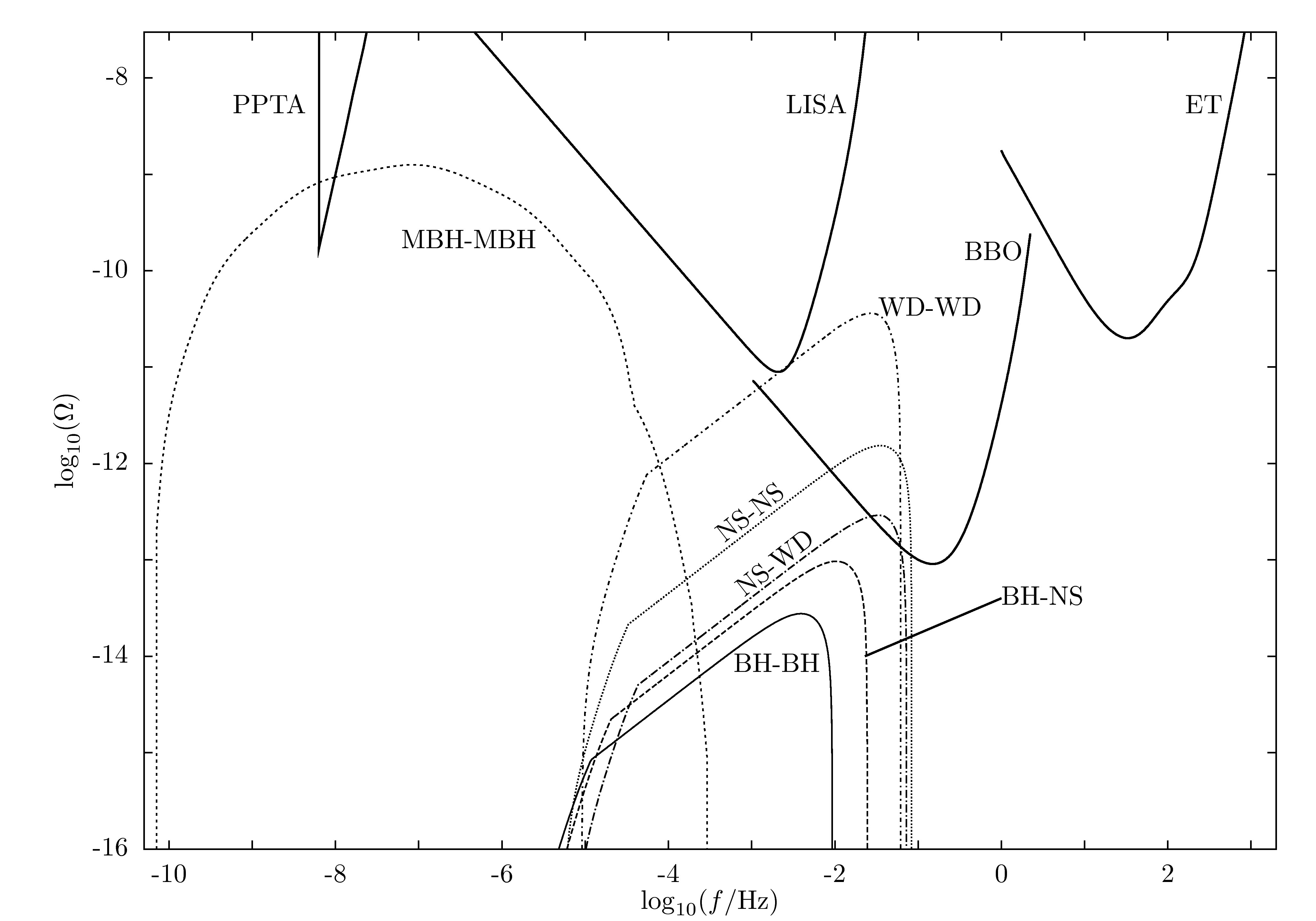}
\caption{Spectral function of the unresolvable background versus observed frequency, using $\mathcal{N}_0=1$ and $\Delta f=1$\,yr$^{-1}$.
The contributions of the different ensembles are calculated with the most likely values of masses and coalescence rates.}
\label{fig:plotsjointunres}
\end{figure*}

\begin{figure*}
\includegraphics{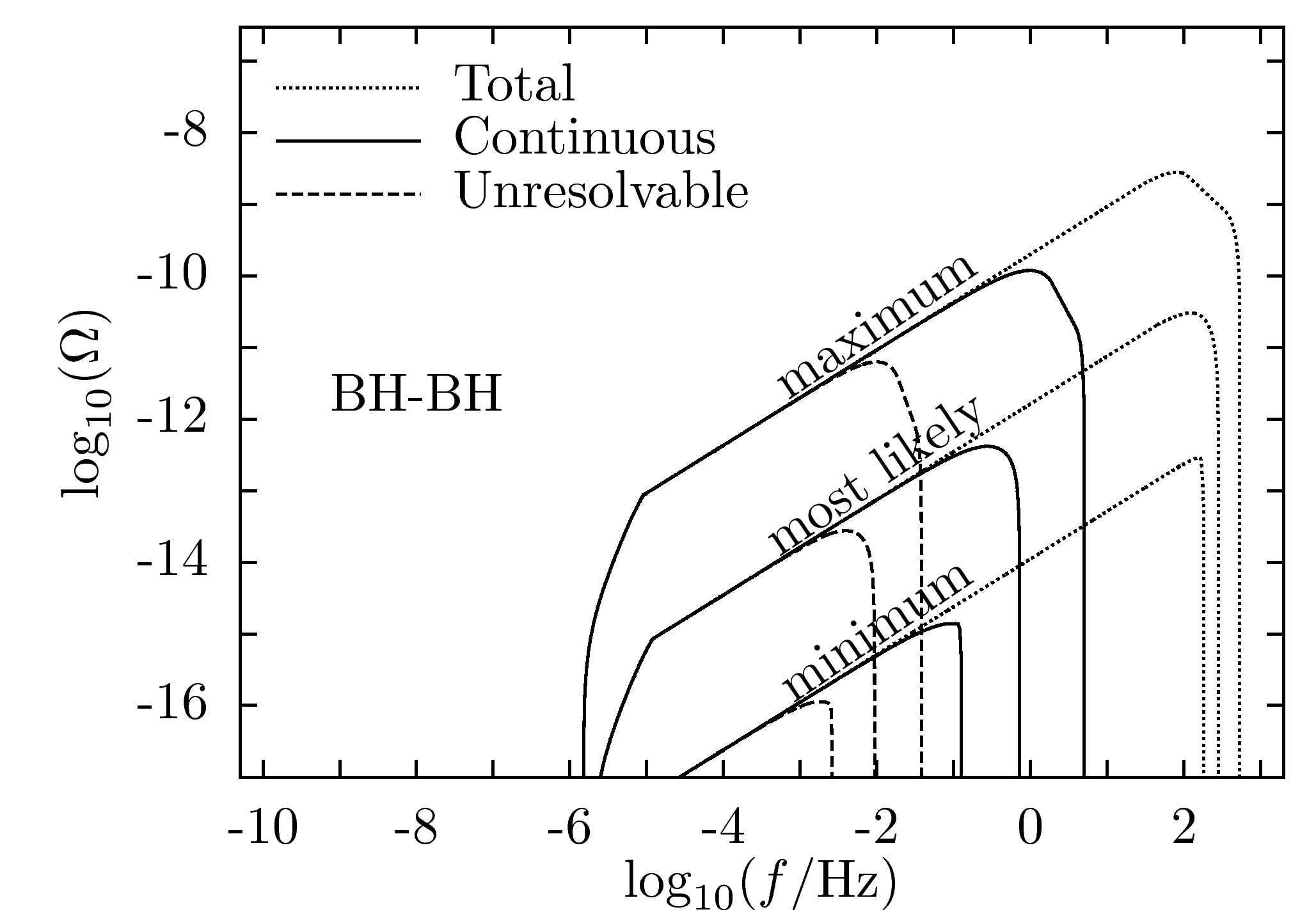}
\includegraphics{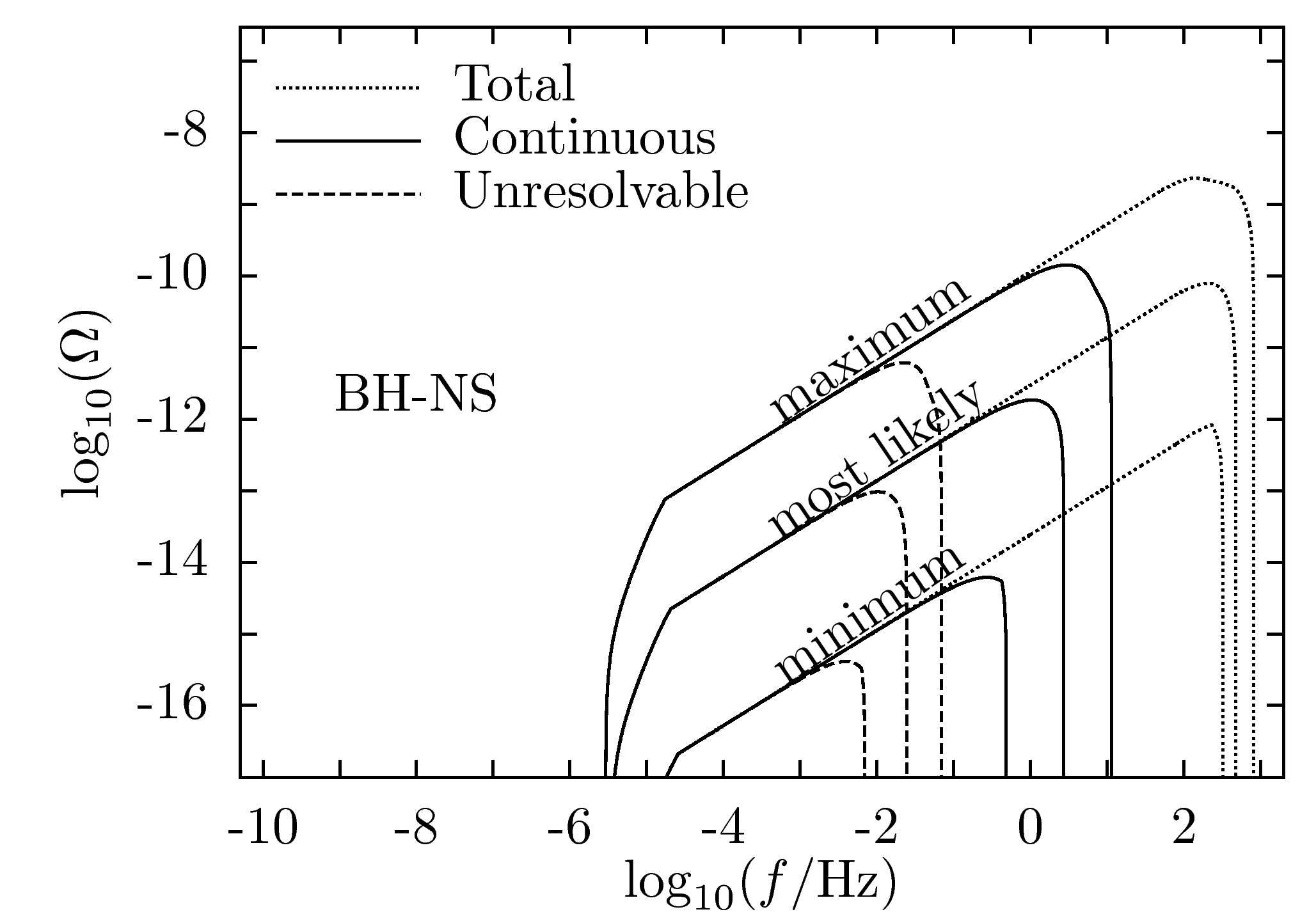}
\includegraphics{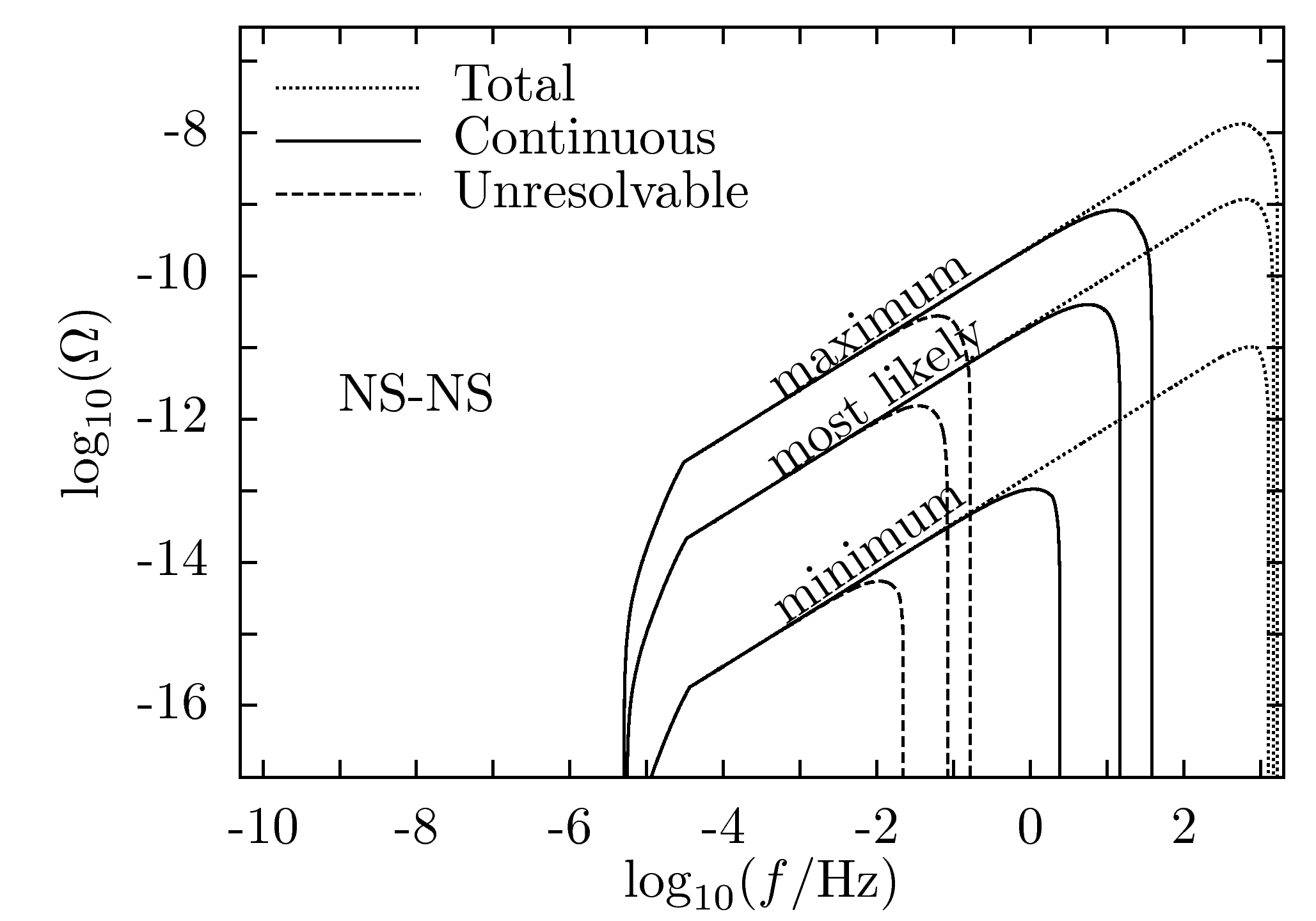}
\includegraphics{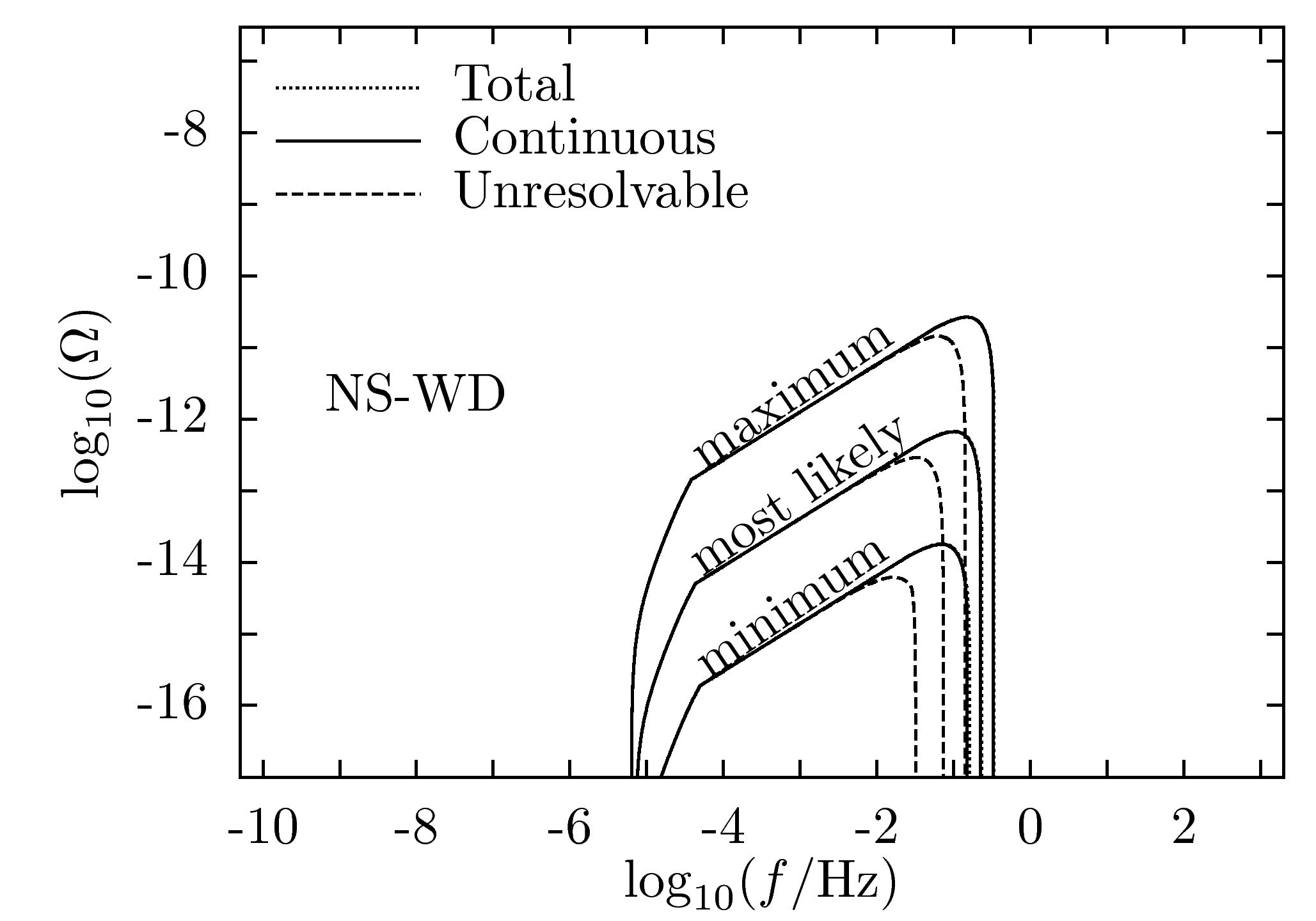}
\includegraphics{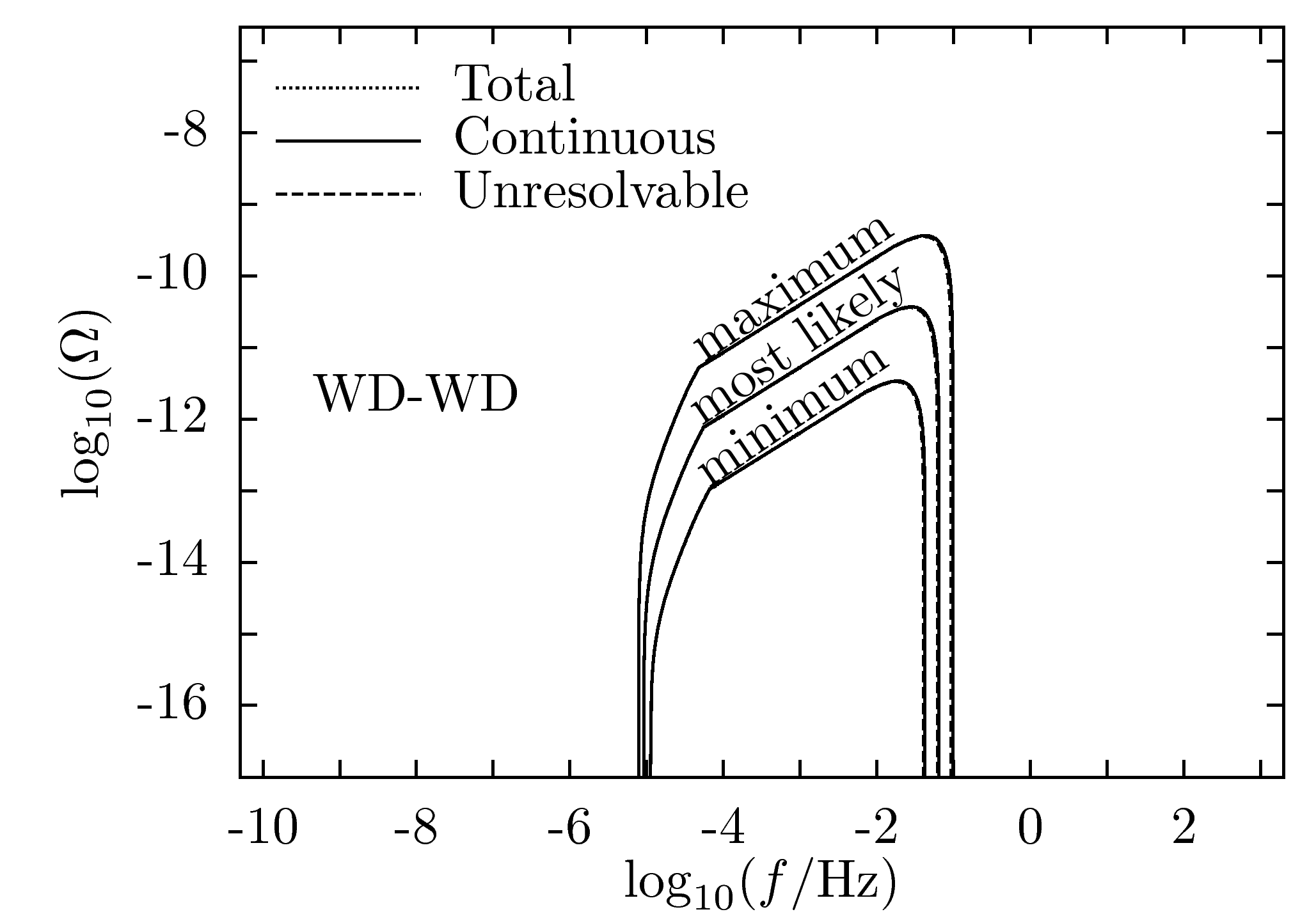}
\includegraphics{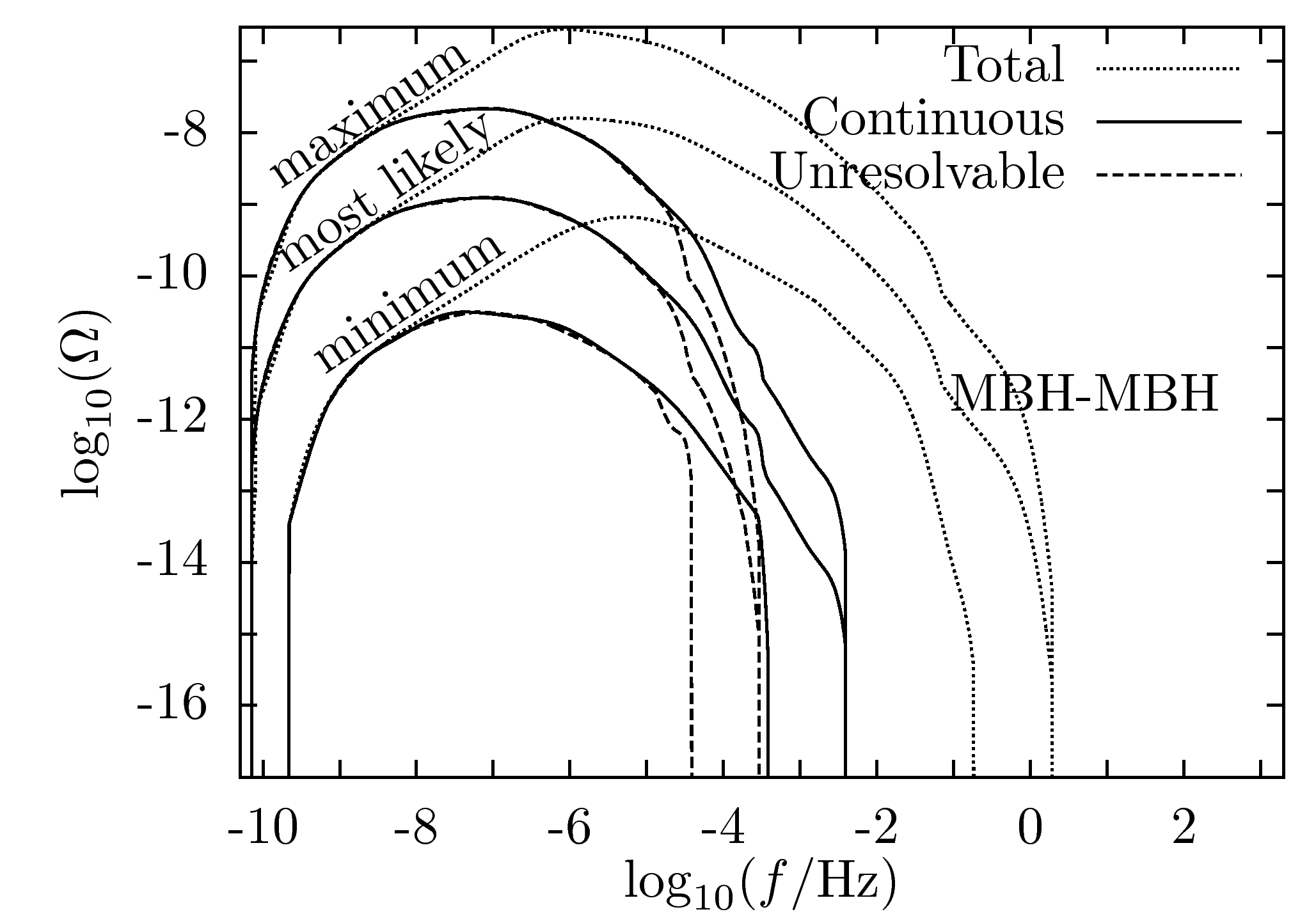}
\caption{Spectral function of the total, continuous and unresolvable backgrounds of the different ensembles, versus observed frequency.
In each plot there are nine curves: three of them are calculated with the highest values of coalescence rates,
three with the most likely, and three with the lowest values possible.
Three of the curves represent the total background, three the continuous part, and three the unresolvable part.
In the case of NS-WD, the total and continuous curves are almost indistinguishable.
The same occurs for WD-WD with the total, continuous and unresolvable curves.}
\label{fig:plotssep}
\end{figure*}

The main results of this paper are presented in Figures \ref{fig:plotsjoint}, \ref{fig:plotsjointcont}, \ref{fig:plotsjointunres}, and \ref{fig:plotssep}.
In Figure \ref{fig:plotsjoint} we show the spectral function of the total background of each ensemble.
In Figure \ref{fig:plotsjointcont}, the spectral function is plotted only in those regions where the background is continuous.
Finally, Figure \ref{fig:plotsjointunres}, which is the most relevant plot of the three, shows the unresolvable background produced by the different ensembles,
assuming $\mathcal{N}_0=1$ and $\Delta f=1$\,yr$^{-1}$.
In these three figures, the values of masses and coalescence rates are the most likely ones.

One can clearly conclude, from Figure \ref{fig:plotsjointunres},
that ground-based detectors operate (and will operate) in a frequency range clean of confusion noise from binary systems.
Without taking into account other possible sources of unresolvable background, this frequency range could be a good scenario for the detection of primordial backgrounds.

In Figure \ref{fig:plotssep} we have plotted the contribution of each ensemble separately.
For each ensemble, there are three different curves of $\omov$: one \textit{maximum}, one \textit{minimum} and one \textit{most likely},
depending on the values of masses and rates.

For stellar binaries, the most likely expectation of $\omov$ is obtained by using the most likely chirp mass and coalescence rate.
The upper curve of $\omov$ is the upper envelope of all curves that are obtained using the maximum rate and sweeping over all possible values of chirp mass.
Similarly, the lower curve is the lower envelope of all curves obtained with the minimum rate and sweeping over all chirp masses.

For massive black hole binaries, the most likely curve is the average of the spectral functions calculated, as explained in Section \ref{sec:mbhb},
for each of the four models considered.
The upper and lower curves are 10 and $1/10$ times the most likely, respectively.
These uncertainties have not been precisely calculated.
Given the lack of observational information about many of the parameters involved, any accurate calculation of the uncertainties would be arbitrary.
More precise errors are calculated in \cite{SesanaEtAl2008}, based on the results of different theoretical models.
The ranges of uncertainty given in \cite{SesanaEtAl2008} are similar to the ones we propose.

One sees in Figure \ref{fig:plotsjointunres} that the unresolvable background is clearly dominated by the contribution of WD-WD, below $\sim10^{-1}$\,Hz,
and of MBH-MBH, below $\sim10^{-4}$\,Hz.

The contribution of galactic binaries, which is believed to produce confusion noise in the frequency window of LISA,
has not been included in the plots, since it cannot be calculated using $\om$.
The spectral function is calculated assuming signals that are distributed homogeneously and isotropically in the universe.
This means, $\om$ is related to the average density of gravitational waves in the universe.
But within the galaxy the density is larger than the average.
Moreover, galactic binaries are distributed anisotropically along the galactic disc.
If one uses the spectral function to plot the contribution of galactic binaries,
one is claiming that the density of gravitational waves in the universe is as large as the one inside the galaxy.
Some papers in the literature which deal with the confusion noise produced by galactic binaries are \cite{NelemansEtAl2001,RuiterEtAl2010}.
The most important contribution to this background is the one by WD-WD.
According to \cite{KosenkoPostnov1998}, galactic WD-WD produce a background about an order of magnitude larger than that of extragalactic ones.


One has to be careful when interpreting Figure \ref{fig:plotsjoint}.
That plot gives us information about the averaged total energy density of gravitational waves produced by each ensemble.
The curve of the total background of NS-NS, for instance, enters the window of ET,
but that does not mean that ET will see a constant noise curve like that.
The signals of NS-NS are, in that frequency range, short signals, that will often (but not constantly) be detected with ET.
The effective sensitivity curve of ET is thus not affected by NS-NS.
On the other hand, an unresolvable background of WD-WD with a rate larger than the most likely one would certainly affect the sensitivity of LISA.
To avoid misunderstandings we have not plotted the sensitivity curves of any detectors together with the total background.
In Figure \ref{fig:plotsjointcont} we have plotted sensitivity curves just to show that the background is discontinuous in the frequency band of ground-based detectors.

At frequencies close to the last stable orbit, the Newtonian spectrum that we have calculated may differ considerably from the real one,
since the assumption of slow orbits made in Section \ref{sec:enerspec} is no longer fulfilled.
Thus, the exact shape of the spectral function at such frequencies is not accurate.
However, the continuous and unresolvable parts of the background lie safely at lower frequencies.

\begin{figure*}
\includegraphics{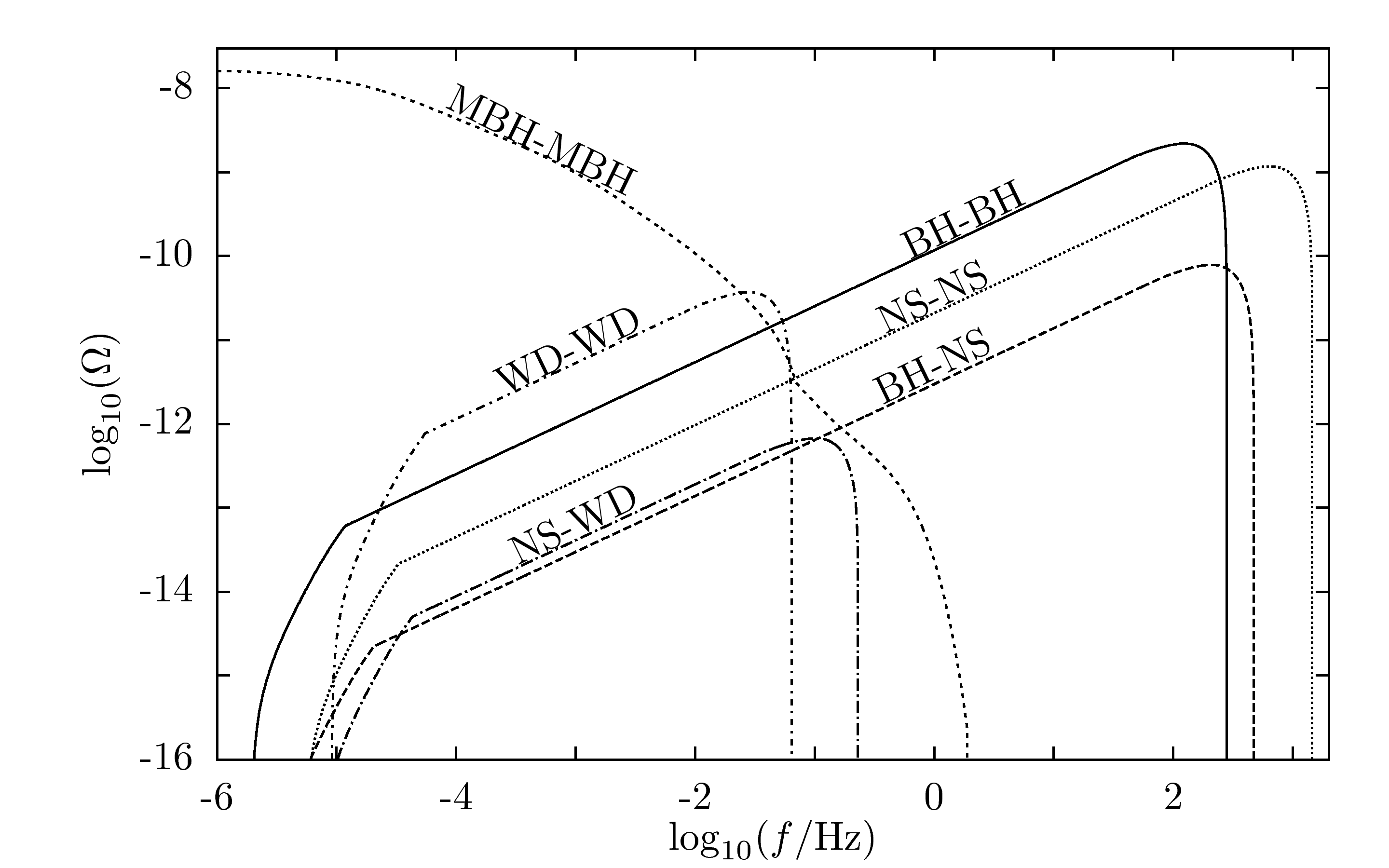}
\includegraphics{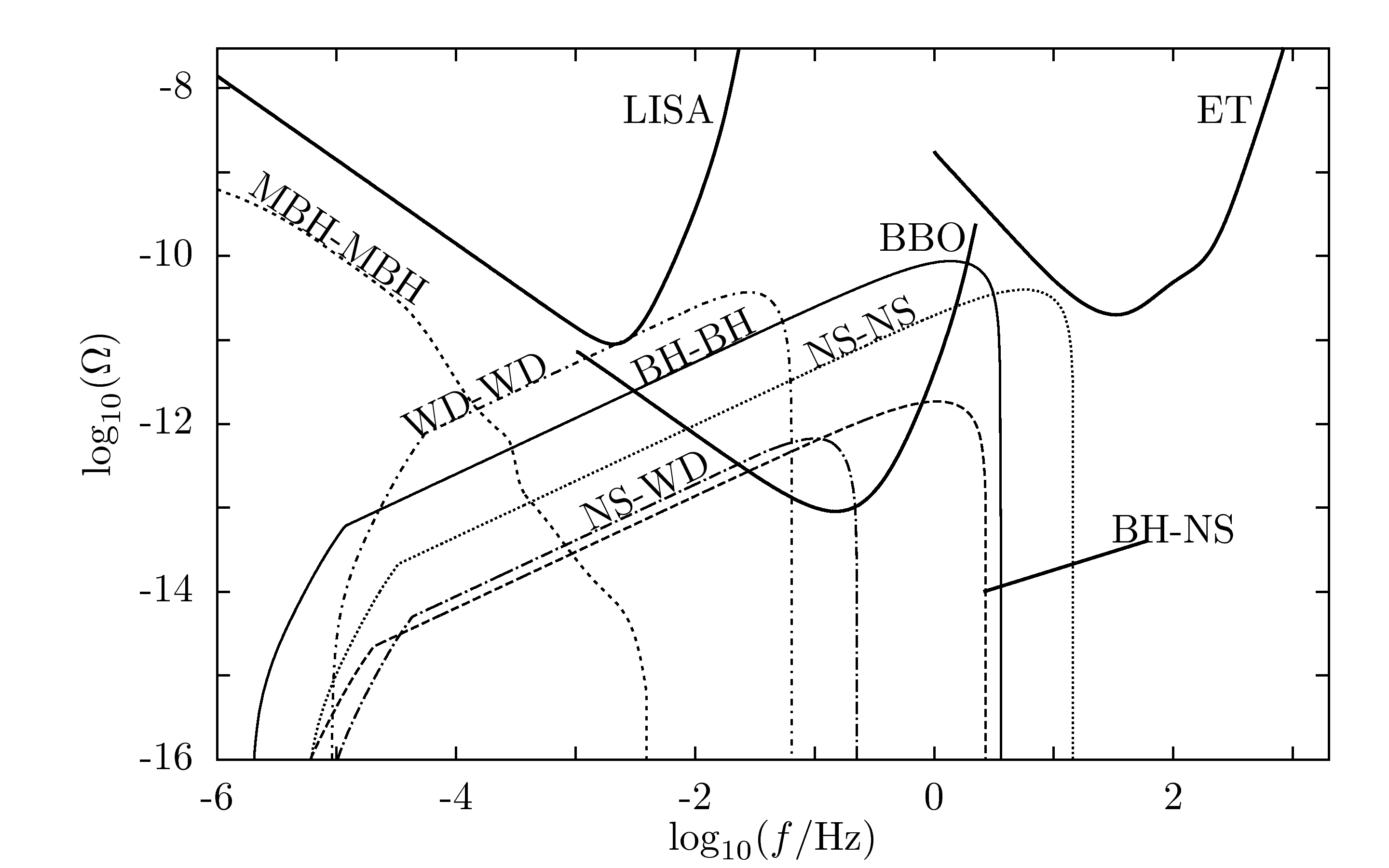}
\includegraphics{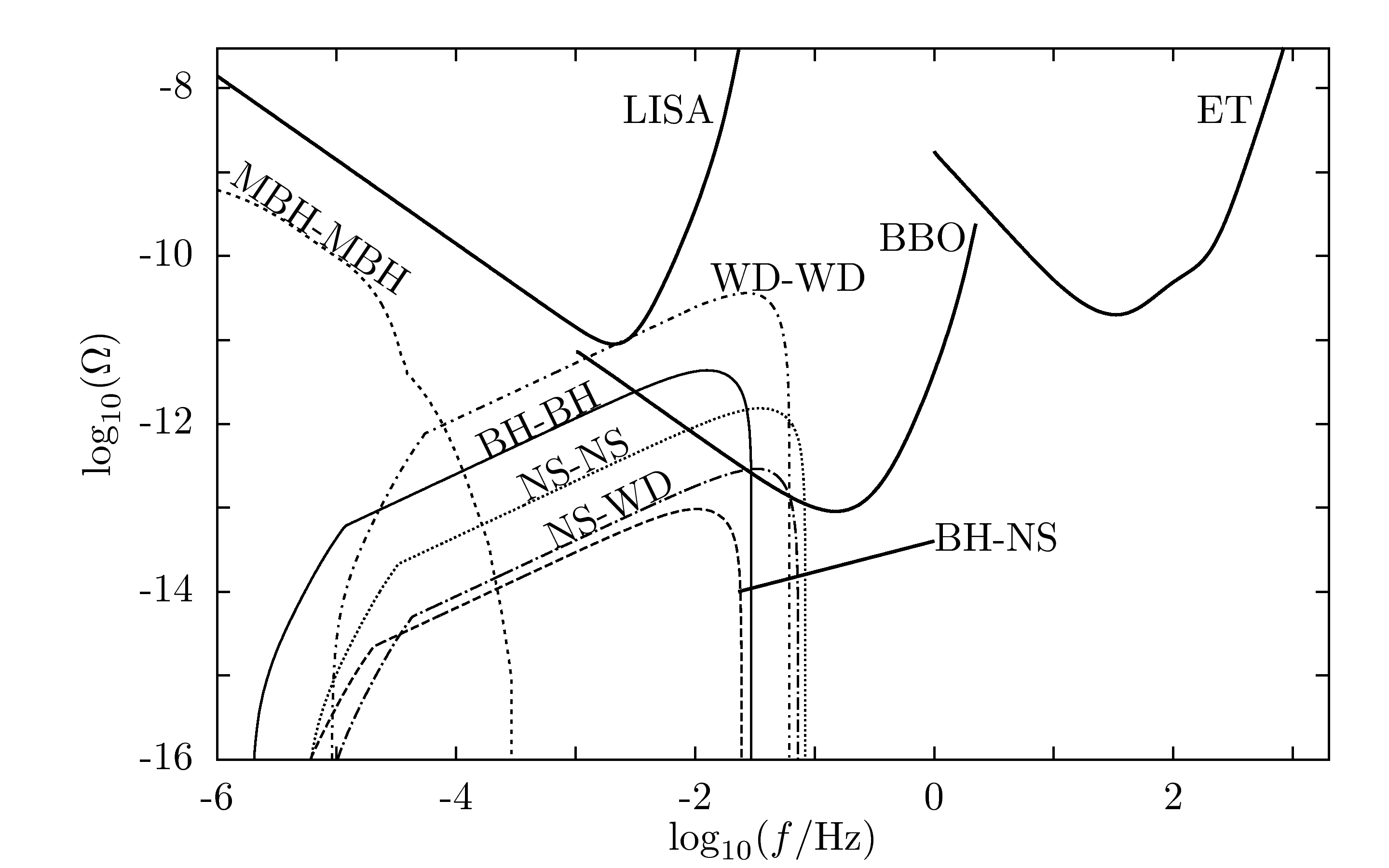}
\caption{Spectral function of the total (top), continuous (middle) and unresolvable (bottom) background versus observed frequency.
The contributions of the different ensembles are calculated with the most likely values of masses and coalescence rates of Table \ref{tb:values},
except for the case of BH-BH.
The high rate of BH-BH, taken from the recent paper \cite{BulikEtAl2011}, is $R=0.36^{+0.50}_{-0.26}$\,Mpc$^{-3}$\,Myr$^{-1}$.}
\label{fig:plotsjoint5}
\end{figure*}

A new prediction on the coalescence rate of BH-BH was published \cite{BulikEtAl2011} during the writing of this paper.
The rate given in that paper, of $R=0.36^{+0.50}_{-0.26}$\,Mpc$^{-3}$\,Myr$^{-1}$, is much larger than the one in Table \ref{tb:values}.
This high estimate is based on the observation of two binaries, both containing a stellar-mass black hole and a Wolf-Rayet star.
Such rates have also been predicted by simulations \cite{BelczynskiEtAl2010c}, considering low-metallicity galaxies.
We show in Figure \ref{fig:plotsjoint5} the total, continuous and unresolvable backgrounds, respectively, that such a rate would produce,
assuming the same mass ranges for black holes given in Section \ref{sec:valuesstellar}.
In Figure \ref{fig:plotsjoint5} we also show the upper and lower limits allowed by the new rate.

\begin{figure}
\includegraphics{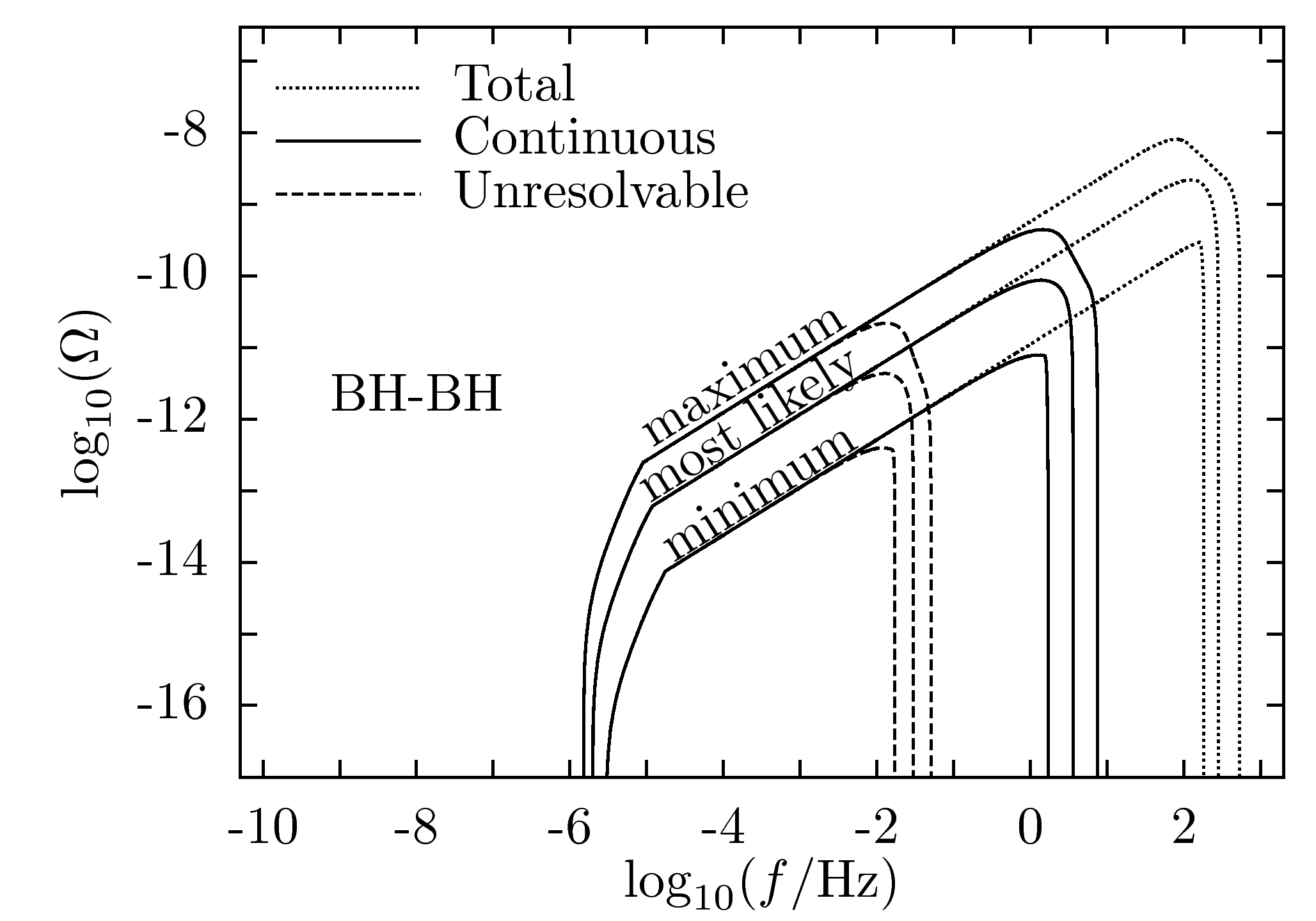}
\caption{Spectral function of the total, continuous, and unresolvable backgrounds of the ensemble of BH-BH, versus observed frequency.
This plot is analogous to that in Figure \ref{fig:plotssep}, but using the rate from \cite{BulikEtAl2011} of $R=0.36^{+0.50}_{-0.26}$\,Mpc$^{-3}$\,Myr$^{-1}$,
instead of the one in Table \ref{tb:values}.}
\label{fig:bhbhnew}
\end{figure}

\section{Discussion}
\label{sec:discussion}

\subsection{\label{sec:discrate} On the coalescence rate of stellar binaries}
In Section \ref{sec:modelscoalrate} we have assumed a coalescence rate that has the same value during all cosmological epochs (see Equation (\ref{eq:ratecons})).
We now justify that this assumption is reasonable, considering the uncertainties in the local coalescence rate.

In \cite{RegimbauHughes2009} the coalescence rates of BH-NS and NS-NS are calculated as a function of the redshift (see Figure 2 of that paper).
The rates peak at around $z\approx 1$ and then decrease, becoming zero between redshift 5 and 6.
The difference between the rate at the peak and the local rate (at $z=0$) is a factor of $\approx 2.1$, for NS-NS, and $\approx 1.7$, for BH-NS.
One can calculate $g(z)$ (using our Equation (\ref{eq:gdef})) and $\overline{g}(z)$ (Equation (\ref{eq:gbardef}))
introducing in the integrals a normalized non-constant rate like the ones of Figure 2 of \cite{RegimbauHughes2009}.
The obtained functions $g(z)$ and $\overline{g}(z)$ differ from the ones calculated with a constant rate by less than a factor of $\sim2$.
The value of this factor would not change significantly if one used other rate functions
(as pointed out in \cite{Phinney2001} regarding the value of $\langle(1+z)^{-1/3}\rangle$).
Since the spectral function and the overlap function are proportional to $g(z)$ and $\overline{g}(z)$, respectively,
the overall difference between using a constant and a non-constant rate would also be less than a factor of 2.
On the other hand, the value of the local rate has an uncertainty of several orders of magnitude (see Table \ref{tb:values}).
We thus consider that a factor of 2 is negligible compared to a factor of (at least) 100.
In addition, we can see in Figure \ref{fig:nsns2} that our estimate of the total background agrees with that of \cite{RegimbauMandic2008},
which was calculated using a non-constant coalescence rate.

Assuming that the rates of other type of stellar binaries have a similar behaviour than those shown in \cite{RegimbauHughes2009},
we can conclude that the use of a constant rate is a good approximation for stellar binaries.

\subsection{\label{sec:minfreq} On the minimum frequency of stellar binaries}
In Section \ref{sec:fmin} we have defined $f_{\text{min}}$ as the gravitational wave frequency such that a binary,
emitting at this frequency, needs an interval of time equal to $T_{\text{max}}$, the maximum inspiral time, to reach coalescence.
But one could in principle find, for each type of binary, a more precise criterion to define $f_{\text{min}}$.

One could define $f_{\text{min}}$, for example, as a function of the velocity kick that the components of the binary experience at formation.
This velocity kick, which can be provoked by a non-symmetrical supernova explosion,
can push one component of the binary with enough energy in a direction opposite to that of the other component and disrupt the binary.
So $f_{\text{min}}$ could be the frequency at which the orbital velocity equals the velocity kick.
With such a criterion, using realistic values of these kicks \cite{WillemsEtAl2005,WillemsEtAl2006,FregeauEtAl2009},
one obtains too long inspiral times, in some cases orders of magnitude longer than the age of the universe.

Our choice of $T_{\text{max}}$ is in fact almost as long as the age of the universe.
Therefore, only binaries that coalesced recently could have had that much time to evolve from their formation, as commented on in Section \ref{sec:maxinsptime}.
However, when considering long inspiral periods, one takes into account part of the contribution from binaries that have not yet coalesced.

In Section \ref{sec:spectralfunction} we point out that the formula of the spectral function assumes short inspiral times,
so that each signal starts and finishes at approximately the same redshift.
But each system needs $\approx$12\,Gyr to complete the process, and the expansion of the universe is indeed relevant during that interval of time.
We now investigate the effect of this apparent inconsistency.

Our rate $R$ accounts for coalescences (and not for births) of binary systems.
This means that we are counting systems that are emitting at frequencies close to $f_{\text{max}}$, the frequency of the coalescence.
What we may be counting wrong are systems emitting at low frequencies.

Suppose a binary, very close to us, that started inspiralling $\approx$12\,Gyr ago and coalesces right now.
We only see the high frequency part of the spectrum, which is not redshifted.
The waves emitted at the beginning of the inspiral (at low frequencies, $\approx$12\,Gyr ago) are now far from us.
But an observer located that far away would observe those waves today highly redshifted.
Our mistake, assuming short inspiral times, is to claim that the distant observer measures that low frequency radiation without any redshift.
So the spectral function should be more redshifted (and thus have lower amplitude) at low frequencies.

We now estimate below which frequency this effects starts to be important.
For that, we assume the following: we assign wrong redshifts as soon as the difference in redshift between birth and coalescence of a signal is larger than 1.
In units of time (using Equation (\ref{eq:dtedzeq})), a difference in redshift of 1 implies timescales larger than $\approx$7\,Gyr at redshifts close to zero
and larger than $\approx$0.4\,Gyr at redshifts close to 5.
To be conservative, we assume that these effects are important when inspiral times are larger than 0.4\,Gyr.
The lifetime of a binary is larger than 0.4\,Gyr if its minimum frequency is lower than $\approx 4\times 10^{-5}$\,Hz for BH-BH, $\approx 7\times 10^{-5}$\,Hz for BH-NS,
$\approx 1\times 10^{-4}$\,Hz for NS-NS, and $\approx 2\times 10^{-4}$\,Hz for NS-WD or WD-WD.
These frequencies are in a range where the spectra of all stellar binaries are covered under an unresolvable background of MBH-MBH.

We thus conclude that the exact values of the minimum frequencies are not relevant in practice.
Furthermore, the assumption of short inspiral times is not fulfilled for stellar binaries at frequencies close to the minimum,
but this does not affect the results.

\subsection{\label{sec:minfreqmbhb} On the minimum frequency of massive black hole binaries}
The minimum frequency of each massive black hole binary, as explained in Section \ref{sec:mbhb},
is assumed to be the frequency $f_R$ at which the slingshot and radiation phases overlap.
This means that we dismiss the gravitational waves radiated during the slingshot phase.

It turns out that the introduction of the slingshot phase in the calculations has a very small effect (well within the uncertainty ranges)
in the spectral function of the superensemble, at frequencies below $\sim 10^{-9}$\,Hz.
The reason is the following:
for each ensemble of masses between $\mathcal{M}$ and $\mathcal{M}+d\mathcal{M}$,
the effect of introducing the slingshot phase is noticeable only at frequencies below $f_{\text{min}}(\mathcal{M})$
(the one calculated using Equation (\ref{eq:deffminapprox}) with $T_{\text{max}}=75$\,Myr).
But the main contribution of each ensemble to the superensemble is at high frequencies, where they have larger spectral functions
(because of the $f^{2/3}$ factor).
In the superensemble, the only appreciable low-frequency contributions are those from ensembles with the largest masses and with non-zero coalescence rates.
Therefore, the effect of introducing the slingshot phase in the superensemble is noticeable only at frequencies close to $f_{\text{min}}(\mathcal{M})$,
when $\mathcal{M}$ is in the range of large masses (of $\sim 10^8-10^9\,M_\odot$).
These frequencies are smaller than $\sim 10^{-9}$\,Hz.

\subsection{\label{sec:unresolvable} On the condition of resolvability}
In Section \ref{sec:resolvability} we state that signals between $f$ and $f+df$ with redshifts larger than $z_*$ such that $z_{\text{low}}(f)<z_*<z_{\text{upp}}(f)$
and $\mathcal{N}(f,\Delta f,z_*)=1$ are unresolvable.
We are hence imposing a one-bin-rule: 
we are not able to distinguish signals if there are more than one per frequency bin.
Other authors suggest other possible criteria, such as the three-bin-rule or the eight-bin-rule \cite{Cornish2003}.
According to these criteria, the condition of unresolvability is reached when each three (or eight) frequency bins are occupied by at least one signal.
We now comment on how using one of these criteria would change our results.

\begin{figure}
\includegraphics{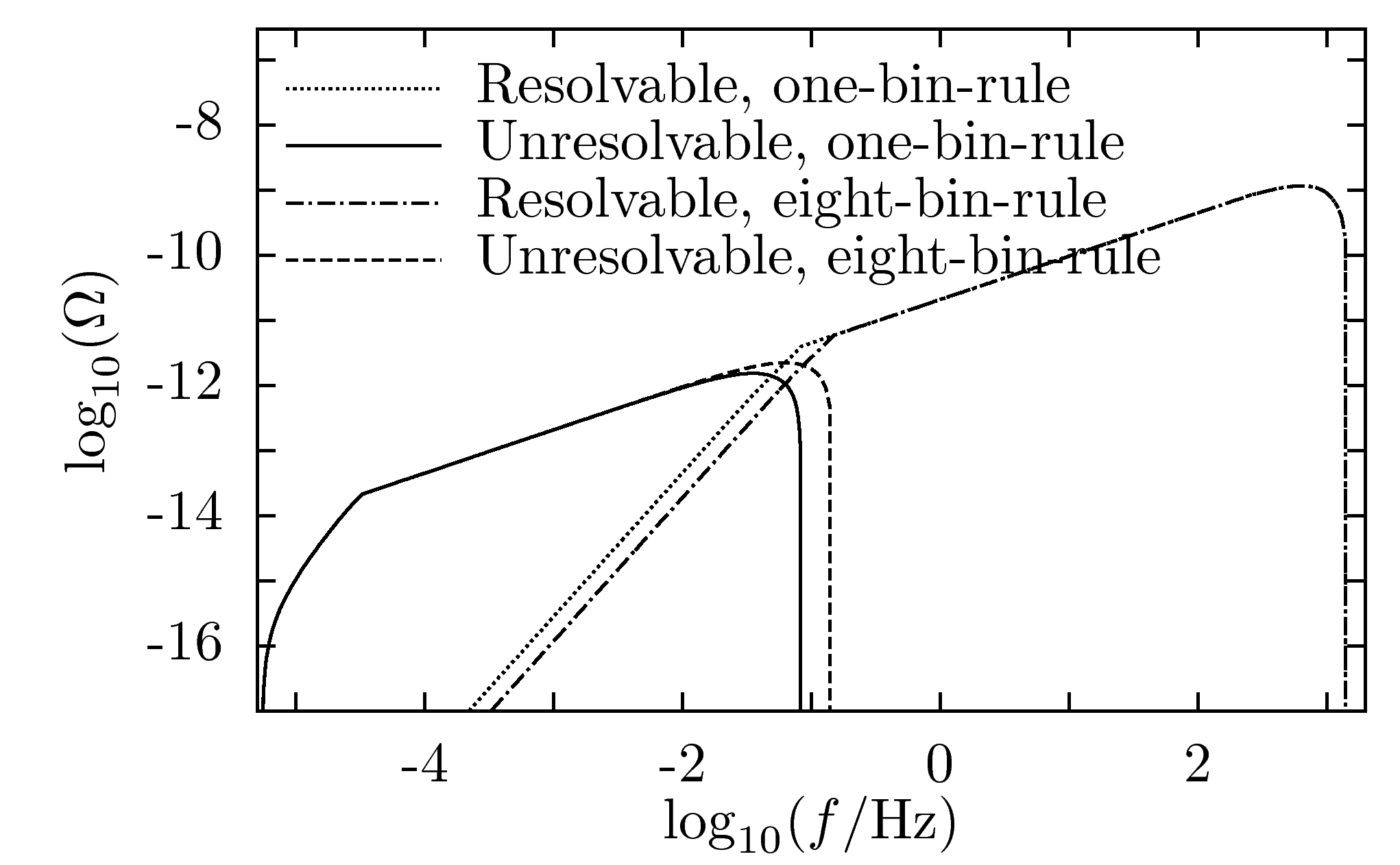}
\caption{Spectral function of the resolvable and unresolvable parts of the background of NS-NS, versus observed frequency.
The values of masses and coalescence rates adopted are the most likely ones.
The resolvable and unresolvable parts calculated with the one- and with the eight-bin rule are compared.}
\label{fig:8binrule}
\end{figure}

Imposing an eight-bin-rule makes the condition of unresolvability less restrictive:
signals become unresolvable at higher frequencies than for the one-bin-rule.
The results would be almost unaffected in the case of WD-WD, since the curve of the unresolvable background is almost as large as that of the total background.
The spectral function of the unresolvable background for the remaining stellar binaries would be slightly extended to higher frequencies.
We can calculate the spectral function with the eight-bin rule, just by changing $\Delta f$ by $8 \times \Delta f$,
so $\Omega_{\text{unresolvable}}(f)=\Omega(f,8\Delta f,\mathcal{N}_0)$.
In Figure \ref{fig:8binrule} we compare the spectral functions of the unresolvable background calculated with the one- and the eight-bin-rule,
for the case of NS-NS with the most likely values of masses and coalescence rates.

One can note that imposing an eight-bin-rule, instead of a one-bin-rule, has the same effect of assuming an observation time of eight years, instead of one.
The expected observation time of LISA is indeed three years; for the PTA, longer observation times are feasible.
So, for MBH-MBH, redoing the calculations with the eight-bin-rule is compensated with the use of longer observation times.
As pointed out in \cite{SesanaEtAl2008}, the unresolvable background changes by less than a factor of 2 for observation times between one and ten years.

\begin{figure*}
\includegraphics{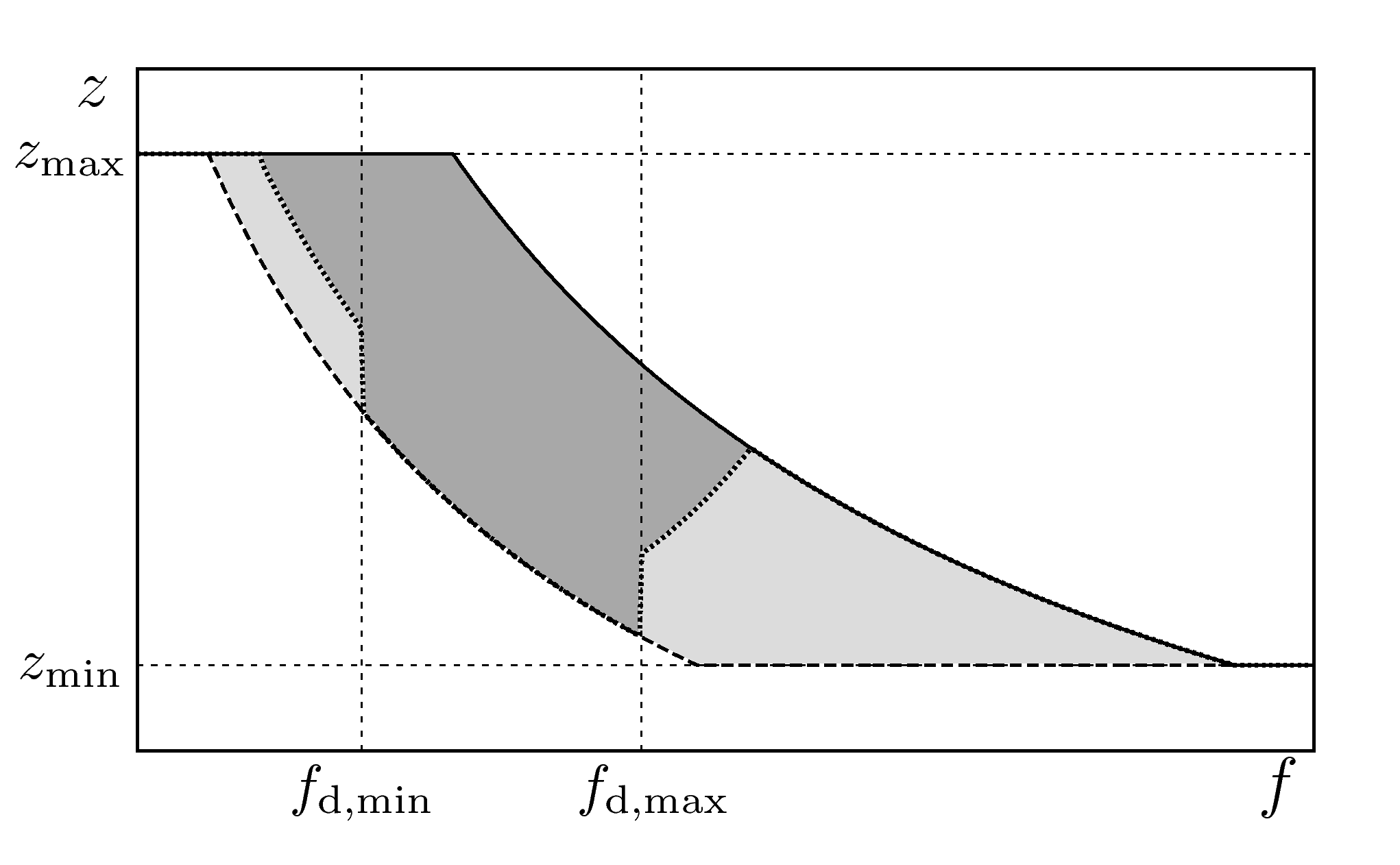}
\includegraphics{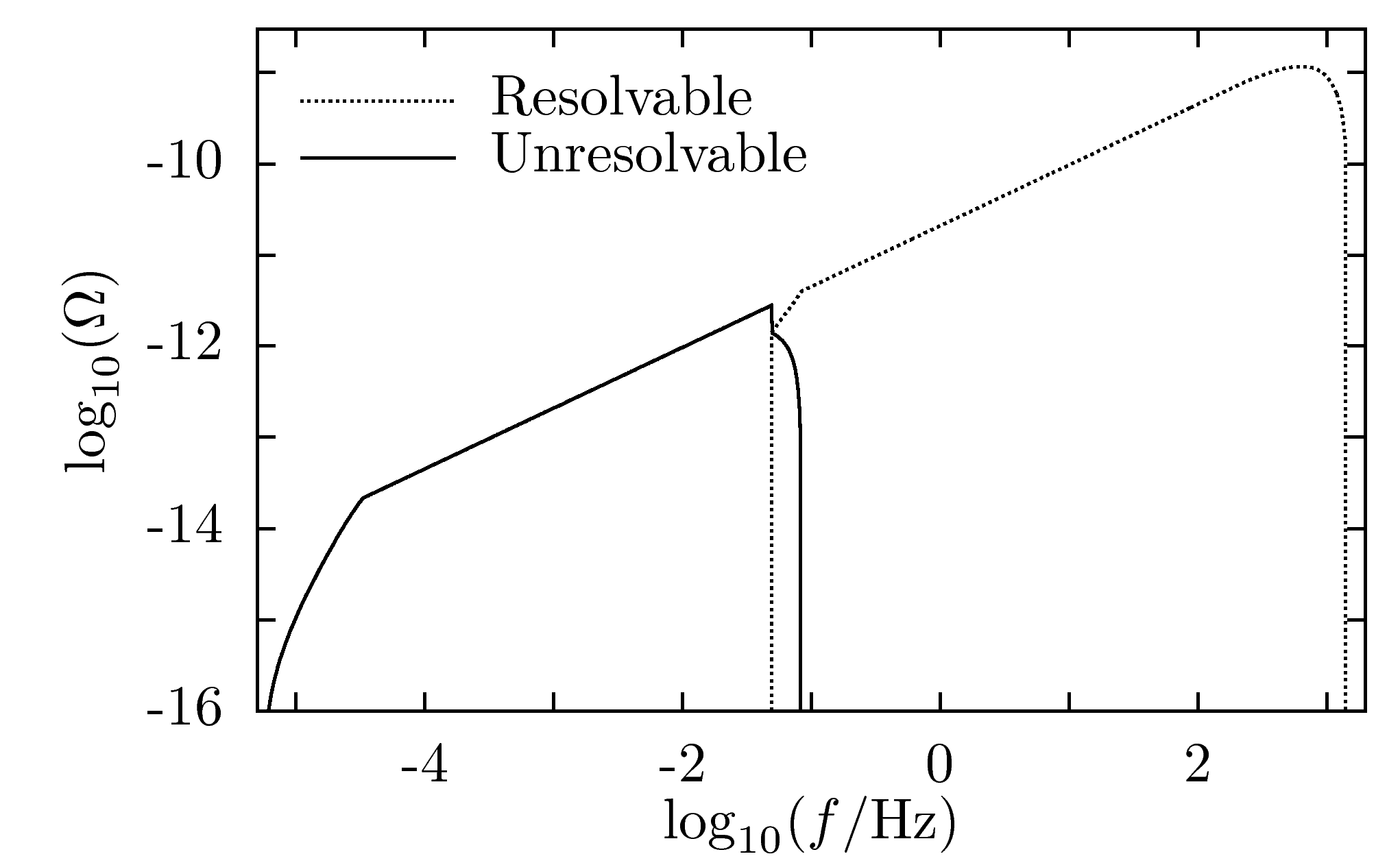}
\caption{Redshift versus observed frequency (left plot, analogous to that in Figure \ref{fig:pfun2})
and spectral function versus observed frequency (right plot, analogous to that in Figure \ref{fig:plotssep},
corresponding to NS-NS with the most likely values of masses and coalescence rates).
The difference between these plots and the ones in Figures \ref{fig:pfun2} and \ref{fig:plotssep} is that here,
as soon as the unresolvable part of the background dominates, all signals become unresolvable.}
\label{fig:pfun5}
\end{figure*}

We now discuss another possible definition of unresolvable background.
In Section \ref{sec:resolvability} we say that, when the unresolvable part of the background dominates, there still exists a resolvable part.
We could consider this resolvable part as also unresolvable.
For that, we could just change the definition of $\overline{z}(f,\Delta f,\mathcal{N}_0)$ in Equation (\ref{eq:zoverlap}) to
\begin{equation}
\label{eq:zoverlapnew}
\overline{z}(f,\Delta f,\mathcal{N}_0)=\left\{ \begin{array}{lc}
	    z_{\text{upp}}(f), & f< f_{\text{p,min}} \\
	    \mathcal{N}^{-1}(f,\Delta f,\mathcal{N}_0), & f_{\text{p,min}}\le f \le f_{\text{d,min}} \\
            z_{\text{low}}(f), & f_{\text{d,min}}< f < f_{\text{d,max}} \\
	    \mathcal{N}^{-1}(f,\Delta f,\mathcal{N}_0), & f_{\text{d,max}}\le f \le f_{\text{p,max}} \\
	    z_{\text{upp}}(f), & f_{\text{p,max}}<f
             \end{array}
\right. .
\end{equation}
On the left side of Figure \ref{fig:pfun5}, a plot of redshifts versus observed frequencies (analogous to that in Figure \ref{fig:pfun2}) is shown,
using the new definition of $\overline{z}(f,\Delta f,\mathcal{N}_0)$.
There we see that, between $f_{\text{d,min}}$ and $f_{\text{d,max}}$, there is no resolvable background.
On the right side of Figure \ref{fig:pfun5} we show the spectral function obtained by inserting (\ref{eq:zoverlapnew}) in (\ref{eq:omgensol}),
for the case of NS-NS with the most likely values of masses and coalescence rates.
The difference between the spectral functions with the old and the new definitions of $\overline{z}(f,\Delta f,\mathcal{N}_0)$
is just a small peak at frequency $f_{\text{d,max}}$.

Since the results are almost unchanged, we prefer the definition of unresolvable background given in Section \ref{sec:resolvability},
because the definition of $\overline{z}(f,\Delta f,\mathcal{N}_0)$ given in Equation (\ref{eq:zoverlap}) is simpler than that in (\ref{eq:zoverlapnew}).

\subsection{\label{sec:discovfun} On the definition of the overlap function}
In Section \ref{sec:resolvability} we mention that the time resolution, $\Delta t$, should be taken into account to define the resolvability,
but in practice it is not important.
We now explain how $\Delta t$ should be introduced in the overlap function and why it is not necessary, in the circumstances considered in this work.

Signals of binaries evolve more rapidly at higher frequencies, and spend therefore less time in a frequency bin.
Eventually, at frequencies and redshifts larger than certain $f$ and $z$,
the interval of time $\tau_e(f,\Delta f,z)$ can become smaller than the time resolution $\Delta t$.
When this happens, all signals spend effectively an interval of time $\Delta t$ in each frequency bin,
and no less than that (since smaller intervals of time cannot be distinguished).
Thus, if the signals are unresolvable at $f$ and $z$, they stay unresolvable for any larger values of frequency and redshift.
Equation (\ref{eq:defoverlapfun}) could be generalized to take into account this effect,
\begin{align}
\label{eq:ovfungen}
 \mathcal{N}(f,\Delta f,\Delta t,z)=&\int_{z_{\text{low}}(f)}^z \max(\tau_e (f,\Delta f,z'),\Delta t [1+z]^{-1}) \nonumber\\
&\times \dot{n}(z') \frac{d\mathcal{V}_c}{dz'} dz'.
\end{align}
The factor $[1+z]^{-1}$ is necessary to compare our time resolution (which is an observed interval of time) with the interval of time at emission $\tau_e(f,\Delta f,z)$.

The effect of introducing $\Delta t$ affects our calculations only if $\tau_e (f,\Delta f,z)\le \Delta t[1+z]^{-1}$ when $\mathcal{N} \ge 1$,
i.e., if there is more than one coalescence every $\Delta t[1+z]^{-1}$.
Taking the highest rate of Table \ref{tb:values} (which is the maximum rate of WD-WD), we see that
\begin{displaymath}
\int_0^5 \Delta t [1+z']^{-1} R \frac{d\mathcal{V}_c}{dz'}dz'
\end{displaymath} 
is greater than one for $\Delta t$ greater than $\sim 1/9$\,s.
A reasonable choice for the time resolution is the inverse of the sampling rate of a detector,
which, in the case of current ground-based detectors, is much smaller than $1/9$\,s.
Therefore, the generalization (\ref{eq:ovfungen}) is not necessary;
the overlap function is well defined by (\ref{eq:defoverlapfun}).

\begin{figure*}
\includegraphics{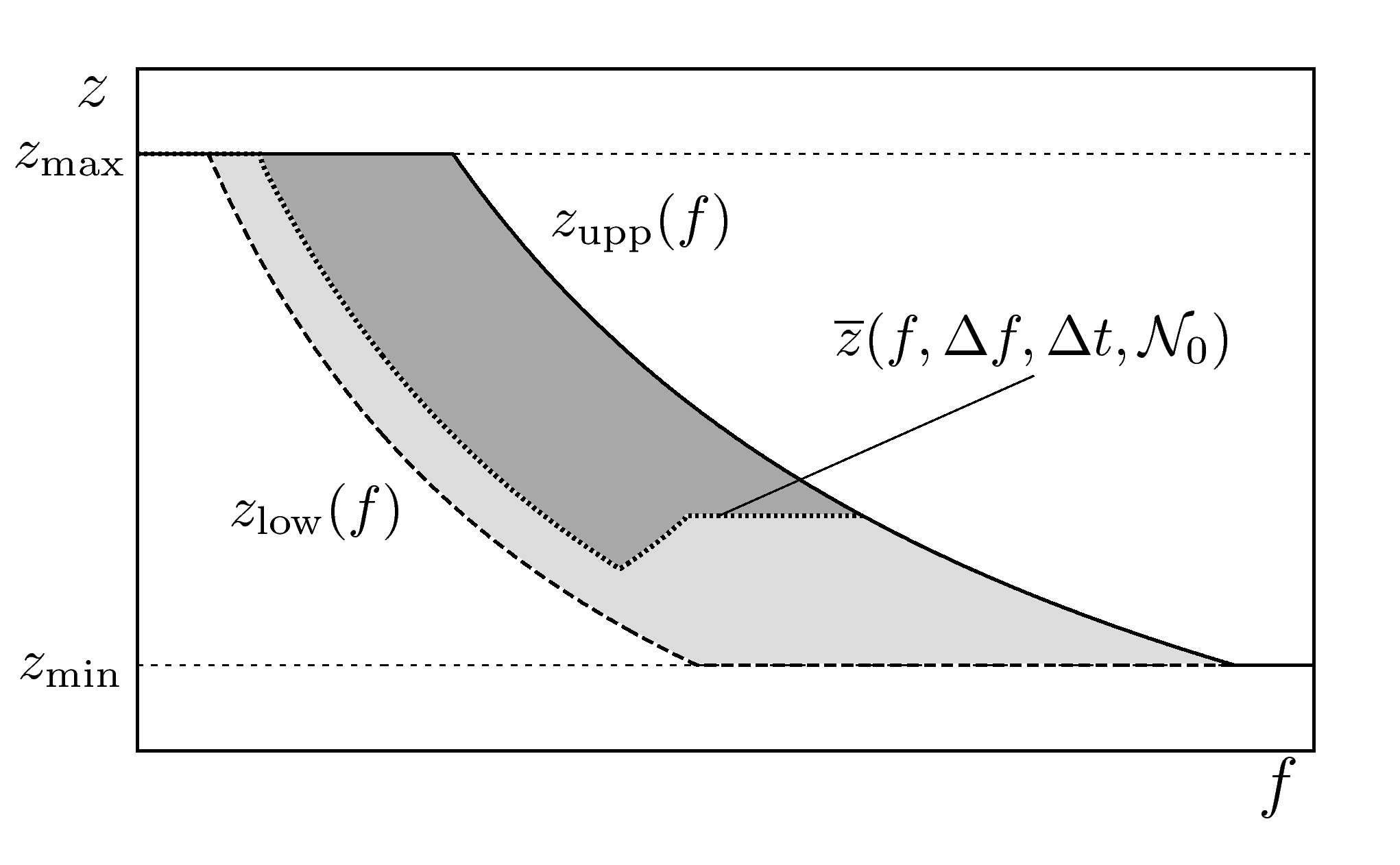}
\includegraphics{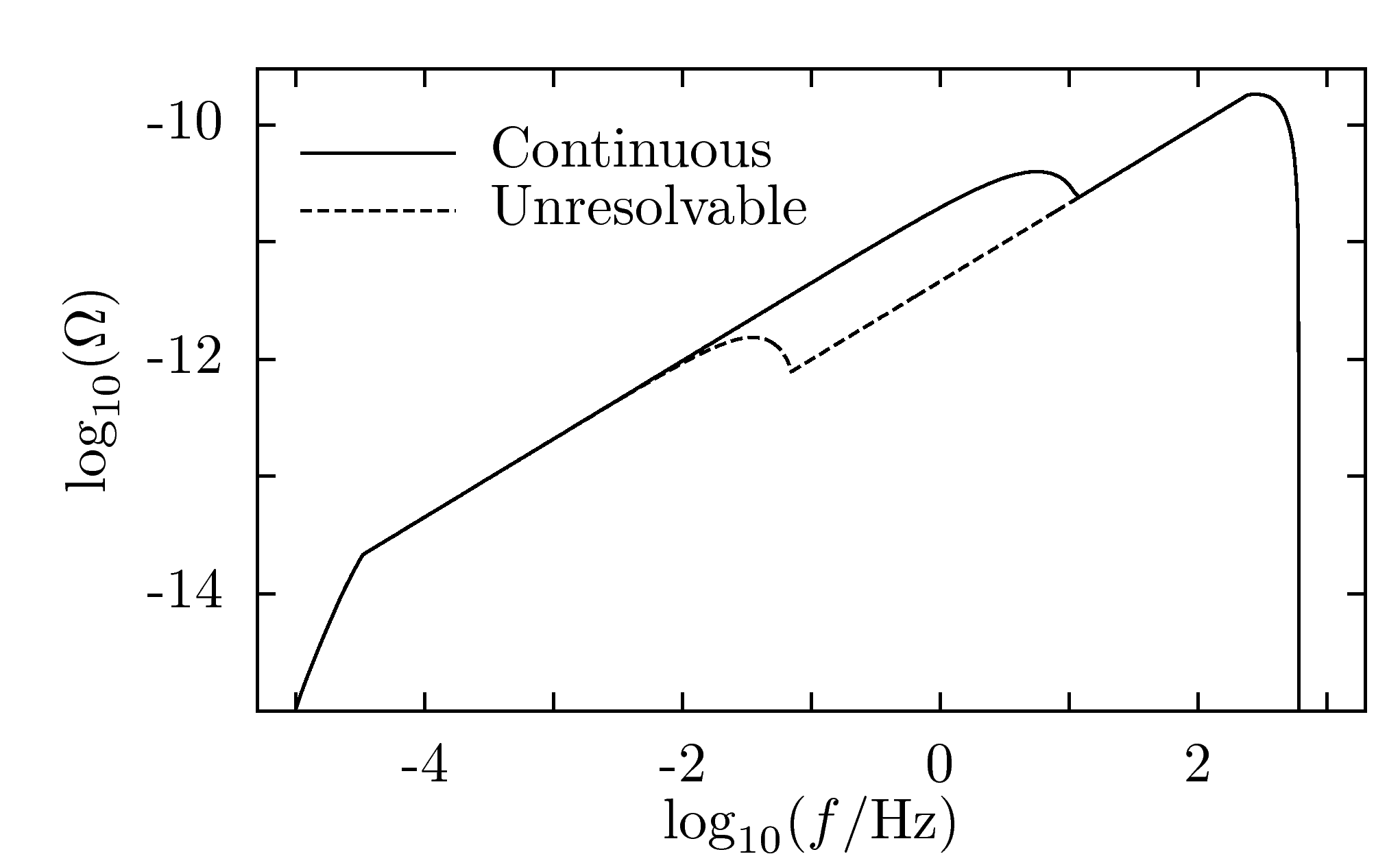}
\caption{Redshift versus observed frequency (left plot, analogous to that in Figure \ref{fig:pfun2})
and spectral function versus observed frequency (right plot, analogous to that in Figure \ref{fig:plotssep},
corresponding to NS-NS with the most likely values of masses and coalescence rates).
These plots (unlike those in Figures \ref{fig:pfun2} and \ref{fig:plotssep}) are calculated assuming an unrealistic time resolution of $\Delta t=600$\,s.
With such a large time resolution, an unresolvable background would be present in the frequency band of ground-based detectors.}
\label{fig:prect}
\end{figure*}

By artificially increasing the time resolution by several orders of magnitude,
we see the effect that the overlap function of Equation (\ref{eq:ovfungen}) produces in $\overline{z}(f,\Delta f,\Delta t,\mathcal{N}_0)$
(which is obtained by inserting Equation (\ref{eq:ovfungen}) in (\ref{eq:zoverlap})) and in $\Omega(f,\Delta f,\Delta t,\mathcal{N}_0)$
(inserting (\ref{eq:ovfungen}) in (\ref{eq:omgensol})).
This effect is plotted in Figure \ref{fig:prect}.
There we see that, above a certain redshift and a certain frequency, all signals contribute to the unresolvable background.
With this example we see that a large time resolution would lead to the existence of an unresolvable background in the frequency band of ground-based detectors.

\subsection{\label{sec:comparison} Comparison with previous work}
\subsubsection{Unresolvable backgrounds}
\begin{figure*}
\includegraphics{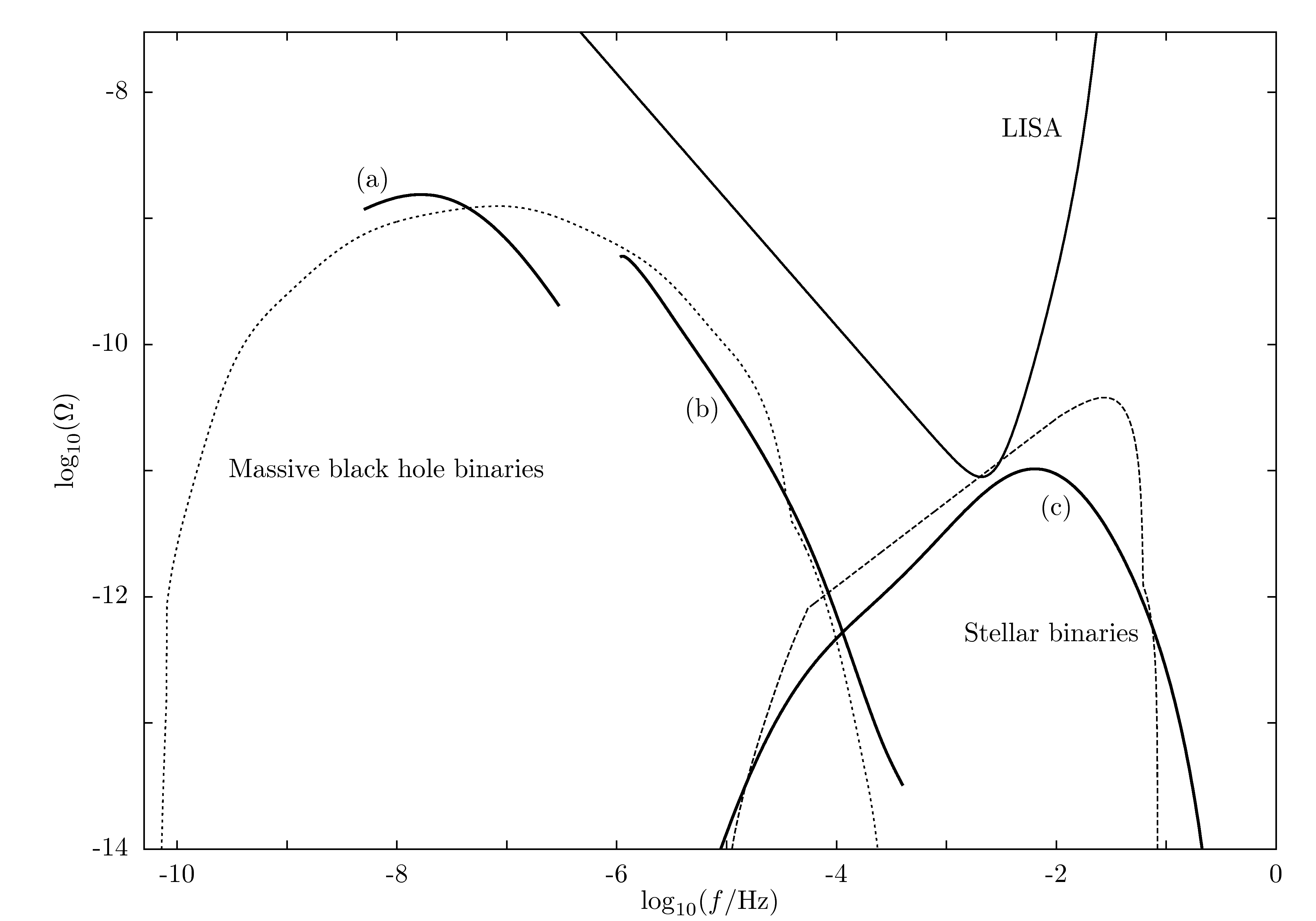}
\caption{Spectral function of the unresolvable background of MBH-MBH (dotted line) and of all stellar binaries (dashed line), versus observed frequency.
These curves are compared with previous predictions from the literature, which correspond to the unresolvable backgrounds of:
(a) MBH-MBH from \cite{SesanaEtAl2008}, (b) MBH-MBH from \cite{SesanaEtAl2005} and (c) extragalactic stellar binaries from \cite{FarmerPhinney2003}.}
\label{fig:plotsjointunres2}
\end{figure*}

In Figure \ref{fig:plotsjointunres2} we show the unresolvable background produced by the superensemble of MBH-MBH,
and the sum of the unresolvable backgrounds of all stellar binaries (which is almost equal to the background made by only WD-WD).
These curves are compared with other predictions from the literature.
The curve (a) is obtained from \cite{SesanaEtAl2008}, using its Equation (14) with the mean values of the parameters in (45), (46) and (47).
That formula is given in terms of the characteristic amplitude, $h_c$, which is related to the spectral function by
\begin{equation}
\label{eq:defcharamp}
\Omega(f)=\frac{\pi f^2}{4 \rho_c G}h_c^2(f).
\end{equation} 
In terms of the strain amplitude $S_h(f)$, the characteristic amplitude is
\begin{equation}
h_c(f)=\sqrt{fS_h(f)}.
\end{equation} 
The curve (b) is taken from Figure 4 of \cite{SesanaEtAl2005}, where $h_{\text{rms}}$ also represents a characteristic amplitude.
Finally, the curve (c) is taken from Figure 16 of \cite{FarmerPhinney2003}.
In all cases we find a good agreement of our predictions with those from the mentioned papers.

\begin{figure}
\includegraphics{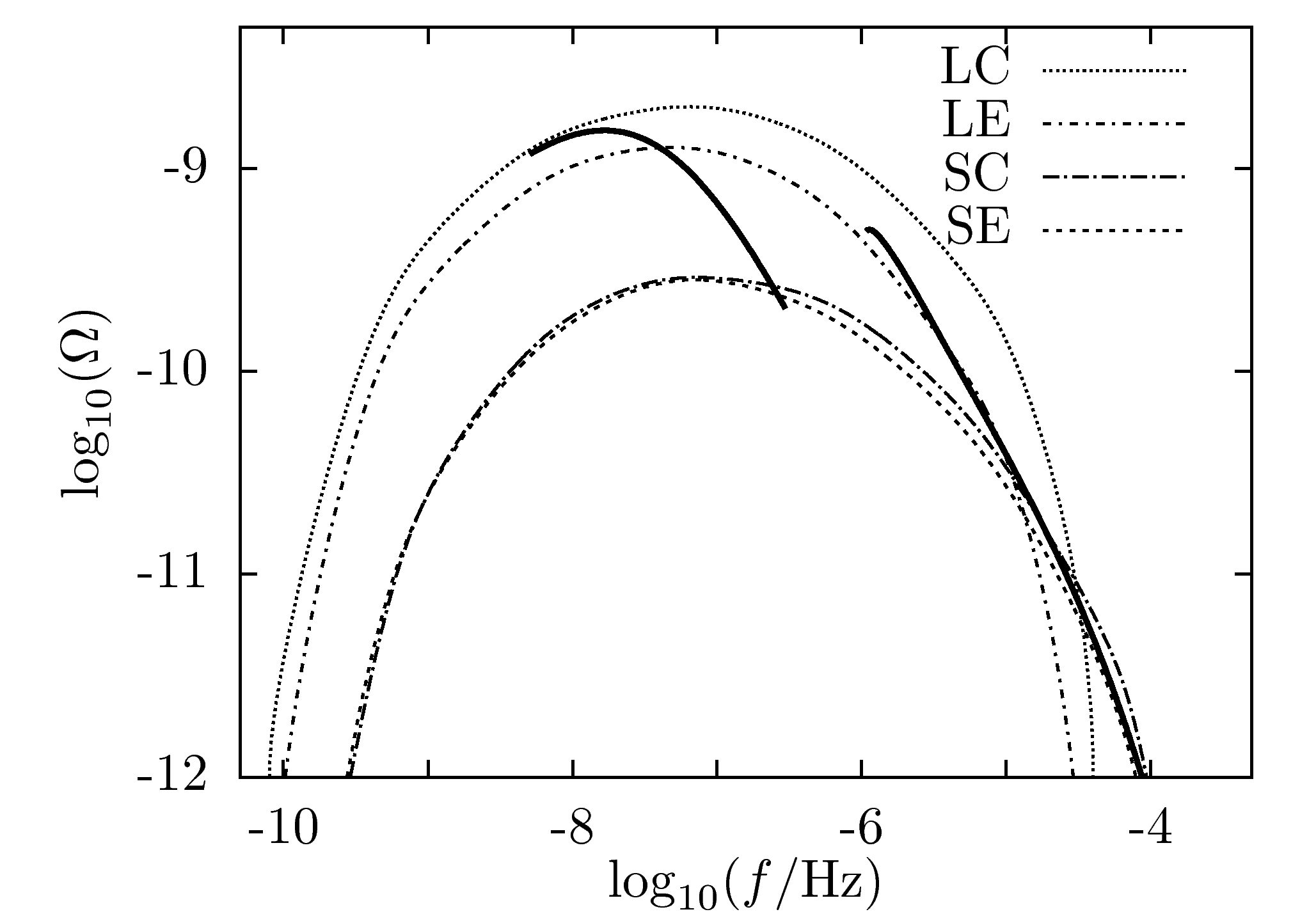}
\caption{Spectral function of the unresolvable background of MBH-MBH, calculated with the four different models (LE, LC, SE and SC), versus observed frequency.
For comparison we include the curves of the unresolvable backgrounds of \cite{SesanaEtAl2005} and \cite{SesanaEtAl2008}.}
\label{fig:mbhbalt}
\end{figure}

In Figure \ref{fig:mbhbalt} the unresolvable background of MBH-MBH is shown, calculated with the four different models (see Section \ref{sec:models}).
The unresolvable backgrounds calculated in \cite{SesanaEtAl2005} and \cite{SesanaEtAl2008} are also plotted for comparison.

\subsubsection{\label{sec:discnsns} Background of neutron star binaries}
\begin{figure}
\includegraphics{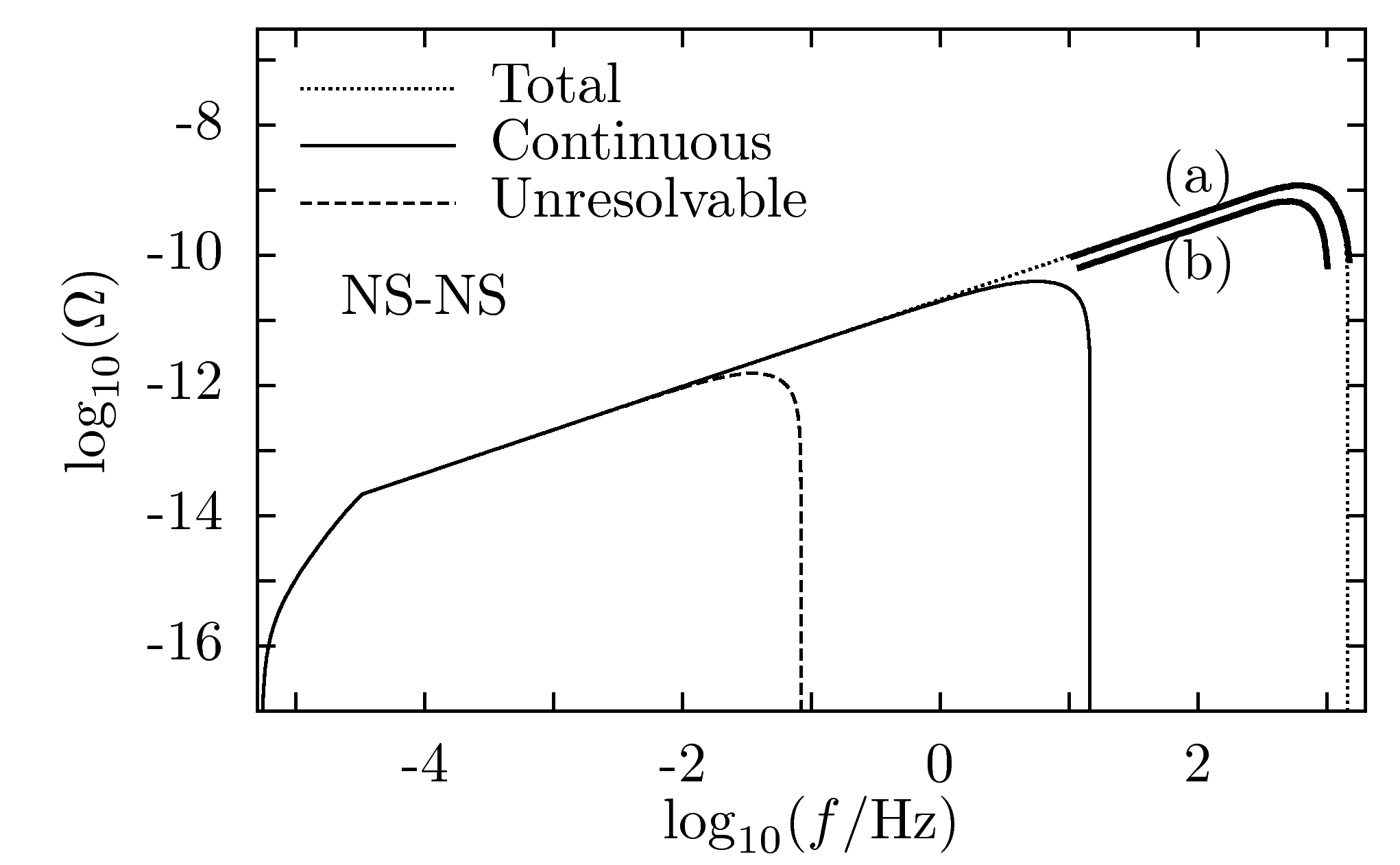}
\caption{Spectral function of the total, continuous and unresolvable backgrounds of the ensemble of NS-NS (with the most likely values of masses and coalescence rates),
versus observed frequency.
We compare these curves with those given in \cite{RegimbauMandic2008}.
Curves (a) and (b) correspond to what in that paper is called \textit{shot noise} and \textit{Gaussian background}, respectively.
In the text we explain the origin of the discrepancy between (b) and our continuous or unresolvable backgrounds.}
\label{fig:nsns2}
\end{figure}

In Figure \ref{fig:nsns2} we see that our estimate of the total background of NS-NS is in good agreement with the one in \cite{RegimbauMandic2008}.
The curve (a) in Figure \ref{fig:nsns2} represents what in \cite{RegimbauMandic2008} is called \textit{shot noise} (see Figure 2 of that paper).
In that work, the existence of a continuous (and Gaussian) background is also asserted; this corresponds to the curve (b) in Figure \ref{fig:nsns2}.
If we compare (b) with either our continuous or our unresolvable curves, we find a big discrepancy.

We now explain the origin of this discrepancy.
In Section \ref{sec:heuristic} we pointed out that signals of equal observed frequency but different redshifts need different intervals of time to coalesce.
However, in Section 3 of \cite{RegimbauMandic2008}, all binaries are assumed to spend the same amount of time in a certain frequency interval,
leading to the conclusion that the background is continuous at high frequencies.
The same has been claimed in similar papers, for example \cite{CowardRegimbau2006,RegimbauChauvineau2007}.

In a later work, \cite{RegimbauHughes2009}, the continuity of the background is calculated in a similar manner as we do;
the redshift of the signals is properly taken into account to measure the interval of time that they spend in the frequency window of the detector.
But in this paper, the continuous background is treated as unresolvable, which is incorrect, as we justify now.
Suppose there is a continuous background of NS-NS in the frequency band of ET,
such that there is an average of a few signals present in the band.
Even if these few signals are observed at the same time, they do not overlap in the frequency domain;
the signals can still be distinguished in frequency, so they are resolvable.

\subsubsection{\label{sec:ovvsdc} Overlap function versus duty cycle}
In the literature, the so-called \textit{duty cycle} is often used.
It is defined by
\begin{equation}
\label{eq:dutycycle}
D(z)=\int_0^z \overline{\tau}_e \dot{n}(z')\frac{d\mathcal{V}_c}{dz'} dz',
\end{equation} 
where $\overline{\tau}_e$ is the duration of a signal in the detector window.
If one assumes that $\overline{\tau}_e$ is constant, as for example in \cite{CowardRegimbau2006,RegimbauChauvineau2007,RegimbauMandic2008},
$D(z)$ does not give any valuable information (this has just been commented on in the previous section).
If $\overline{\tau}_e=\tau_e(f_1,f_2-f_1,z)$, i.e.,
if $\overline{\tau}_e$ is the time that each signal of redshift $z$ spends in the frequency window $[f_1,f_2]$ of a certain detector,
$D(z)$ characterizes the continuity of the background.
But the property of the background that is indeed relevant is the resolvability,
which is measured by the overlap function, defined in Equation (\ref{eq:defoverlapfun}).

An overlap function like the one in Equation (\ref{eq:defoverlapfun}) is useful for quantifying the resolvability of long signals.
Now suppose that there is an ensemble of systems that do not emit gravitational waves during a long period of time, but rather in a \textit{burst}.
For such systems one cannot obtain an accurate function $\tau_e(f,\Delta f,z)$.
In this case, the resolvability can be quantified using the overlap function,
by changing $\tau_e(f,\Delta f,z)$ to $\overline{\tau}_e$, the typical duration of a burst.
This overlap function would then coincide with the duty cycle.
In the case that $\overline{\tau}_e$ could be smaller than the time resolution $\Delta t$, one should rather use the generalized overlap function of (\ref{eq:ovfungen}).

The overlap function is therefore a generalization of the duty cycle, that can be used for short or long signals.

\section{\label{sec:summary}Summary and conclusions}
We have reviewed basic aspects of the gravitational wave background.
We have derived a formula (Equation (\ref{eq:omegalim})) for the spectral function, $\om$,
for an ensemble of many similar systems emitting gravitational radiation at different times and locations.
This formula has been generalized to account for the duration of the signals and the observation time.
With the generalized spectral function, $\omov$ (in Equation (\ref{eq:omgensol})),
one can distinguish between unresolvable and resolvable backgrounds (Equations (\ref{eq:omunres}) and (\ref{eq:omres}), respectively),
and between continuous and discontinuous backgrounds (Equations (\ref{eq:omcont}) and (\ref{eq:omdisc}), respectively).

The resolvability is a fundamental property of the background.
An unresolvable background (often called confusion noise or stochastic background) is fully characterized by $\omov$.
A resolvable background is composed of signals whose waveforms can be distinguished and in some circumstances subtracted out of the data.
Precise definitions of resolvable and unresolvable backgrounds can be found in Section \ref{sec:resolvability}.
Figure \ref{fig:pfun2} illustrates the different regimes of the background.

The resolvability is characterized by the overlap function, $\mathcal{N}(f,\Delta f,z)$,
which gives the average number of signals, with frequency $f$ and redshifts smaller than $z$, per frequency bin $\Delta f$ (the frequency resolution).
A formula for the overlap function is given in Equation (\ref{eq:defoverlapfun}).
In Section \ref{sec:comparison} we have shown that the overlap function is a generalization of the duty cycle.
The latter has been often used in the literature to quantify the continuity and even the resolvability of the background, leading in some cases to incorrect results.

The continuity is a secondary property of the background,
which just gives an idea of how often the signals are present in the frequency window of a detector.
The overlap function can also be used to characterize the continuity of the background, as explained in Section \ref{sec:continuity}.

We have calculated the spectral functions of the backgrounds of stellar binaries (those containing white dwarfs, neutron stars or stellar-mass black holes)
and of massive black hole binaries.
In Table \ref{tb:values} we have summarized the values, taken from the literature, of the coalescence rates of each ensemble.
The ranges of masses assumed for neutron stars, white dwarfs and stellar-mass black holes are in Equations (\ref{eq:nsmasses}),
(\ref{eq:wdmasses}) and (\ref{eq:bhmasses}), respectively.
A semi-analytical solution of the generalized spectral function has been derived for stellar binaries (Equation (\ref{eq:omsemian})).
The calculations involving massive black hole binaries have been performed numerically, using the coalescence rates obtained with the four models of \cite{ArunEtAl2009}.

The spectral functions of the backgrounds produced by the different ensembles are plotted in Section \ref{sec:results},
over the frequencies of all present and planned detectors.
The total, continuous and unresolvable backgrounds are plotted in Figures \ref{fig:plotsjoint}, \ref{fig:plotsjointcont} and \ref{fig:plotsjointunres}, respectively,
with the most likely values of masses and coalescence rates.
In Figure \ref{fig:plotssep} the same curves are plotted separately for each ensemble, with their uncertainties.
The total, continuous and unresolvable backgrounds, using the rate of BH-BH recently predicted in \cite{BulikEtAl2011},
are plotted in Figures \ref{fig:plotsjoint5} and \ref{fig:bhbhnew}.

The unresolvable background is dominated by white dwarf binaries, below $\sim 10^{-1}$\,Hz, and by massive black hole binaries, below $\sim 10^{-4}$\,Hz.
These backgrounds could enter the frequency window of LISA, PPTA and BBO.
The continuous background of BH-BH, using the recent coalescence rate predicted in \cite{BulikEtAl2011}, becomes more important than the one made by NS-NS,
especially in the band of BBO.
The confusion noise produced by galactic binaries has not been shown in the figures,
since it cannot be calculated using the spectral function.
Some papers in the literature which cover this issue are \cite{KosenkoPostnov1998,NelemansEtAl2001,RuiterEtAl2010}.

Finally, with Figures \ref{fig:plotsjointcont} and \ref{fig:plotsjointunres},
we conclude that present and planned ground-based detectors are in a frequency range where no continuous or unresolvable backgrounds from binary systems are present.
Therefore, without considering other possible sources of confusion noise, this band could be suitable for searching for primordial backgrounds.

\begin{acknowledgments}
I thank Bruce Allen, Reinhard Prix and Karl Wette for their help, suggestions and corrections throughout the entire paper.
I thank Pau Amaro-Seoane, Vladimir Dergachev and Alberto Sesana for their friendly help and for the fruitful discussions.
Finally, I acknowledge Marta Volonteri and again Alberto Sesana for providing me with the numerical models of the coalescence rates of massive black hole binaries.

This work was supported by the IMPRS on Gravitational Wave Astronomy.
\end{acknowledgments}

\bibliography{Rosado2011b}

\begin{thebibliography}{98}%
\makeatletter
\providecommand \@ifxundefined [1]{%
 \@ifx{#1\undefined}
}%
\providecommand \@ifnum [1]{%
 \ifnum #1\expandafter \@firstoftwo
 \else \expandafter \@secondoftwo
 \fi
}%
\providecommand \@ifx [1]{%
 \ifx #1\expandafter \@firstoftwo
 \else \expandafter \@secondoftwo
 \fi
}%
\providecommand \natexlab [1]{#1}%
\providecommand \enquote  [1]{``#1''}%
\providecommand \bibnamefont  [1]{#1}%
\providecommand \bibfnamefont [1]{#1}%
\providecommand \citenamefont [1]{#1}%
\providecommand \href@noop [0]{\@secondoftwo}%
\providecommand \href [0]{\begingroup \@sanitize@url \@href}%
\providecommand \@href[1]{\@@startlink{#1}\@@href}%
\providecommand \@@href[1]{\endgroup#1\@@endlink}%
\providecommand \@sanitize@url [0]{\catcode `\\12\catcode `\$12\catcode
  `\&12\catcode `\#12\catcode `\^12\catcode `\_12\catcode `\%12\relax}%
\providecommand \@@startlink[1]{}%
\providecommand \@@endlink[0]{}%
\providecommand \url  [0]{\begingroup\@sanitize@url \@url }%
\providecommand \@url [1]{\endgroup\@href {#1}{\urlprefix }}%
\providecommand \urlprefix  [0]{URL }%
\providecommand \Eprint [0]{\href }%
\providecommand \doibase [0]{http://dx.doi.org/}%
\providecommand \selectlanguage [0]{\@gobble}%
\providecommand \bibinfo  [0]{\@secondoftwo}%
\providecommand \bibfield  [0]{\@secondoftwo}%
\providecommand \translation [1]{[#1]}%
\providecommand \BibitemOpen [0]{}%
\providecommand \bibitemStop [0]{}%
\providecommand \bibitemNoStop [0]{.\EOS\space}%
\providecommand \EOS [0]{\spacefactor3000\relax}%
\providecommand \BibitemShut  [1]{\csname bibitem#1\endcsname}%
\let\auto@bib@innerbib\@empty
\bibitem [{\citenamefont {Allen}\ and\ \citenamefont
  {Romano}(1999)}]{AllenRomano1999}%
  \BibitemOpen
  \bibfield  {author} {\bibinfo {author} {\bibfnamefont {B.}~\bibnamefont
  {Allen}}\ and\ \bibinfo {author} {\bibfnamefont {J.~D.}\ \bibnamefont
  {Romano}},\ }\href@noop {} {\bibfield  {journal} {\bibinfo  {journal} {Phys.
  Rev. D}\ }\textbf {\bibinfo {volume} {59}},\ \bibinfo {pages} {102001}
  (\bibinfo {year} {1999})}\BibitemShut {NoStop}%
\bibitem [{\citenamefont {Maggiore}(2000)}]{Maggiore2000}%
  \BibitemOpen
  \bibfield  {author} {\bibinfo {author} {\bibfnamefont {M.}~\bibnamefont
  {Maggiore}},\ }\href@noop {} {\bibfield  {journal} {\bibinfo  {journal}
  {Phys. Reports}\ }\textbf {\bibinfo {volume} {331}},\ \bibinfo {pages} {283}
  (\bibinfo {year} {2000})}\BibitemShut {NoStop}%
\bibitem [{\citenamefont {Barack}\ and\ \citenamefont
  {Cutler}(2004)}]{BarackCutler2004}%
  \BibitemOpen
  \bibfield  {author} {\bibinfo {author} {\bibfnamefont {L.}~\bibnamefont
  {Barack}}\ and\ \bibinfo {author} {\bibfnamefont {C.}~\bibnamefont
  {Cutler}},\ }\href@noop {} {\bibfield  {journal} {\bibinfo  {journal} {Phys.
  Rev. D}\ }\textbf {\bibinfo {volume} {70}},\ \bibinfo {pages} {122002}
  (\bibinfo {year} {2004})}\BibitemShut {NoStop}%
\bibitem [{\citenamefont {Cutler}\ and\ \citenamefont
  {Harms}(2006)}]{CutlerHarms2006}%
  \BibitemOpen
  \bibfield  {author} {\bibinfo {author} {\bibfnamefont {C.}~\bibnamefont
  {Cutler}}\ and\ \bibinfo {author} {\bibfnamefont {J.}~\bibnamefont {Harms}},\
  }\href@noop {} {\bibfield  {journal} {\bibinfo  {journal} {Phys. Rev. D}\
  }\textbf {\bibinfo {volume} {73}},\ \bibinfo {pages} {042001} (\bibinfo
  {year} {2006})}\BibitemShut {NoStop}%
\bibitem [{\citenamefont {Ferrari}\ \emph
  {et~al.}(1999{\natexlab{a}})\citenamefont {Ferrari}, \citenamefont
  {Matarrese},\ and\ \citenamefont {Schneider}}]{FerrariEtAl1999}%
  \BibitemOpen
  \bibfield  {author} {\bibinfo {author} {\bibfnamefont {V.}~\bibnamefont
  {Ferrari}}, \bibinfo {author} {\bibfnamefont {S.}~\bibnamefont {Matarrese}},
  \ and\ \bibinfo {author} {\bibfnamefont {R.}~\bibnamefont {Schneider}},\
  }\href@noop {} {\bibfield  {journal} {\bibinfo  {journal} {MNRAS}\ }\textbf
  {\bibinfo {volume} {303}},\ \bibinfo {pages} {247} (\bibinfo {year}
  {1999}{\natexlab{a}})}\BibitemShut {NoStop}%
\bibitem [{\citenamefont {Ferrari}\ \emph
  {et~al.}(1999{\natexlab{b}})\citenamefont {Ferrari}, \citenamefont
  {Matarrese},\ and\ \citenamefont {Schneider}}]{FerrariEtAl1999b}%
  \BibitemOpen
  \bibfield  {author} {\bibinfo {author} {\bibfnamefont {V.}~\bibnamefont
  {Ferrari}}, \bibinfo {author} {\bibfnamefont {S.}~\bibnamefont {Matarrese}},
  \ and\ \bibinfo {author} {\bibfnamefont {R.}~\bibnamefont {Schneider}},\
  }\href@noop {} {\bibfield  {journal} {\bibinfo  {journal} {MNRAS}\ }\textbf
  {\bibinfo {volume} {303}},\ \bibinfo {pages} {258} (\bibinfo {year}
  {1999}{\natexlab{b}})}\BibitemShut {NoStop}%
\bibitem [{\citenamefont {Regimbau}\ and\ \citenamefont
  {de~Freitas~Pacheco}(2001)}]{RegimbauFreitas2001}%
  \BibitemOpen
  \bibfield  {author} {\bibinfo {author} {\bibfnamefont {T.}~\bibnamefont
  {Regimbau}}\ and\ \bibinfo {author} {\bibfnamefont {J.~A.}\ \bibnamefont
  {de~Freitas~Pacheco}},\ }\href@noop {} {\bibfield  {journal} {\bibinfo
  {journal} {A\&A}\ }\textbf {\bibinfo {volume} {376}},\ \bibinfo {pages} {381}
  (\bibinfo {year} {2001})}\BibitemShut {NoStop}%
\bibitem [{\citenamefont {Coward}\ \emph {et~al.}(2001)\citenamefont {Coward},
  \citenamefont {Burman},\ and\ \citenamefont {Blair}}]{CowardEtAl2001}%
  \BibitemOpen
  \bibfield  {author} {\bibinfo {author} {\bibfnamefont {D.}~\bibnamefont
  {Coward}}, \bibinfo {author} {\bibfnamefont {R.}~\bibnamefont {Burman}}, \
  and\ \bibinfo {author} {\bibfnamefont {D.}~\bibnamefont {Blair}},\
  }\href@noop {} {\bibfield  {journal} {\bibinfo  {journal} {MNRAS}\ }\textbf
  {\bibinfo {volume} {324}},\ \bibinfo {pages} {1015} (\bibinfo {year}
  {2001})}\BibitemShut {NoStop}%
\bibitem [{\citenamefont {Coward}\ \emph {et~al.}(2002)\citenamefont {Coward},
  \citenamefont {Burman},\ and\ \citenamefont {Blair}}]{CowardEtAl2002}%
  \BibitemOpen
  \bibfield  {author} {\bibinfo {author} {\bibfnamefont {D.}~\bibnamefont
  {Coward}}, \bibinfo {author} {\bibfnamefont {R.}~\bibnamefont {Burman}}, \
  and\ \bibinfo {author} {\bibfnamefont {D.}~\bibnamefont {Blair}},\
  }\href@noop {} {\bibfield  {journal} {\bibinfo  {journal} {MNRAS}\ }\textbf
  {\bibinfo {volume} {329}},\ \bibinfo {pages} {411} (\bibinfo {year}
  {2002})}\BibitemShut {NoStop}%
\bibitem [{\citenamefont {Schneider}\ \emph {et~al.}(2001)\citenamefont
  {Schneider} \emph {et~al.}}]{SchneiderEtAl2001}%
  \BibitemOpen
  \bibfield  {author} {\bibinfo {author} {\bibfnamefont {R.}~\bibnamefont
  {Schneider}} \emph {et~al.},\ }\href@noop {} {\bibfield  {journal} {\bibinfo
  {journal} {MNRAS}\ }\textbf {\bibinfo {volume} {324}},\ \bibinfo {pages}
  {797} (\bibinfo {year} {2001})}\BibitemShut {NoStop}%
\bibitem [{\citenamefont {Farmer}\ and\ \citenamefont
  {Phinney}(2003)}]{FarmerPhinney2003}%
  \BibitemOpen
  \bibfield  {author} {\bibinfo {author} {\bibfnamefont {A.~J.}\ \bibnamefont
  {Farmer}}\ and\ \bibinfo {author} {\bibfnamefont {E.~S.}\ \bibnamefont
  {Phinney}},\ }\href@noop {} {\bibfield  {journal} {\bibinfo  {journal}
  {MNRAS}\ }\textbf {\bibinfo {volume} {346}},\ \bibinfo {pages} {1197}
  (\bibinfo {year} {2003})}\BibitemShut {NoStop}%
\bibitem [{\citenamefont {Regimbau}\ and\ \citenamefont
  {de~Freitas~Pacheco}(2006{\natexlab{a}})}]{RegimbauFreitas2006}%
  \BibitemOpen
  \bibfield  {author} {\bibinfo {author} {\bibfnamefont {T.}~\bibnamefont
  {Regimbau}}\ and\ \bibinfo {author} {\bibfnamefont {J.~A.}\ \bibnamefont
  {de~Freitas~Pacheco}},\ }\href@noop {} {\bibfield  {journal} {\bibinfo
  {journal} {ApJ}\ }\textbf {\bibinfo {volume} {642}},\ \bibinfo {pages} {455}
  (\bibinfo {year} {2006}{\natexlab{a}})}\BibitemShut {NoStop}%
\bibitem [{\citenamefont {Regimbau}\ and\ \citenamefont
  {Mandic}(2008)}]{RegimbauMandic2008}%
  \BibitemOpen
  \bibfield  {author} {\bibinfo {author} {\bibfnamefont {T.}~\bibnamefont
  {Regimbau}}\ and\ \bibinfo {author} {\bibfnamefont {V.}~\bibnamefont
  {Mandic}},\ }\href@noop {} {\bibfield  {journal} {\bibinfo  {journal} {Clas.
  Quantum Grav.}\ }\textbf {\bibinfo {volume} {25}},\ \bibinfo {pages} {184018}
  (\bibinfo {year} {2008})}\BibitemShut {NoStop}%
\bibitem [{\citenamefont {Regimbau}\ and\ \citenamefont
  {Hughes}(2009)}]{RegimbauHughes2009}%
  \BibitemOpen
  \bibfield  {author} {\bibinfo {author} {\bibfnamefont {T.}~\bibnamefont
  {Regimbau}}\ and\ \bibinfo {author} {\bibfnamefont {S.~A.}\ \bibnamefont
  {Hughes}},\ }\href@noop {} {\bibfield  {journal} {\bibinfo  {journal} {Phys.
  Rev. D}\ }\textbf {\bibinfo {volume} {79}},\ \bibinfo {pages} {062002}
  (\bibinfo {year} {2009})}\BibitemShut {NoStop}%
\bibitem [{\citenamefont {Sesana}\ \emph {et~al.}(2005)\citenamefont {Sesana}
  \emph {et~al.}}]{SesanaEtAl2005}%
  \BibitemOpen
  \bibfield  {author} {\bibinfo {author} {\bibfnamefont {A.}~\bibnamefont
  {Sesana}} \emph {et~al.},\ }\href@noop {} {\bibfield  {journal} {\bibinfo
  {journal} {ApJ}\ }\textbf {\bibinfo {volume} {623}},\ \bibinfo {pages} {23}
  (\bibinfo {year} {2005})}\BibitemShut {NoStop}%
\bibitem [{\citenamefont {Sesana}\ \emph {et~al.}(2008)\citenamefont {Sesana},
  \citenamefont {Vecchio},\ and\ \citenamefont {Colacino}}]{SesanaEtAl2008}%
  \BibitemOpen
  \bibfield  {author} {\bibinfo {author} {\bibfnamefont {A.}~\bibnamefont
  {Sesana}}, \bibinfo {author} {\bibfnamefont {A.}~\bibnamefont {Vecchio}}, \
  and\ \bibinfo {author} {\bibfnamefont {C.~N.}\ \bibnamefont {Colacino}},\
  }\href@noop {} {\bibfield  {journal} {\bibinfo  {journal} {MNRAS}\ }\textbf
  {\bibinfo {volume} {390}},\ \bibinfo {pages} {192} (\bibinfo {year}
  {2008})}\BibitemShut {NoStop}%
\bibitem [{\citenamefont {Regimbau}\ and\ \citenamefont
  {de~Freitas~Pacheco}(2006{\natexlab{b}})}]{RegimbauFreitas2006b}%
  \BibitemOpen
  \bibfield  {author} {\bibinfo {author} {\bibfnamefont {T.}~\bibnamefont
  {Regimbau}}\ and\ \bibinfo {author} {\bibfnamefont {J.~A.}\ \bibnamefont
  {de~Freitas~Pacheco}},\ }\href@noop {} {\bibfield  {journal} {\bibinfo
  {journal} {A\&A}\ }\textbf {\bibinfo {volume} {447}},\ \bibinfo {pages} {1}
  (\bibinfo {year} {2006}{\natexlab{b}})}\BibitemShut {NoStop}%
\bibitem [{\citenamefont {de~Araujo}\ and\ \citenamefont
  {Miranda}(2005)}]{AraujoMiranda2005}%
  \BibitemOpen
  \bibfield  {author} {\bibinfo {author} {\bibfnamefont {J.}~\bibnamefont
  {de~Araujo}}\ and\ \bibinfo {author} {\bibfnamefont {O.}~\bibnamefont
  {Miranda}},\ }\href@noop {} {\bibfield  {journal} {\bibinfo  {journal} {Phys.
  Rev. D}\ }\textbf {\bibinfo {volume} {71}},\ \bibinfo {pages} {127503}
  (\bibinfo {year} {2005})}\BibitemShut {NoStop}%
\bibitem [{\citenamefont {de~Araujo}\ \emph {et~al.}(2004)\citenamefont
  {de~Araujo}, \citenamefont {Miranda},\ and\ \citenamefont
  {Aguiar}}]{AraujoEtAl2004}%
  \BibitemOpen
  \bibfield  {author} {\bibinfo {author} {\bibfnamefont {J.}~\bibnamefont
  {de~Araujo}}, \bibinfo {author} {\bibfnamefont {O.}~\bibnamefont {Miranda}},
  \ and\ \bibinfo {author} {\bibfnamefont {O.}~\bibnamefont {Aguiar}},\
  }\href@noop {} {\bibfield  {journal} {\bibinfo  {journal} {MNRAS}\ }\textbf
  {\bibinfo {volume} {348}},\ \bibinfo {pages} {1373} (\bibinfo {year}
  {2004})}\BibitemShut {NoStop}%
\bibitem [{\citenamefont {Zhu}\ \emph {et~al.}(2011)\citenamefont {Zhu},
  \citenamefont {Fan},\ and\ \citenamefont {Zhu}}]{ZhuEtAl2011b}%
  \BibitemOpen
  \bibfield  {author} {\bibinfo {author} {\bibfnamefont {X.-J.}\ \bibnamefont
  {Zhu}}, \bibinfo {author} {\bibfnamefont {X.-L.}\ \bibnamefont {Fan}}, \ and\
  \bibinfo {author} {\bibfnamefont {Z.-H.}\ \bibnamefont {Zhu}},\ }\href@noop
  {} {\bibfield  {journal} {\bibinfo  {journal} {ApJ}\ }\textbf {\bibinfo
  {volume} {729}},\ \bibinfo {pages} {59} (\bibinfo {year} {2011})}\BibitemShut
  {NoStop}%
\bibitem [{\citenamefont {Grishchuk}\ \emph {et~al.}(2001)\citenamefont
  {Grishchuk} \emph {et~al.}}]{GrishchukEtAl2001}%
  \BibitemOpen
  \bibfield  {author} {\bibinfo {author} {\bibfnamefont {L.}~\bibnamefont
  {Grishchuk}} \emph {et~al.},\ }\href@noop {} {\bibfield  {journal} {\bibinfo
  {journal} {Physics-Uspekhi}\ }\textbf {\bibinfo {volume} {44}},\ \bibinfo
  {pages} {1} (\bibinfo {year} {2001})}\BibitemShut {NoStop}%
\bibitem [{\citenamefont {Ando}\ \emph {et~al.}(2001)\citenamefont {Ando} \emph
  {et~al.}}]{AndoEtAl2001}%
  \BibitemOpen
  \bibfield  {author} {\bibinfo {author} {\bibfnamefont {M.}~\bibnamefont
  {Ando}} \emph {et~al.},\ }\href@noop {} {\bibfield  {journal} {\bibinfo
  {journal} {Phys. Rev. Lett.}\ }\textbf {\bibinfo {volume} {86}},\ \bibinfo
  {pages} {3950} (\bibinfo {year} {2001})}\BibitemShut {NoStop}%
\bibitem [{\citenamefont {L{\"u}ck}\ and\ \citenamefont {{GEO600
  team}}(1997)}]{LueckEtAl1997}%
  \BibitemOpen
  \bibfield  {author} {\bibinfo {author} {\bibfnamefont {H.}~\bibnamefont
  {L{\"u}ck}}\ and\ \bibinfo {author} {\bibnamefont {{GEO600 team}}},\
  }\href@noop {} {\bibfield  {journal} {\bibinfo  {journal} {Clas. Quantum
  Grav.}\ }\textbf {\bibinfo {volume} {14}},\ \bibinfo {pages} {1471} (\bibinfo
  {year} {1997})}\BibitemShut {NoStop}%
\bibitem [{\citenamefont {Caron}\ \emph {et~al.}(1997)\citenamefont {Caron}
  \emph {et~al.}}]{CaronEtAl1997}%
  \BibitemOpen
  \bibfield  {author} {\bibinfo {author} {\bibfnamefont {B.}~\bibnamefont
  {Caron}} \emph {et~al.},\ }\href@noop {} {\bibfield  {journal} {\bibinfo
  {journal} {Clas. Quantum Grav.}\ }\textbf {\bibinfo {volume} {14}},\ \bibinfo
  {pages} {1461} (\bibinfo {year} {1997})}\BibitemShut {NoStop}%
\bibitem [{\citenamefont {{LIGO Scientific Collaboration}}(2009)}]{LIGO2009}%
  \BibitemOpen
  \bibfield  {author} {\bibinfo {author} {\bibnamefont {{LIGO Scientific
  Collaboration}}},\ }\href@noop {} {\bibfield  {journal} {\bibinfo  {journal}
  {RPP}\ }\textbf {\bibinfo {volume} {72}},\ \bibinfo {pages} {076901}
  (\bibinfo {year} {2009})}\BibitemShut {NoStop}%
\bibitem [{\citenamefont {Flaminio}\ \emph {et~al.}(2005)\citenamefont
  {Flaminio} \emph {et~al.}}]{FlaminioEtAl2005}%
  \BibitemOpen
  \bibfield  {author} {\bibinfo {author} {\bibfnamefont {R.}~\bibnamefont
  {Flaminio}} \emph {et~al.},\ }\href@noop {} {\bibfield  {journal} {\bibinfo
  {journal} {VIR-NOT-DIR-1390-304}\ } (\bibinfo {year} {2005})}\BibitemShut
  {NoStop}%
\bibitem [{\citenamefont {Kuroda}\ and\ \citenamefont {{LCGT
  Collaboration}}(2006)}]{KurodaEtAl2006}%
  \BibitemOpen
  \bibfield  {author} {\bibinfo {author} {\bibfnamefont {K.}~\bibnamefont
  {Kuroda}}\ and\ \bibinfo {author} {\bibnamefont {{LCGT Collaboration}}},\
  }\href@noop {} {\bibfield  {journal} {\bibinfo  {journal} {Clas. Quantum
  Grav.}\ }\textbf {\bibinfo {volume} {23}},\ \bibinfo {pages} {S215} (\bibinfo
  {year} {2006})}\BibitemShut {NoStop}%
\bibitem [{\citenamefont {Punturo}(2010)}]{PunturoEtAl2010}%
  \BibitemOpen
  \bibfield  {author} {\bibinfo {author} {\bibfnamefont {M.}~\bibnamefont
  {Punturo}},\ }\href@noop {} {\bibfield  {journal} {\bibinfo  {journal} {Clas.
  Quantum Grav.}\ }\textbf {\bibinfo {volume} {27}},\ \bibinfo {pages} {194002}
  (\bibinfo {year} {2010})}\BibitemShut {NoStop}%
\bibitem [{\citenamefont {Danzmann}\ and\ \citenamefont {{LISA study
  team}}(1996)}]{DanzmannEtAl1996}%
  \BibitemOpen
  \bibfield  {author} {\bibinfo {author} {\bibfnamefont {K.}~\bibnamefont
  {Danzmann}}\ and\ \bibinfo {author} {\bibnamefont {{LISA study team}}},\
  }\href@noop {} {\bibfield  {journal} {\bibinfo  {journal} {Clas. Quantum
  Grav.}\ }\textbf {\bibinfo {volume} {13}},\ \bibinfo {pages} {A247} (\bibinfo
  {year} {1996})}\BibitemShut {NoStop}%
\bibitem [{\citenamefont {Phinney}\ \emph {et~al.}(2004)\citenamefont {Phinney}
  \emph {et~al.}}]{PhinneyEtAl2004}%
  \BibitemOpen
  \bibfield  {author} {\bibinfo {author} {\bibfnamefont {E.~S.}\ \bibnamefont
  {Phinney}} \emph {et~al.},\ }\href@noop {} {\bibfield  {journal} {\bibinfo
  {journal} {NASA OSS Vision Missions Program, Proposal VM03-0021-0021}\ }
  (\bibinfo {year} {2004})}\BibitemShut {NoStop}%
\bibitem [{\citenamefont {Hobbs}\ \emph {et~al.}(2010)\citenamefont {Hobbs}
  \emph {et~al.}}]{HobbsEtAl2010}%
  \BibitemOpen
  \bibfield  {author} {\bibinfo {author} {\bibfnamefont {G.}~\bibnamefont
  {Hobbs}} \emph {et~al.},\ }\href@noop {} {\bibfield  {journal} {\bibinfo
  {journal} {Clas. Quantum Grav.}\ }\textbf {\bibinfo {volume} {27}},\ \bibinfo
  {pages} {084013} (\bibinfo {year} {2010})}\BibitemShut {NoStop}%
\bibitem [{\citenamefont {Manchester}(2008)}]{Manchester2008}%
  \BibitemOpen
  \bibfield  {author} {\bibinfo {author} {\bibfnamefont {R.~N.}\ \bibnamefont
  {Manchester}},\ }\href@noop {} {\bibfield  {journal} {\bibinfo  {journal}
  {AIP Conf. Proc.}\ }\textbf {\bibinfo {volume} {983}},\ \bibinfo {pages}
  {584} (\bibinfo {year} {2008})}\BibitemShut {NoStop}%
\bibitem [{\citenamefont {Phinney}(2001)}]{Phinney2001}%
  \BibitemOpen
  \bibfield  {author} {\bibinfo {author} {\bibfnamefont {E.~S.}\ \bibnamefont
  {Phinney}},\ }\href@noop {} {\bibfield  {journal} {\bibinfo  {journal}
  {arXiv: astro-ph/0108028v1}\ } (\bibinfo {year} {2001})}\BibitemShut
  {NoStop}%
\bibitem [{\citenamefont {Allen}(1996)}]{Allen1996}%
  \BibitemOpen
  \bibfield  {author} {\bibinfo {author} {\bibfnamefont {B.}~\bibnamefont
  {Allen}},\ }\href@noop {} {\bibfield  {journal} {\bibinfo  {journal} {arXiv:
  gr-qc/9604033v3}\ } (\bibinfo {year} {1996})}\BibitemShut {NoStop}%
\bibitem [{\citenamefont {Penzias}\ and\ \citenamefont
  {Wilson}(1965)}]{PenziasWilson1965}%
  \BibitemOpen
  \bibfield  {author} {\bibinfo {author} {\bibfnamefont {A.~A.}\ \bibnamefont
  {Penzias}}\ and\ \bibinfo {author} {\bibfnamefont {R.~W.}\ \bibnamefont
  {Wilson}},\ }\href@noop {} {\bibfield  {journal} {\bibinfo  {journal} {ApJ}\
  }\textbf {\bibinfo {volume} {142}},\ \bibinfo {pages} {419} (\bibinfo {year}
  {1965})}\BibitemShut {NoStop}%
\bibitem [{\citenamefont {Boggess}\ \emph {et~al.}(1992)\citenamefont {Boggess}
  \emph {et~al.}}]{BoggessEtAl1992}%
  \BibitemOpen
  \bibfield  {author} {\bibinfo {author} {\bibfnamefont {N.}~\bibnamefont
  {Boggess}} \emph {et~al.},\ }\href@noop {} {\bibfield  {journal} {\bibinfo
  {journal} {ApJ}\ }\textbf {\bibinfo {volume} {397}},\ \bibinfo {pages} {420}
  (\bibinfo {year} {1992})}\BibitemShut {NoStop}%
\bibitem [{\citenamefont {Bennett}\ \emph {et~al.}(2003)\citenamefont {Bennett}
  \emph {et~al.}}]{BennettEtAl2003}%
  \BibitemOpen
  \bibfield  {author} {\bibinfo {author} {\bibfnamefont {C.}~\bibnamefont
  {Bennett}} \emph {et~al.},\ }\href@noop {} {\bibfield  {journal} {\bibinfo
  {journal} {ApJ}\ }\textbf {\bibinfo {volume} {583}},\ \bibinfo {pages} {1}
  (\bibinfo {year} {2003})}\BibitemShut {NoStop}%
\bibitem [{\citenamefont {Hawking}\ and\ \citenamefont
  {Israel}(1987)}]{HawkingIsrael1987}%
  \BibitemOpen
  \bibfield  {author} {\bibinfo {author} {\bibfnamefont {S.~W.}\ \bibnamefont
  {Hawking}}\ and\ \bibinfo {author} {\bibfnamefont {W.}~\bibnamefont
  {Israel}},\ }\href@noop {} {\emph {\bibinfo {title} {300 Years of
  Gravitation}}}\ (\bibinfo  {publisher} {Cambridge University Press},\
  \bibinfo {address} {Cambridge, England},\ \bibinfo {year} {1987})\BibitemShut
  {NoStop}%
\bibitem [{\citenamefont {Allen}(1988)}]{Allen1988}%
  \BibitemOpen
  \bibfield  {author} {\bibinfo {author} {\bibfnamefont {B.}~\bibnamefont
  {Allen}},\ }\href@noop {} {\bibfield  {journal} {\bibinfo  {journal} {Phys.
  Rev. D}\ }\textbf {\bibinfo {volume} {37}},\ \bibinfo {pages} {2078}
  (\bibinfo {year} {1988})}\BibitemShut {NoStop}%
\bibitem [{\citenamefont {Vilenkin}(1985)}]{Vilenkin1985}%
  \BibitemOpen
  \bibfield  {author} {\bibinfo {author} {\bibfnamefont {A.}~\bibnamefont
  {Vilenkin}},\ }\href@noop {} {\bibfield  {journal} {\bibinfo  {journal}
  {Phys. Reports}\ }\textbf {\bibinfo {volume} {121}},\ \bibinfo {pages} {263}
  (\bibinfo {year} {1985})}\BibitemShut {NoStop}%
\bibitem [{\citenamefont {Schneider}\ \emph {et~al.}(1999)\citenamefont
  {Schneider}, \citenamefont {Ferrari},\ and\ \citenamefont
  {Matarrese}}]{SchneiderEtAl1999}%
  \BibitemOpen
  \bibfield  {author} {\bibinfo {author} {\bibfnamefont {R.}~\bibnamefont
  {Schneider}}, \bibinfo {author} {\bibfnamefont {V.}~\bibnamefont {Ferrari}},
  \ and\ \bibinfo {author} {\bibfnamefont {S.}~\bibnamefont {Matarrese}},\
  }\href@noop {} {\bibfield  {journal} {\bibinfo  {journal} {arXiv:
  astro-ph/9903470v1}\ } (\bibinfo {year} {1999})}\BibitemShut {NoStop}%
\bibitem [{\citenamefont {Schneider}\ \emph {et~al.}(2010)\citenamefont
  {Schneider}, \citenamefont {Marassi},\ and\ \citenamefont
  {Ferrari}}]{SchneiderEtAl2010}%
  \BibitemOpen
  \bibfield  {author} {\bibinfo {author} {\bibfnamefont {R.}~\bibnamefont
  {Schneider}}, \bibinfo {author} {\bibfnamefont {S.}~\bibnamefont {Marassi}},
  \ and\ \bibinfo {author} {\bibfnamefont {V.}~\bibnamefont {Ferrari}},\
  }\href@noop {} {\bibfield  {journal} {\bibinfo  {journal} {Clas. Quantum
  Grav.}\ }\textbf {\bibinfo {volume} {27}},\ \bibinfo {pages} {194007}
  (\bibinfo {year} {2010})}\BibitemShut {NoStop}%
\bibitem [{\citenamefont {Maggiore}(2008)}]{Maggiore2008}%
  \BibitemOpen
  \bibfield  {author} {\bibinfo {author} {\bibfnamefont {M.}~\bibnamefont
  {Maggiore}},\ }\href@noop {} {\emph {\bibinfo {title} {Gravitational Waves
  Volume 1: Theory and Experiments}}}\ (\bibinfo  {publisher} {Oxford
  University Press},\ \bibinfo {address} {Oxford},\ \bibinfo {year}
  {2008})\BibitemShut {NoStop}%
\bibitem [{\citenamefont {Misner}\ \emph {et~al.}(1973)\citenamefont {Misner},
  \citenamefont {Thorne},\ and\ \citenamefont {Wheeler}}]{MisnerEtAl1973}%
  \BibitemOpen
  \bibfield  {author} {\bibinfo {author} {\bibfnamefont {C.~W.}\ \bibnamefont
  {Misner}}, \bibinfo {author} {\bibfnamefont {K.~S.}\ \bibnamefont {Thorne}},
  \ and\ \bibinfo {author} {\bibfnamefont {J.~A.}\ \bibnamefont {Wheeler}},\
  }\href@noop {} {\emph {\bibinfo {title} {Gravitation}}}\ (\bibinfo
  {publisher} {W. H. Freeman and Co.},\ \bibinfo {address} {San Francisco},\
  \bibinfo {year} {1973})\BibitemShut {NoStop}%
\bibitem [{\citenamefont {Jarosik}\ \emph {et~al.}(2011)\citenamefont {Jarosik}
  \emph {et~al.}}]{JarosikEtAl2011}%
  \BibitemOpen
  \bibfield  {author} {\bibinfo {author} {\bibfnamefont {N.}~\bibnamefont
  {Jarosik}} \emph {et~al.},\ }\href@noop {} {\bibfield  {journal} {\bibinfo
  {journal} {ApJ Suppl. Ser.}\ }\textbf {\bibinfo {volume} {192}},\ \bibinfo
  {pages} {14} (\bibinfo {year} {2011})}\BibitemShut {NoStop}%
\bibitem [{\citenamefont {Riess}\ \emph {et~al.}(2009)\citenamefont {Riess}
  \emph {et~al.}}]{RiessEtAl2009}%
  \BibitemOpen
  \bibfield  {author} {\bibinfo {author} {\bibfnamefont {A.~G.}\ \bibnamefont
  {Riess}} \emph {et~al.},\ }\href@noop {} {\bibfield  {journal} {\bibinfo
  {journal} {ApJ}\ }\textbf {\bibinfo {volume} {699}},\ \bibinfo {pages} {539}
  (\bibinfo {year} {2009})}\BibitemShut {NoStop}%
\bibitem [{\citenamefont {Hawking}\ and\ \citenamefont
  {Ellis}(1973)}]{HawkingEllis1973}%
  \BibitemOpen
  \bibfield  {author} {\bibinfo {author} {\bibfnamefont {S.~W.}\ \bibnamefont
  {Hawking}}\ and\ \bibinfo {author} {\bibfnamefont {G.~F.~R.}\ \bibnamefont
  {Ellis}},\ }\href@noop {} {\emph {\bibinfo {title} {The Large Scale Structure
  of Space-Time}}}\ (\bibinfo  {publisher} {Cambridge University Press},\
  \bibinfo {address} {Cambridge, England},\ \bibinfo {year} {1973})\BibitemShut
  {NoStop}%
\bibitem [{\citenamefont {L{\"a}mmerzahl}\ \emph {et~al.}(2001)\citenamefont
  {L{\"a}mmerzahl}, \citenamefont {Everitt},\ and\ \citenamefont
  {Hehl}}]{LaemmerzahlEtAl2001}%
  \BibitemOpen
  \bibfield  {author} {\bibinfo {author} {\bibfnamefont {C.}~\bibnamefont
  {L{\"a}mmerzahl}}, \bibinfo {author} {\bibfnamefont {C.~W.~F.}\ \bibnamefont
  {Everitt}}, \ and\ \bibinfo {author} {\bibfnamefont {F.~W.}\ \bibnamefont
  {Hehl}},\ }\href@noop {} {\emph {\bibinfo {title} {Gyros, Clocks,
  Interferometers...: Testing Relativistic Gravity in Space}}}\ (\bibinfo
  {publisher} {Springer Verlag},\ \bibinfo {address} {Berlin},\ \bibinfo {year}
  {2001})\BibitemShut {NoStop}%
\bibitem [{Note1()}]{Note1}%
  \BibitemOpen
  \bibinfo {note} {Notice also that $P_e(f[1+z])$ has been defined differently
  in the first version of this paper, in arXiv:1106.5795v1. The equations for
  the spectral function are, nevertheless, equivalent in the two
  versions.}\BibitemShut {Stop}%
\bibitem [{\citenamefont {Connolly}\ \emph {et~al.}(1997)\citenamefont
  {Connolly} \emph {et~al.}}]{ConnollyEtAl1997}%
  \BibitemOpen
  \bibfield  {author} {\bibinfo {author} {\bibfnamefont {A.}~\bibnamefont
  {Connolly}} \emph {et~al.},\ }\href@noop {} {\bibfield  {journal} {\bibinfo
  {journal} {ApJ}\ }\textbf {\bibinfo {volume} {486}},\ \bibinfo {pages} {L11}
  (\bibinfo {year} {1997})}\BibitemShut {NoStop}%
\bibitem [{\citenamefont {Porciani}\ and\ \citenamefont
  {Madau}(2001)}]{PorcianiMadau2001}%
  \BibitemOpen
  \bibfield  {author} {\bibinfo {author} {\bibfnamefont {C.}~\bibnamefont
  {Porciani}}\ and\ \bibinfo {author} {\bibfnamefont {P.}~\bibnamefont
  {Madau}},\ }\href@noop {} {\bibfield  {journal} {\bibinfo  {journal} {ApJ}\
  }\textbf {\bibinfo {volume} {548}},\ \bibinfo {pages} {522} (\bibinfo {year}
  {2001})}\BibitemShut {NoStop}%
\bibitem [{\citenamefont {2dFGRS Team}(2001)}]{2dFGRS2001}%
  \BibitemOpen
  \bibfield  {author} {\bibinfo {author} {\bibnamefont {2dFGRS Team}},\
  }\href@noop {} {\bibfield  {journal} {\bibinfo  {journal} {MNRAS}\ }\textbf
  {\bibinfo {volume} {326}},\ \bibinfo {pages} {255} (\bibinfo {year}
  {2001})}\BibitemShut {NoStop}%
\bibitem [{\citenamefont {Hopkins}\ and\ \citenamefont
  {Beacom}(2006)}]{HopkinsBeacom2006}%
  \BibitemOpen
  \bibfield  {author} {\bibinfo {author} {\bibfnamefont {A.~M.}\ \bibnamefont
  {Hopkins}}\ and\ \bibinfo {author} {\bibfnamefont {J.~F.}\ \bibnamefont
  {Beacom}},\ }\href@noop {} {\bibfield  {journal} {\bibinfo  {journal} {ApJ}\
  }\textbf {\bibinfo {volume} {651}},\ \bibinfo {pages} {142} (\bibinfo {year}
  {2006})}\BibitemShut {NoStop}%
\bibitem [{\citenamefont {Nagamine}\ \emph {et~al.}(2006)\citenamefont
  {Nagamine} \emph {et~al.}}]{NagamineEtAl2006}%
  \BibitemOpen
  \bibfield  {author} {\bibinfo {author} {\bibfnamefont {K.}~\bibnamefont
  {Nagamine}} \emph {et~al.},\ }\href@noop {} {\bibfield  {journal} {\bibinfo
  {journal} {ApJ}\ }\textbf {\bibinfo {volume} {653}},\ \bibinfo {pages} {881}
  (\bibinfo {year} {2006})}\BibitemShut {NoStop}%
\bibitem [{\citenamefont {Fardal}\ \emph {et~al.}(2007)\citenamefont {Fardal}
  \emph {et~al.}}]{FardalEtAl2007}%
  \BibitemOpen
  \bibfield  {author} {\bibinfo {author} {\bibfnamefont {M.~A.}\ \bibnamefont
  {Fardal}} \emph {et~al.},\ }\href@noop {} {\bibfield  {journal} {\bibinfo
  {journal} {MNRAS}\ }\textbf {\bibinfo {volume} {379}},\ \bibinfo {pages}
  {985} (\bibinfo {year} {2007})}\BibitemShut {NoStop}%
\bibitem [{\citenamefont {Wilkins}\ \emph {et~al.}(2008)\citenamefont
  {Wilkins}, \citenamefont {Trentham},\ and\ \citenamefont
  {Hopkins}}]{WilkinsEtAl2008}%
  \BibitemOpen
  \bibfield  {author} {\bibinfo {author} {\bibfnamefont {S.~M.}\ \bibnamefont
  {Wilkins}}, \bibinfo {author} {\bibfnamefont {N.}~\bibnamefont {Trentham}}, \
  and\ \bibinfo {author} {\bibfnamefont {A.~M.}\ \bibnamefont {Hopkins}},\
  }\href@noop {} {\bibfield  {journal} {\bibinfo  {journal} {MNRAS}\ }\textbf
  {\bibinfo {volume} {385}},\ \bibinfo {pages} {687} (\bibinfo {year}
  {2008})}\BibitemShut {NoStop}%
\bibitem [{\citenamefont {Kormendy}\ and\ \citenamefont
  {Richstone}(1995)}]{KormendyRichstone1995}%
  \BibitemOpen
  \bibfield  {author} {\bibinfo {author} {\bibfnamefont {J.}~\bibnamefont
  {Kormendy}}\ and\ \bibinfo {author} {\bibfnamefont {D.}~\bibnamefont
  {Richstone}},\ }\href@noop {} {\bibfield  {journal} {\bibinfo  {journal}
  {Annu. Rev. Astron. Astrophys.}\ }\textbf {\bibinfo {volume} {33}},\ \bibinfo
  {pages} {581} (\bibinfo {year} {1995})}\BibitemShut {NoStop}%
\bibitem [{\citenamefont {Magorrian}\ \emph {et~al.}(1998)\citenamefont
  {Magorrian} \emph {et~al.}}]{MagorrianEtAl1998}%
  \BibitemOpen
  \bibfield  {author} {\bibinfo {author} {\bibfnamefont {J.}~\bibnamefont
  {Magorrian}} \emph {et~al.},\ }\href@noop {} {\bibfield  {journal} {\bibinfo
  {journal} {Astronomical Journal}\ }\textbf {\bibinfo {volume} {115}},\
  \bibinfo {pages} {2285} (\bibinfo {year} {1998})}\BibitemShut {NoStop}%
\bibitem [{\citenamefont {Arun}\ \emph {et~al.}(2009)\citenamefont {Arun} \emph
  {et~al.}}]{ArunEtAl2009}%
  \BibitemOpen
  \bibfield  {author} {\bibinfo {author} {\bibfnamefont {K.~G.}\ \bibnamefont
  {Arun}} \emph {et~al.},\ }\href@noop {} {\bibfield  {journal} {\bibinfo
  {journal} {Clas. Quantum Grav.}\ }\textbf {\bibinfo {volume} {26}},\ \bibinfo
  {pages} {094027} (\bibinfo {year} {2009})}\BibitemShut {NoStop}%
\bibitem [{\citenamefont {Nauenberg}(1972)}]{Nauenberg1972}%
  \BibitemOpen
  \bibfield  {author} {\bibinfo {author} {\bibfnamefont {M.}~\bibnamefont
  {Nauenberg}},\ }\href@noop {} {\bibfield  {journal} {\bibinfo  {journal}
  {ApJ}\ }\textbf {\bibinfo {volume} {175}},\ \bibinfo {pages} {417} (\bibinfo
  {year} {1972})}\BibitemShut {NoStop}%
\bibitem [{\citenamefont {Sesana}\ \emph {et~al.}(2004)\citenamefont {Sesana}
  \emph {et~al.}}]{SesanaEtAl2004}%
  \BibitemOpen
  \bibfield  {author} {\bibinfo {author} {\bibfnamefont {A.}~\bibnamefont
  {Sesana}} \emph {et~al.},\ }\href@noop {} {\bibfield  {journal} {\bibinfo
  {journal} {ApJ}\ }\textbf {\bibinfo {volume} {611}},\ \bibinfo {pages} {623}
  (\bibinfo {year} {2004})}\BibitemShut {NoStop}%
\bibitem [{\citenamefont {Quinlan}(1996)}]{Quinlan1996}%
  \BibitemOpen
  \bibfield  {author} {\bibinfo {author} {\bibfnamefont {G.~D.}\ \bibnamefont
  {Quinlan}},\ }\href@noop {} {\bibfield  {journal} {\bibinfo  {journal} {New
  Astronomy}\ }\textbf {\bibinfo {volume} {1}},\ \bibinfo {pages} {35}
  (\bibinfo {year} {1996})}\BibitemShut {NoStop}%
\bibitem [{\citenamefont {Milosavljevi{\'c}}\ and\ \citenamefont
  {Merritt}(2003)}]{MilosavljeviMerritt2003}%
  \BibitemOpen
  \bibfield  {author} {\bibinfo {author} {\bibfnamefont {M.}~\bibnamefont
  {Milosavljevi{\'c}}}\ and\ \bibinfo {author} {\bibfnamefont {D.}~\bibnamefont
  {Merritt}},\ }\href@noop {} {\bibfield  {journal} {\bibinfo  {journal} {AIP
  Conf. Proc.}\ }\textbf {\bibinfo {volume} {686}},\ \bibinfo {pages} {201}
  (\bibinfo {year} {2003})}\BibitemShut {NoStop}%
\bibitem [{\citenamefont {Dotti}\ \emph {et~al.}(2006)\citenamefont {Dotti},
  \citenamefont {Colpi},\ and\ \citenamefont {Haardt}}]{DottiEtAl2006}%
  \BibitemOpen
  \bibfield  {author} {\bibinfo {author} {\bibfnamefont {M.}~\bibnamefont
  {Dotti}}, \bibinfo {author} {\bibfnamefont {M.}~\bibnamefont {Colpi}}, \ and\
  \bibinfo {author} {\bibfnamefont {F.}~\bibnamefont {Haardt}},\ }\href@noop {}
  {\bibfield  {journal} {\bibinfo  {journal} {MNRAS}\ }\textbf {\bibinfo
  {volume} {367}},\ \bibinfo {pages} {103} (\bibinfo {year}
  {2006})}\BibitemShut {NoStop}%
\bibitem [{\citenamefont {Escala}\ \emph {et~al.}(2005)\citenamefont {Escala}
  \emph {et~al.}}]{EscalaEtAl2005}%
  \BibitemOpen
  \bibfield  {author} {\bibinfo {author} {\bibfnamefont {A.}~\bibnamefont
  {Escala}} \emph {et~al.},\ }\href@noop {} {\bibfield  {journal} {\bibinfo
  {journal} {ApJ}\ }\textbf {\bibinfo {volume} {630}},\ \bibinfo {pages} {152}
  (\bibinfo {year} {2005})}\BibitemShut {NoStop}%
\bibitem [{\citenamefont {Kocsis}\ and\ \citenamefont
  {Sesana}(2010)}]{KocsisEtAl2010}%
  \BibitemOpen
  \bibfield  {author} {\bibinfo {author} {\bibfnamefont {B.}~\bibnamefont
  {Kocsis}}\ and\ \bibinfo {author} {\bibfnamefont {A.}~\bibnamefont
  {Sesana}},\ }\href@noop {} {\bibfield  {journal} {\bibinfo  {journal} {arXiv:
  astro-ph.CO/1002.0584v2}\ } (\bibinfo {year} {2010})}\BibitemShut {NoStop}%
\bibitem [{\citenamefont {Ferrarese}\ and\ \citenamefont
  {Merritt}(2000)}]{FerrareseMerritt2000}%
  \BibitemOpen
  \bibfield  {author} {\bibinfo {author} {\bibfnamefont {L.}~\bibnamefont
  {Ferrarese}}\ and\ \bibinfo {author} {\bibfnamefont {D.}~\bibnamefont
  {Merritt}},\ }\href@noop {} {\bibfield  {journal} {\bibinfo  {journal} {ApJ}\
  }\textbf {\bibinfo {volume} {539}},\ \bibinfo {pages} {L9} (\bibinfo {year}
  {2000})}\BibitemShut {NoStop}%
\bibitem [{\citenamefont {Gebhardt}\ \emph {et~al.}(2000)\citenamefont
  {Gebhardt} \emph {et~al.}}]{GebhardtEtAl2000}%
  \BibitemOpen
  \bibfield  {author} {\bibinfo {author} {\bibfnamefont {K.}~\bibnamefont
  {Gebhardt}} \emph {et~al.},\ }\href@noop {} {\bibfield  {journal} {\bibinfo
  {journal} {ApJ}\ }\textbf {\bibinfo {volume} {539}},\ \bibinfo {pages} {L13}
  (\bibinfo {year} {2000})}\BibitemShut {NoStop}%
\bibitem [{\citenamefont {G{\"u}ltekin}\ \emph {et~al.}(2009)\citenamefont
  {G{\"u}ltekin} \emph {et~al.}}]{GueltekinEtAl2009}%
  \BibitemOpen
  \bibfield  {author} {\bibinfo {author} {\bibfnamefont {K.}~\bibnamefont
  {G{\"u}ltekin}} \emph {et~al.},\ }\href@noop {} {\bibfield  {journal}
  {\bibinfo  {journal} {ApJ}\ }\textbf {\bibinfo {volume} {698}},\ \bibinfo
  {pages} {198} (\bibinfo {year} {2009})}\BibitemShut {NoStop}%
\bibitem [{\citenamefont {de~Freitas~Pacheco}(1997)}]{Freitas1997}%
  \BibitemOpen
  \bibfield  {author} {\bibinfo {author} {\bibfnamefont {J.~A.}\ \bibnamefont
  {de~Freitas~Pacheco}},\ }\href@noop {} {\bibfield  {journal} {\bibinfo
  {journal} {Astroparticle Physics}\ }\textbf {\bibinfo {volume} {8}},\
  \bibinfo {pages} {21} (\bibinfo {year} {1997})}\BibitemShut {NoStop}%
\bibitem [{\citenamefont {Phinney}(1991)}]{Phinney1991}%
  \BibitemOpen
  \bibfield  {author} {\bibinfo {author} {\bibfnamefont {E.~S.}\ \bibnamefont
  {Phinney}},\ }\href@noop {} {\bibfield  {journal} {\bibinfo  {journal} {ApJ}\
  }\textbf {\bibinfo {volume} {380}},\ \bibinfo {pages} {L17} (\bibinfo {year}
  {1991})}\BibitemShut {NoStop}%
\bibitem [{\citenamefont {{LIGO Scientific Collaboration}}(2010)}]{LIGO2010}%
  \BibitemOpen
  \bibfield  {author} {\bibinfo {author} {\bibnamefont {{LIGO Scientific
  Collaboration}}},\ }\href@noop {} {\bibfield  {journal} {\bibinfo  {journal}
  {Clas. Quantum Grav.}\ }\textbf {\bibinfo {volume} {27}},\ \bibinfo {pages}
  {173001} (\bibinfo {year} {2010})}\BibitemShut {NoStop}%
\bibitem [{\citenamefont {Kopparapu}\ \emph {et~al.}(2008)\citenamefont
  {Kopparapu} \emph {et~al.}}]{KopparapuEtAl2008}%
  \BibitemOpen
  \bibfield  {author} {\bibinfo {author} {\bibfnamefont {R.~K.}\ \bibnamefont
  {Kopparapu}} \emph {et~al.},\ }\href@noop {} {\bibfield  {journal} {\bibinfo
  {journal} {ApJ}\ }\textbf {\bibinfo {volume} {675}},\ \bibinfo {pages} {1459}
  (\bibinfo {year} {2008})}\BibitemShut {NoStop}%
\bibitem [{\citenamefont {Postnov}\ and\ \citenamefont
  {Yungelson}(2006)}]{PostnovYungelson2006}%
  \BibitemOpen
  \bibfield  {author} {\bibinfo {author} {\bibfnamefont {K.~A.}\ \bibnamefont
  {Postnov}}\ and\ \bibinfo {author} {\bibfnamefont {L.~R.}\ \bibnamefont
  {Yungelson}},\ }\href@noop {} {\bibfield  {journal} {\bibinfo  {journal}
  {Living Rev. Relativity}\ }\textbf {\bibinfo {volume} {9}},\ \bibinfo {pages}
  {6} (\bibinfo {year} {2006})},\ \bibinfo {note}
  {[http://www.livingreviews.org/lrr-2006-6]}\BibitemShut {NoStop}%
\bibitem [{\citenamefont {Kalogera}\ \emph {et~al.}(2001)\citenamefont
  {Kalogera} \emph {et~al.}}]{KalogeraEtAl2001}%
  \BibitemOpen
  \bibfield  {author} {\bibinfo {author} {\bibfnamefont {V.}~\bibnamefont
  {Kalogera}} \emph {et~al.},\ }\href@noop {} {\bibfield  {journal} {\bibinfo
  {journal} {ApJ}\ }\textbf {\bibinfo {volume} {556}},\ \bibinfo {pages} {340}
  (\bibinfo {year} {2001})}\BibitemShut {NoStop}%
\bibitem [{\citenamefont {Mandel}\ and\ \citenamefont
  {O'Shaughnessy}(2010)}]{MandelShaughnessy2010}%
  \BibitemOpen
  \bibfield  {author} {\bibinfo {author} {\bibfnamefont {I.}~\bibnamefont
  {Mandel}}\ and\ \bibinfo {author} {\bibfnamefont {R.}~\bibnamefont
  {O'Shaughnessy}},\ }\href@noop {} {\bibfield  {journal} {\bibinfo  {journal}
  {Clas. Quantum Grav.}\ }\textbf {\bibinfo {volume} {27}},\ \bibinfo {pages}
  {114007} (\bibinfo {year} {2010})}\BibitemShut {NoStop}%
\bibitem [{\citenamefont {Kalogera}\ \emph {et~al.}(2007)\citenamefont
  {Kalogera} \emph {et~al.}}]{KalogeraEtAl2007}%
  \BibitemOpen
  \bibfield  {author} {\bibinfo {author} {\bibfnamefont {V.}~\bibnamefont
  {Kalogera}} \emph {et~al.},\ }\href@noop {} {\bibfield  {journal} {\bibinfo
  {journal} {Phys. Reports}\ }\textbf {\bibinfo {volume} {442}},\ \bibinfo
  {pages} {75} (\bibinfo {year} {2007})}\BibitemShut {NoStop}%
\bibitem [{\citenamefont {O'Shaughnessy}\ \emph {et~al.}(2008)\citenamefont
  {O'Shaughnessy} \emph {et~al.}}]{ShaughnessyEtAl2008}%
  \BibitemOpen
  \bibfield  {author} {\bibinfo {author} {\bibfnamefont {R.}~\bibnamefont
  {O'Shaughnessy}} \emph {et~al.},\ }\href@noop {} {\bibfield  {journal}
  {\bibinfo  {journal} {ApJ}\ }\textbf {\bibinfo {volume} {672}},\ \bibinfo
  {pages} {479} (\bibinfo {year} {2008})}\BibitemShut {NoStop}%
\bibitem [{\citenamefont {Kalogera}\ \emph {et~al.}(2004)\citenamefont
  {Kalogera} \emph {et~al.}}]{KalogeraEtAl2004Erratum}%
  \BibitemOpen
  \bibfield  {author} {\bibinfo {author} {\bibfnamefont {V.}~\bibnamefont
  {Kalogera}} \emph {et~al.},\ }\href@noop {} {\bibfield  {journal} {\bibinfo
  {journal} {ApJ}\ }\textbf {\bibinfo {volume} {614}},\ \bibinfo {pages} {L137}
  (\bibinfo {year} {2004})}\BibitemShut {NoStop}%
\bibitem [{\citenamefont {Nelemans}(2003)}]{Nelemans2003}%
  \BibitemOpen
  \bibfield  {author} {\bibinfo {author} {\bibfnamefont {G.}~\bibnamefont
  {Nelemans}},\ }\href@noop {} {\bibfield  {journal} {\bibinfo  {journal} {AIP
  Conf. Proc.}\ }\textbf {\bibinfo {volume} {686}},\ \bibinfo {pages} {263}
  (\bibinfo {year} {2003})}\BibitemShut {NoStop}%
\bibitem [{\citenamefont {Bulik}\ \emph {et~al.}(2011)\citenamefont {Bulik},
  \citenamefont {Belczynski},\ and\ \citenamefont {Prestwich}}]{BulikEtAl2011}%
  \BibitemOpen
  \bibfield  {author} {\bibinfo {author} {\bibfnamefont {T.}~\bibnamefont
  {Bulik}}, \bibinfo {author} {\bibfnamefont {K.}~\bibnamefont {Belczynski}}, \
  and\ \bibinfo {author} {\bibfnamefont {A.}~\bibnamefont {Prestwich}},\
  }\href@noop {} {\bibfield  {journal} {\bibinfo  {journal} {ApJ}\ }\textbf
  {\bibinfo {volume} {730}},\ \bibinfo {pages} {140} (\bibinfo {year}
  {2011})}\BibitemShut {NoStop}%
\bibitem [{\citenamefont {Strobel}\ and\ \citenamefont
  {Weigel}(2001)}]{StrobelWeigel2001}%
  \BibitemOpen
  \bibfield  {author} {\bibinfo {author} {\bibfnamefont {K.}~\bibnamefont
  {Strobel}}\ and\ \bibinfo {author} {\bibfnamefont {M.}~\bibnamefont
  {Weigel}},\ }\href@noop {} {\bibfield  {journal} {\bibinfo  {journal} {A\&A}\
  }\textbf {\bibinfo {volume} {367}},\ \bibinfo {pages} {582} (\bibinfo {year}
  {2001})}\BibitemShut {NoStop}%
\bibitem [{\citenamefont {Limoges}\ and\ \citenamefont
  {Bergeron}(2010)}]{LimogesBergeron2010}%
  \BibitemOpen
  \bibfield  {author} {\bibinfo {author} {\bibfnamefont {M.-M.}\ \bibnamefont
  {Limoges}}\ and\ \bibinfo {author} {\bibfnamefont {P.}~\bibnamefont
  {Bergeron}},\ }\href@noop {} {\bibfield  {journal} {\bibinfo  {journal}
  {ApJ}\ }\textbf {\bibinfo {volume} {714}},\ \bibinfo {pages} {1037} (\bibinfo
  {year} {2010})}\BibitemShut {NoStop}%
\bibitem [{\citenamefont {Kepler}\ \emph {et~al.}(2007)\citenamefont {Kepler}
  \emph {et~al.}}]{KeplerEtAl2007}%
  \BibitemOpen
  \bibfield  {author} {\bibinfo {author} {\bibfnamefont {S.~O.}\ \bibnamefont
  {Kepler}} \emph {et~al.},\ }\href@noop {} {\bibfield  {journal} {\bibinfo
  {journal} {MNRAS}\ }\textbf {\bibinfo {volume} {375}},\ \bibinfo {pages}
  {1315} (\bibinfo {year} {2007})}\BibitemShut {NoStop}%
\bibitem [{\citenamefont {Zi{\'o}{\l}kowski}(2008)}]{Ziolkowski2008}%
  \BibitemOpen
  \bibfield  {author} {\bibinfo {author} {\bibfnamefont {J.}~\bibnamefont
  {Zi{\'o}{\l}kowski}},\ }\href@noop {} {\bibfield  {journal} {\bibinfo
  {journal} {Chin. J. Astron. Astrophys. Suplement}\ }\textbf {\bibinfo
  {volume} {8}},\ \bibinfo {pages} {273} (\bibinfo {year} {2008})}\BibitemShut
  {NoStop}%
\bibitem [{\citenamefont {Casares}(2007)}]{Casares2007}%
  \BibitemOpen
  \bibfield  {author} {\bibinfo {author} {\bibfnamefont {J.}~\bibnamefont
  {Casares}},\ }\href@noop {} {\bibfield  {journal} {\bibinfo  {journal}
  {Proceedings IAU Symposium}\ }\textbf {\bibinfo {volume} {2}},\ \bibinfo
  {pages} {3} (\bibinfo {year} {2007})}\BibitemShut {NoStop}%
\bibitem [{Note2()}]{Note2}%
  \BibitemOpen
  \bibinfo {note} {The numerical values of the functions $\protect \mathaccentV
  {dot}05F{\protect \overline {n}}(\protect \mathcal {M},z)$ (for each of the
  four models mentioned at the beginning of Section \ref {sec:models}) were
  kindly provided by A. Sesana and M. Volonteri in a private
  communication}\BibitemShut {NoStop}%
\bibitem [{\citenamefont {Regimbau}(2011)}]{Regimbau2011}%
  \BibitemOpen
  \bibfield  {author} {\bibinfo {author} {\bibfnamefont {T.}~\bibnamefont
  {Regimbau}},\ }\href@noop {} {\bibfield  {journal} {\bibinfo  {journal}
  {Research in Astron. Astrophys.}\ }\textbf {\bibinfo {volume} {11}},\
  \bibinfo {pages} {369} (\bibinfo {year} {2011})}\BibitemShut {NoStop}%
\bibitem [{\citenamefont {Nelemans}\ \emph {et~al.}(2001)\citenamefont
  {Nelemans}, \citenamefont {Yungelson},\ and\ \citenamefont {{Portegies
  Zwart}}}]{NelemansEtAl2001}%
  \BibitemOpen
  \bibfield  {author} {\bibinfo {author} {\bibfnamefont {G.}~\bibnamefont
  {Nelemans}}, \bibinfo {author} {\bibfnamefont {L.~R.}\ \bibnamefont
  {Yungelson}}, \ and\ \bibinfo {author} {\bibfnamefont {S.~F.}\ \bibnamefont
  {{Portegies Zwart}}},\ }\href@noop {} {\bibfield  {journal} {\bibinfo
  {journal} {A\&A}\ }\textbf {\bibinfo {volume} {375}},\ \bibinfo {pages} {890}
  (\bibinfo {year} {2001})}\BibitemShut {NoStop}%
\bibitem [{\citenamefont {Ruiter}\ \emph {et~al.}(2010)\citenamefont {Ruiter}
  \emph {et~al.}}]{RuiterEtAl2010}%
  \BibitemOpen
  \bibfield  {author} {\bibinfo {author} {\bibfnamefont {A.~J.}\ \bibnamefont
  {Ruiter}} \emph {et~al.},\ }\href@noop {} {\bibfield  {journal} {\bibinfo
  {journal} {ApJ}\ }\textbf {\bibinfo {volume} {717}},\ \bibinfo {pages} {1006}
  (\bibinfo {year} {2010})}\BibitemShut {NoStop}%
\bibitem [{\citenamefont {Kosenko}\ and\ \citenamefont
  {Postnov}(1998)}]{KosenkoPostnov1998}%
  \BibitemOpen
  \bibfield  {author} {\bibinfo {author} {\bibfnamefont {D.}~\bibnamefont
  {Kosenko}}\ and\ \bibinfo {author} {\bibfnamefont {K.}~\bibnamefont
  {Postnov}},\ }\href@noop {} {\bibfield  {journal} {\bibinfo  {journal}
  {A\&A}\ }\textbf {\bibinfo {volume} {336}},\ \bibinfo {pages} {786} (\bibinfo
  {year} {1998})}\BibitemShut {NoStop}%
\bibitem [{\citenamefont {Belczynski}\ \emph {et~al.}(2010)\citenamefont
  {Belczynski} \emph {et~al.}}]{BelczynskiEtAl2010c}%
  \BibitemOpen
  \bibfield  {author} {\bibinfo {author} {\bibfnamefont {K.}~\bibnamefont
  {Belczynski}} \emph {et~al.},\ }\href@noop {} {\bibfield  {journal} {\bibinfo
   {journal} {ApJ Letters}\ }\textbf {\bibinfo {volume} {715}},\ \bibinfo
  {pages} {L138} (\bibinfo {year} {2010})}\BibitemShut {NoStop}%
\bibitem [{\citenamefont {Willems}\ \emph {et~al.}(2005)\citenamefont {Willems}
  \emph {et~al.}}]{WillemsEtAl2005}%
  \BibitemOpen
  \bibfield  {author} {\bibinfo {author} {\bibfnamefont {B.}~\bibnamefont
  {Willems}} \emph {et~al.},\ }\href@noop {} {\bibfield  {journal} {\bibinfo
  {journal} {ApJ}\ }\textbf {\bibinfo {volume} {625}},\ \bibinfo {pages} {324}
  (\bibinfo {year} {2005})}\BibitemShut {NoStop}%
\bibitem [{\citenamefont {Willems}\ \emph {et~al.}(2006)\citenamefont {Willems}
  \emph {et~al.}}]{WillemsEtAl2006}%
  \BibitemOpen
  \bibfield  {author} {\bibinfo {author} {\bibfnamefont {B.}~\bibnamefont
  {Willems}} \emph {et~al.},\ }\href@noop {} {\bibfield  {journal} {\bibinfo
  {journal} {Phys. Rev. D}\ }\textbf {\bibinfo {volume} {74}},\ \bibinfo
  {pages} {043003} (\bibinfo {year} {2006})}\BibitemShut {NoStop}%
\bibitem [{\citenamefont {Fregeau}\ \emph {et~al.}(2009)\citenamefont {Fregeau}
  \emph {et~al.}}]{FregeauEtAl2009}%
  \BibitemOpen
  \bibfield  {author} {\bibinfo {author} {\bibfnamefont {J.~M.}\ \bibnamefont
  {Fregeau}} \emph {et~al.},\ }\href@noop {} {\bibfield  {journal} {\bibinfo
  {journal} {ApJ}\ }\textbf {\bibinfo {volume} {695}},\ \bibinfo {pages} {L20}
  (\bibinfo {year} {2009})}\BibitemShut {NoStop}%
\bibitem [{\citenamefont {Cornish}(2003)}]{Cornish2003}%
  \BibitemOpen
  \bibfield  {author} {\bibinfo {author} {\bibfnamefont {N.~J.}\ \bibnamefont
  {Cornish}},\ }\href@noop {} {\bibfield  {journal} {\bibinfo  {journal}
  {arXiv: gr-qc/0304020v1}\ } (\bibinfo {year} {2003})}\BibitemShut {NoStop}%
\bibitem [{\citenamefont {Coward}\ and\ \citenamefont
  {Regimbau}(2006)}]{CowardRegimbau2006}%
  \BibitemOpen
  \bibfield  {author} {\bibinfo {author} {\bibfnamefont {D.}~\bibnamefont
  {Coward}}\ and\ \bibinfo {author} {\bibfnamefont {T.}~\bibnamefont
  {Regimbau}},\ }\href@noop {} {\bibfield  {journal} {\bibinfo  {journal} {New
  Astronomy Reviews}\ }\textbf {\bibinfo {volume} {50}},\ \bibinfo {pages}
  {461} (\bibinfo {year} {2006})}\BibitemShut {NoStop}%
\bibitem [{\citenamefont {Regimbau}\ and\ \citenamefont
  {Chauvineau}(2007)}]{RegimbauChauvineau2007}%
  \BibitemOpen
  \bibfield  {author} {\bibinfo {author} {\bibfnamefont {T.}~\bibnamefont
  {Regimbau}}\ and\ \bibinfo {author} {\bibfnamefont {B.}~\bibnamefont
  {Chauvineau}},\ }\href@noop {} {\bibfield  {journal} {\bibinfo  {journal}
  {Clas. Quantum Grav.}\ }\textbf {\bibinfo {volume} {24}},\ \bibinfo {pages}
  {S627} (\bibinfo {year} {2007})}\BibitemShut {NoStop}%
\end{thebibliography}%

\end{document}